\documentclass[aps]{revtex4} 

\usepackage{amsmath,amssymb,bm,epsfig}
\usepackage{color}
\usepackage{natbib}
\usepackage{hyperref} 
\usepackage{ulem}
\usepackage{graphicx}
\usepackage{xcolor,colortbl}
\usepackage{pifont}

\def\nn{\nonumber}

\begin{document}
\title{Thermal effects on $\rho$ meson properties in an external magnetic field}
\author{Snigdha Ghosh$^{a,c}$}
\email{snigdha.physics@gmail.com, snigdha@vecc.gov.in}
\author{Arghya Mukherjee$^{b,c}$}
\email{arghya.mukherjee@saha.ac.in}
\author{Mahatsab Mandal$^d$}
\email{mahatsab@gmail.com}
\author{Sourav Sarkar$^{a,c}$}
\email{sourav@vecc.gov.in}
\author{ Pradip Roy$^{b,c}$}
\email{pradipk.roy@saha.ac.in}
\affiliation{$^a$Variable Energy Cyclotron Centre, 1/AF Bidhannagar, Kolkata 700 064, India}
\affiliation{$^b$Saha Institute of Nuclear Physics, 1/AF Bidhannagar, Kolkata - 700064, India}
\affiliation{$^c$Homi Bhabha National Institute, Training School Complex, Anushaktinagar, Mumbai - 400085, India}
\affiliation{$^d$Government General Degree College at Kalna-I, Burdwan, West Bengal - 713405, India}
%+++++++++++++++++++++++++++++++++++++++++++++++++++++++++++++++++++++++++++++++++++++++++++++++++++++++++++++++++++++++++++++++++
\begin{abstract}
A detailed study of the analytic structure of 1-loop self energy graphs for neutral and charged $\rho$ mesons is presented 
at finite temperature and arbitrary magnetic field using the real time formalism of thermal field theory. 
The imaginary part of the self energy is obtained from the discontinuities of these graphs across the Unitary and Landau cuts, which 
is seen to be different for $\rho^0$ and $\rho^\pm$. 
The magnetic field dependent vacuum contribution to the real part of the self energy, which is usually ignored,  
is found to be appreciable.
A significant effect of temperature and magnetic field is seen in the self energy, spectral function, 
effective mass and dispersion relation of $\rho^0$ as well as of $\rho^\pm$ relative to its trivial Landau shift.
However, for charged $\rho$ mesons, on account of the dominance of the Landau term, the effective mass appears to be independent 
of temperature.	The trivial coupling of magnetic moment of $\rho^\pm$ with external magnetic field, when incorporated in the calculation, 
makes the $\rho^\pm$ to condense at high magnetic field.
\end{abstract}
\maketitle
%+++++++++++++++++++++++++++++++++++++++++++++++++++++++++++++++++++++++++++++++++++++++++++++++++++++++++++++++++++++++++++++++++
%
\section{Introduction}
Recent researches in Quantum Chromodynamics(QCD) in presence of  magnetic background have revealed many remarkable 
properties of strong interaction \cite{Kharzeev:2012ph}. From 
anomalous Chiral Magnetic Effect (CME), Chiral Vortical Effect to Magnetic Catalysis(MC), Inverse Magnetic Catalysis(IMC) and vacuum superconductivity, these 
non-trivial interplay between the topology, symmetry and anomaly structure
\cite{Kharzeev:2007tn,Kharzeev:2007jp,Fukushima:2008xe,Gusynin:1995nb, Gusynin:1999pq,Bali:2011qj,Chernodub:2010qx,Chernodub:2012tf} have enriched the fundamental aspects of QCD to a great extent.  On one hand, noticeable influence on the 
strongly interacting sector can be achieved only when the background magnetic field is strong enough to be comparable to QCD scale i.e $eB\approx m_\pi^2$, on the other hand, 
non-central heavy-ion collisions in Relativistic Heavy-Ion Collider(RHIC) and Large Hadron Collider(LHC) can generate magnetic fields of $eB\approx15 m_\pi^2$
\cite{Skokov:2009qp}. Apart from
its own theoretical intricacies, this promising platform for experimental manifestations has been one of the key reasons for ensuing  interests in this field of
research. Moreover, a similar environment inside the core of  magnetars adds to  its astrophysical and 
cosmological importance
 \cite{Duncan:1992hi,Ferrer:2005vd,Ferrer:2006vw,Ferrer:2007iw, Fukushima:2007fc,Feng:2009vt,Fayazbakhsh:2010gc, Fayazbakhsh:2010bh}.

A large amount of progress has been achieved in solving the so called puzzle of MC and IMC using the effective models, most of which are focused on considering
magnetic field dependent coupling constants or other magnetic field dependent parameters of the model (see for example \cite{Andersen_review}). One of the important 
methods for extracting the information of  $eB$ dependencies of the chiral  phase transition parameters is to study the modifications of hadronic, in particular 
mesonic properties 
in presence of medium/density along with external magnetic field since they are  more directly related to the chiral phase transition \cite{wangprd86}. 
In this paper we mainly focus  on the  temperature modifications of $\rho$ meson properties in presence of static homogeneous magnetic background. 
The study of the $\rho$ meson properties like the  effective mass and dispersion relations are important in the context of magnetic field induced vacuum 
superconductivity~\cite{Chernodub:2012tf,Vafa:1983tf,Hidaka:2012mz,Chernodub:2012zx, Li:2013aa,Chernodub:2013uja,Liu:2014uwa,Kawaguchi:2015gpt}. 
It 
should be mentioned here that  the dilepton production rate in heavy-ion collisions is directly proportional to the in-medium $\rho$ spectral function and  is well studied
at vanishing magnetic field in Ref.~\cite{Rapp:1999ej,Alam:1999sc,Mallik:2016anp,Ghosh:2009bt}. However, the existence of such high external magnetic field 
in  non central collisions does affect the 
spectral function of $\rho$ \cite{Chernodub:2010qx,rhopaper}. Thus the detailed study of the in-medium spectral 
properties of $\rho$ meson in presence of  $eB$ may prove to be indispensable for analyzing  the results of heavy-ion collision experiments.  

Most of the calculations of one loop self energy functions at finite temperature under external magnetic field present in the
literature employ either strong or weak magnetic field approximation \cite{Navarro:2010eu,Ayala:2004dx,Bandyopadhyay:2016fyd}. 
A few of them have relaxed this approximation and calculations are presented
for arbitrary value of magnetic field \cite{DOlivo:2002omk,Hattori:2012je}. In the later case, 
even though the full Schwinger propagator for the loop particles
is considered,  the real part of the self energy neglects the magnetic field dependent vacuum contribution. 
In this work we have taken the full Schwinger propagator for the loop particles and do not make any approximations 
on the magnitude of the magnetic field. We have also included the magnetic field dependent vacuum contributions to 
the real part of the self energy.
In addition to these novelties, we have explicitly worked out the analytic structure of the 
self energy at finite temperature and non-zero magnetic field which to the best of our knowledge has not been discussed 
elsewhere. Discontinuities of the self energy graphs across the Unitary and Landau cut are seen to be different for the charged 
and neutral $\rho$ mesons.
It should be noted here that, if the external boson is charged, its momentum transverse to the external magnetic field is 
never zero due to Landau quantization. To show the importance of the loop correction of $\rho^\pm$ at finite temperature and 
non-zero magnetic field, we first present various properties of $\rho^\pm$ by neglecting this trivial Landau shift. 
Later, we also show results incorporating the Landau quantization of transverse momenta as well as including the trivial 
coupling of the magnetic moment of $\rho^\pm$ with external magnetic field. In this case, we will show that the loop-correction 
to effective mass is subleading at all temperatures.

A few comments on the applicability of our calculation are in order. All the results in this paper, are presented 
for temperatures in the range 100 MeV $\le T \le $160 MeV where, the degrees of freedom of strongly interacting matter 
are basically hadrons. However, if there exists a strong magnetic field (order of typical QCD scale), then the system may undergo 
a phase transition even in this temperature range (so called IMC effect~\cite{Bali:2011qj}) to the deconfined phase where the degrees of freedom are quarks and gluons. In that case, the hadronic description will not be applicable. However, in this work we have not considered these possibilities.
%complications and assume hadronic description throughout the entire range of magnetic field.

The article is organized as follows.
In Sec.~\ref{sec.vac_self} the vacuum self energy of $\rho$ is discussed followed by the evaluation of the in-medium $\rho$ self-energy
at zero magnetic field in Sec.~\ref{sec.med_self}. Next in Sec.~\ref{sec.med.eb_self}, the in-medium self energy at 
non-zero external magnetic field is presented. Sec.~\ref{sec.analytic} is devoted to the discussion of the analytic structure of the in-medium  self energy functions
in a magnetic field.
In Sec.~\ref{sec.results}, the numerical results are shown and discussed. Finally we summarize and conclude in Sec.~\ref{sec.summary}.
Some of the relevant calculational details are provided in the Appendix.

% %##################################################################################################################

\section{$\rho$ self energy in the vacuum}\label{sec.vac_self}
The lowest order (LO) Lagrangian for effective $\rho\pi\pi$ interaction is given by~\cite{Ghosh:2009bt}
\begin{eqnarray}
\mathcal{L}_{int} = -g_{\rho\pi\pi}\partial_\mu\vec{\rho}_\nu\cdot\partial^\mu\vec{\pi}\times\partial^\nu\vec{\pi}, \nn
\end{eqnarray} 
with the effective coupling constant $g_{\rho\pi\pi}$ = 20.72 GeV$^{-2}$, which is fixed from the vacuum 
$\rho\rightarrow\pi\pi$ decay width $\Gamma_{\rho\rightarrow\pi\pi}$ = 150 MeV. 

\begin{figure}[h]
\begin{center}
\includegraphics[angle=0, scale=0.32]{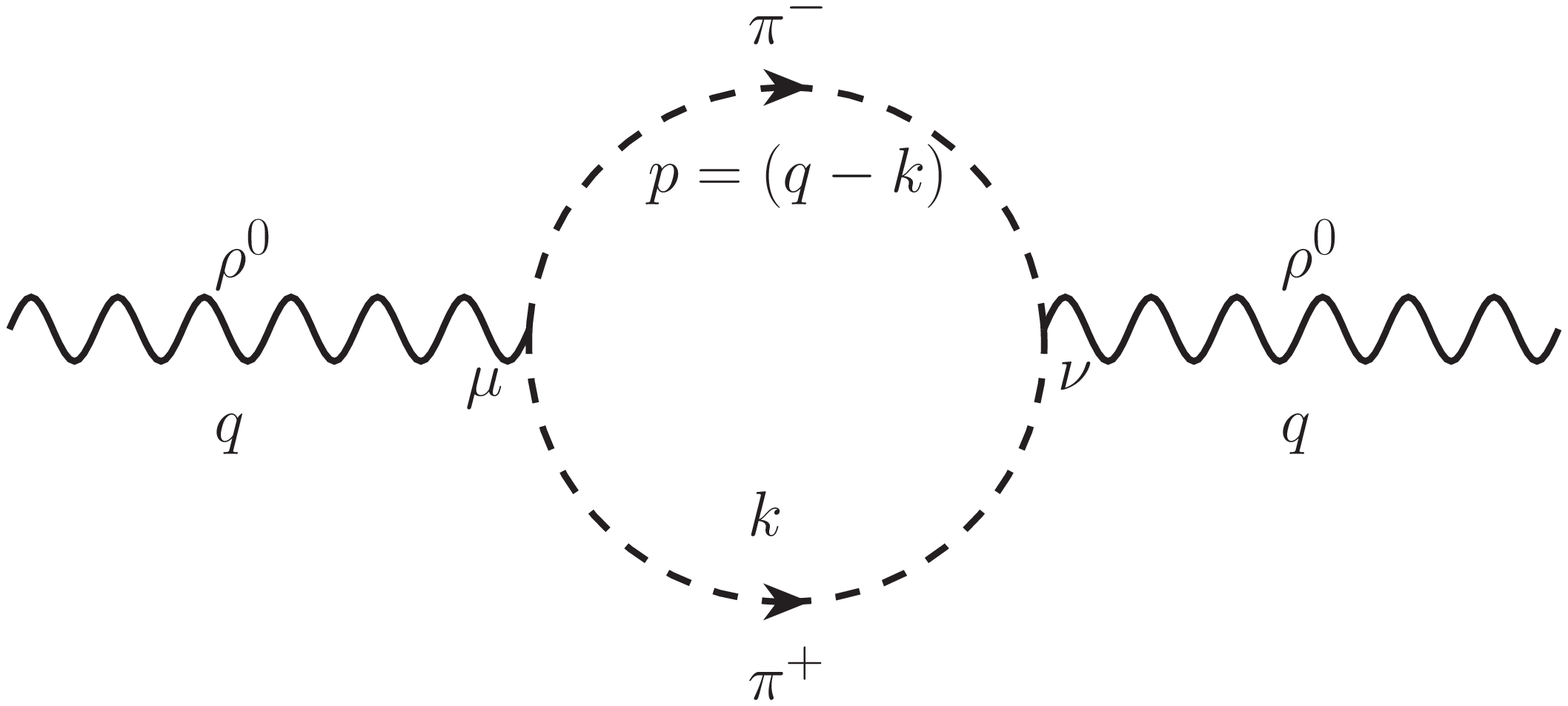} ~~~~~ \includegraphics[angle=0, scale=0.32]{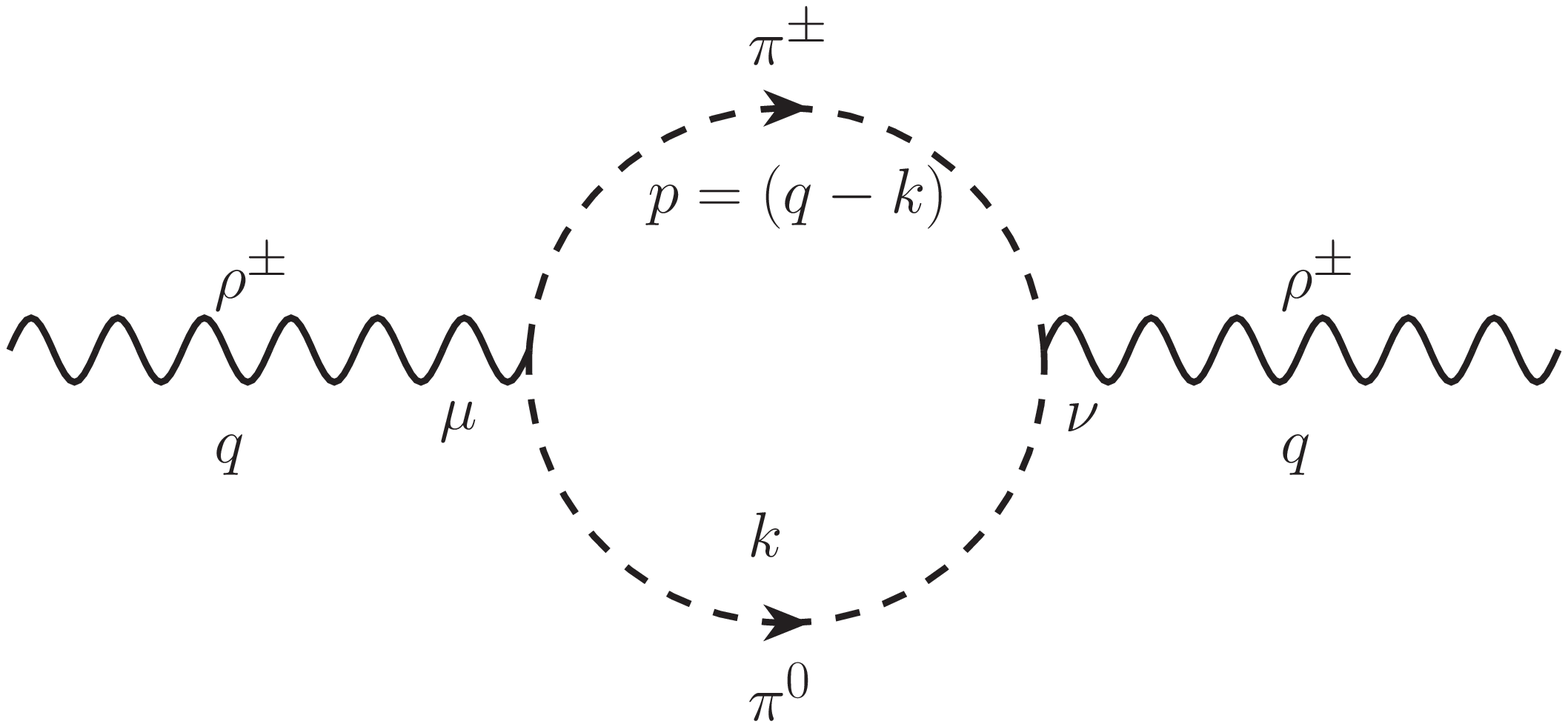}
\end{center}
\caption{Feynman diagrams for $\rho$ self energy.}
\label{fig.feynman}
\end{figure}

Using $\mathcal{L}_{int}$, the vacuum self energies of $\rho^0$ and $\rho^{\pm}$ for Feynman diagrams shown in Fig.~\ref{fig.feynman}, 
can be written as
\begin{eqnarray}
\left(\Pi^{\mu\nu}_0(q)\right)_{vac} &=& i\int\frac{d^4k}{(2\pi)^4}\mathcal{N}^{\mu\nu}(q,k)\Delta_\pm(k)\Delta_\pm(p) \label{eq.vacself.rho0} \\
\left(\Pi^{\mu\nu}_\pm(q)\right)_{vac} &=& i\int\frac{d^4k}{(2\pi)^4}\mathcal{N}^{\mu\nu}(q,k)\Delta_0(k)\Delta_\pm(p) \label{eq.vacself.rhopm}
\end{eqnarray} 
respectively. Here $\Delta_0(k)=\left(\frac{-1}{k^2-m_0^2+i\epsilon}\right)$ 
and $\Delta_\pm(k)=\left(\frac{-1}{k^2-m_\pm^2+i\epsilon}\right)$ are the vacuum Feynman propagators 
of $\pi^0$ and $\pi^{\pm}$ with masses $m_0$ and $m_\pm$ respectively and $\mathcal{N}^{\mu\nu}(q,k)$ is given by,
\begin{eqnarray}
\mathcal{N}^{\mu\nu}(q,k)=g_{\rho\pi\pi}^2\left[q^4k^\mu k^\nu + (q.k)^2q^\mu q^\nu-q^2(q.k)(q^\mu k^\nu+k^\mu q^\nu)\right], \nn
\end{eqnarray}
which contains the factors coming from the interaction vertices. The momentum integrations in 
Eqs.~(\ref{eq.vacself.rho0}) and (\ref{eq.vacself.rhopm}) can be evaluated using standard Feynman parametrization 
followed by dimensional regularization. If we take $m_0=m_\pm=m_\pi$, then the vacuum self energies of 
$\rho^0$ and $\rho^\pm$ are identical and is given by,
\begin{eqnarray}
\Pi^{\mu\nu}_{vac}(q) = \left(\frac{q^2g_{\rho\pi\pi}^2}{32\pi^2}\right)\left( q^2g^{\mu\nu}-q^\mu q^\nu \right)\int\limits_{0}^{1}dx \Delta\left(\ln\frac{\Delta}{\mu_0}-1\right) \label{eq.vacself.rho},
\end{eqnarray}
where $\Delta = m^2-x(1-x)q^2-i\epsilon$ and $\mu_0$ is a scale of dimension GeV$^2$. The metric 
tensor in this work is taken as $g^{\mu\nu}=diag(1,-1,-1,-1)$.
%+++++++++++++++++++++++++++++++++++++++++++++++++++++++++++++++++++++++++++++++++++++++++++++++++++++++++++++++++++++++++++++++++
\section{$\rho$ self energy in the medium}\label{sec.med_self}
In the real time formalism of thermal field theory, the thermal propagators as well as the self energies 
become 2$\times$2 matrices~\cite{bellac,Mallik:2016anp}. However, they can be diagonalized in terms of a single analytic function
 which is related to any one component of the corresponding 2$\times$2 matrix, say the 11-component. 
The 11-component of the $\pi^0$ and $\pi^\pm$ thermal propagators are given by,
\begin{eqnarray}
D_0^{11}(k) &=& \Delta_0(k) + 2i\eta^k\text{Im}~\Delta_0(k) \label{eq.pi0.11.propagator} \\
D_\pm^{11}(k) &=& \Delta_\pm(k) + 2i\eta^k\text{Im}~\Delta_\pm(k) \label{eq.pipm.11.propagator}
\end{eqnarray}
where, $\eta^k=\left[ e^{k.u/T}-1 \right]^{-1}$ is the Bose-Einstein distribution function of the pions 
with $u^\mu$ being the medium four-velocity. In local rest frame of the medium, $u^\mu\equiv(1,\vec{0})$.
The complete in-medium propagator matrix $\bf{D}^{\mu\nu}$ satisfies the following Dyson-Schwinger equation,
\begin{eqnarray}
\bf{D}^{\mu\nu} = \bf{\Delta}^{\mu\nu} - \bf{\Delta}^{\mu\alpha}\bf{\Pi}_{\alpha\beta} \bf{D}^{\beta\nu} \label{eq.dyson}
\end{eqnarray}
where, $\bf{\Delta}^{\mu\nu}$ is the free thermal vector propagator matrix and $\bf{\Pi}_{\alpha\beta}$ is 
the 1-loop thermal self energy matrix. Each of the quantities in Eq.~(\ref{eq.dyson}) can be expressed in diagonal form in terms of 
analytic functions denoted by a bar, so that it can be written as,
\begin{eqnarray}
\bar{D}^{\mu\nu} = \bar{\Delta}^{\mu\nu} - \bar{\Delta}^{\mu\alpha}\bar{\Pi}_{\alpha\beta} \bar{D}^{\beta\nu}. \label{eq.dyson.bar}
\end{eqnarray}
The self energy function $\bar{\Pi}_{\alpha\beta}$ is related to the 11-component of $\bf{\Pi}_{\alpha\beta}$ by the following relations,
\begin{eqnarray}
\text{Re}~\bar{\Pi}_{\alpha\beta}(q) &=& \text{Re}~\Pi_{\alpha\beta}^{11}(q) \label{eq.real.11-bar}\\
\text{Im}~\bar{\Pi}_{\alpha\beta}(q) &=& \epsilon(q^0)\tanh\left(\frac{q^0}{2T}\right)\text{Im}~\Pi_{\alpha\beta}^{11} \label{eq.imag.11-bar}
\end{eqnarray}
where, $\epsilon(q^0)=\Theta\left(q^0\right)-\Theta\left(-q^0\right)$ is the sign function. 
In order to obtain the 11-component of the $\rho^0$ and $\rho^\pm$ self energies, one has to replace the vacuum $\pi^0$ and $\pi^\pm$ propagators in Eq.~(\ref{eq.vacself.rho0}) and (\ref{eq.vacself.rhopm}) by their corresponding 11-components as given in Eqs.~(\ref{eq.pi0.11.propagator}) and (\ref{eq.pipm.11.propagator}),
\begin{eqnarray}
\left(\Pi^{\mu\nu}_0(q)\right)^{11} &=& i\int\frac{d^4k}{(2\pi)^4}\mathcal{N}^{\mu\nu}(q,k)D_\pm^{11}(k)D_\pm^{11}(p) \label{eq.therself.rho0} \\
\left(\Pi^{\mu\nu}_\pm(q)\right)^{11} &=& i\int\frac{d^4k}{(2\pi)^4}\mathcal{N}^{\mu\nu}(q,k)D_0^{11}(k)D_\pm^{11}(p) \label{eq.therelf.rhopm}.
\end{eqnarray} 
Performing the $k^0$ integral and using Eqs.~(\ref{eq.real.11-bar}) and (\ref{eq.imag.11-bar}) we get the thermal
self energy functions for $\rho^0$ and $\rho^\pm$ which are identical if we take $m_0=m_\pm=m_\pi$,
\begin{eqnarray}
\text{Re}~\bar{\Pi}^{\mu\nu}(q) &=& \text{Re}~\Pi^{\mu\nu}_{vac}(q) + \int\frac{d^3k}{(2\pi)^3}\frac{1}{2\omega_k\omega_p}\mathcal{P}\left[
\left(\frac{\eta^k\omega_p \mathcal{N}^{\mu\nu}(k^0=\omega_k)}{(q_0-\omega_k)^2-\omega_p^2}\right) +
\left(\frac{\eta^k\omega_p \mathcal{N}^{\mu\nu}(k^0=-\omega_k)}{(q_0+\omega_k)^2-\omega_p^2}\right)+\right. \nn\\ 
&& \hspace{5cm} \left.\left(\frac{\eta^p\omega_k \mathcal{N}^{\mu\nu}(k^0=q_0-\omega_p)}{(q_0-\omega_p)^2-\omega_k^2}\right) +
\left(\frac{\eta^p\omega_k \mathcal{N}^{\mu\nu}(k^0=q_0+\omega_p)}{(q_0+\omega_p)^2-\omega_k^2}\right) \right] \nn
\end{eqnarray}
and
\begin{eqnarray}
\text{Im}~\bar{\Pi}^{\mu\nu}(q) &=& -\pi\epsilon(q_0)\int\frac{d^3k}{(2\pi)^3}\frac{1}{4\omega_k\omega_p}
\left[\frac{}{}\mathcal{N}^{\mu\nu}(k^0=\omega_k)\left\{(1+\eta^k+\eta^p)\delta(q_0-\omega_k-\omega_p)+(-\eta^k+\eta^p) \delta(q_0-\omega_k+\omega_p)\frac{}{}\right\} \right. \nonumber \\
&&\hspace{2cm} \left. +~\mathcal{N}^{\mu\nu}(k^0=-\omega_k)\left\{(-1-\eta^k-\eta^p)\delta(q_0+\omega_k+\omega_p)+
(\eta^k-\eta^p)\delta(q_0+\omega_k-\omega_p)\frac{}{}\right\} \frac{}{} \right],
\label{eq.im.pibar.therm}
\end{eqnarray}
where, $\omega_k=\sqrt{\vec{k}^2+m^2_\pi}$, $\omega_p=\sqrt{\vec{p}^2+m^2_\pi} =\sqrt{(\vec{q}-\vec{k})^2+m^2_\pi}$ and 
$\mathcal{P}$ denotes the Cauchy principal value integration.
%+++++++++++++++++++++++++++++++++++++++++++++++++++++++++++++++++++++++++++++++++++++++++++++++++++++++++++++++++++++++++++++++++
\section{$\rho$ self energy in the medium under external magnetic field}\label{sec.med.eb_self}
In presence of external magnetic field (in addition to finite temperature), the $\pi^0$ propagator remains unaffected whereas 
the 11-component of $\pi^{\pm}$ propagator becomes~\cite{DOlivo:2002omk},
\begin{eqnarray}
D_B^{11}(k) &=& \Delta_B(k) + 2i\eta^k\text{Im}~\Delta_B(k), \nn
\end{eqnarray} 
where, $\Delta_B(k)$ is the Schwinger proper time propagator for a charged scalar field~\cite{Ayala:2004dx,Schwinger:1951nm} in momentum
spece,
\begin{eqnarray}
\Delta_B(k) = i\int\limits_{0}^{\infty}\frac{ds}{\cos(eBs)}\exp\left[ is\left( k_\parallel^2+k_\perp^2\frac{\tan(eBs)}{eBs}-m^2_\pi+i\epsilon \right) \right]~.\label{eq.schwinger}
\end{eqnarray}
 The corresponding coordinate space propagator 
contains a phase factor which is not translationally invariant. 
However with a suitable choice of the gauge, the phase factor 
can be removed~\cite{Ayala:2015qwa}, and one can work with the momentum space propagator.
In Eq.~(\ref{eq.schwinger}), $e=|e|$ is the absolute electronic charge; the external magnetic field is taken along 
the +ve z-direction ($\vec{B}=B\hat{z}$) and correspondingly any four-vector $a$ is decomposed as $a=(a_\parallel+a_\perp)$, where
$a_\parallel^\mu\equiv(a^0,0,0,a_z)$ and $a_\perp^\mu\equiv(0,a_x,a_y,0)$. The metric tensor $g^{\mu\nu}$ is also decomposed as 
$g^{\mu\nu}=g^{\mu\nu}_\parallel+g^{\mu\nu}_\perp$, where $g^{\mu\nu}_\parallel = diag(1,0,0,-1)$ and
$g^{\mu\nu}_\perp=diag(0,-1,-1,0)$. Performing the proper time integration in Eq.~(\ref{eq.schwinger}), one gets,
\begin{eqnarray}
\Delta_B(k) = \sum\limits_{l=0}^{\infty}\frac{-\phi_l(\alpha_k)}{k_\parallel^2-m_l^2+i\epsilon}~, \nn
\end{eqnarray}
where 
\begin{eqnarray}
m_l = \sqrt{m_\pi^2+(2l+1)eB}, \label{eq.ml}
\end{eqnarray}
$\phi_l(\alpha_k) = 2(-1)^lL_l(2\alpha_k)e^{-\alpha_k}$, 
$\alpha_k=-k_\perp^2/eB$ and $L_l(x)$ is the Laguerre polynomial of order $l$ with $L_{-1}(x)=0$. 
Replacing $D_\pm^{11}\rightarrow D_B^{11}$ in Eqs.~(\ref{eq.therself.rho0}) and (\ref{eq.therelf.rhopm}), 
and following Eqs.~(\ref{eq.real.11-bar}) and (\ref{eq.imag.11-bar}), 
one gets the $\rho^0$ and $\rho^\pm$ self energy functions at finite temperature in external magnetic field as,
\begin{eqnarray}
\bar{\Pi}^{\mu\nu}_0 &=& (\Pi^{\mu\nu}_0)_B + (\Pi^{\mu\nu}_0)_{BT} \label{eq.self0.eb.t}\\
\bar{\Pi}^{\mu\nu}_\pm &=& (\Pi^{\mu\nu}_\pm)_B + (\Pi^{\mu\nu}_\pm)_{BT}~. \label{eq.selfpm.eb.t}
\end{eqnarray}
In Eqs.~(\ref{eq.self0.eb.t}) and (\ref{eq.selfpm.eb.t}), subscript ``$B$" and ``$BT$" denote purely magnetic field dependent and 
both magnetic field as well as temperature dependent contributions respectively. 

It is well known that the momenta of charged bosons transverse to the direction of external magnetic field
are Landau quantized so that $q_\perp^2=-(2n+1)eB$ with $n$ = 0, 1, 2 .... . So for $\rho^\pm$, we present results for arbitrary four-momentum 
$q^\mu\equiv(q^0,q_x,q_y,q_z)$, whereas for $\rho^0$ we take for simplicity $q^\mu\equiv(q^0,0,0,q_z)$.
The calculations of the real parts of the ``$B$" terms i.e. the magnetic field dependent vacuum contributions 
are rather involved for which some relevant intermediate steps are provided in the Appendices (\ref{appendix.a}) and (\ref{appendix.b}). 
In comparison, the calculations of the real parts of the ``$BT$" terms as well as the imaginary parts of the self energies 
are relatively straight forward and similar to the $eB=0$ case. We summarize the explicit forms of different terms 
in Eqs.~(\ref{eq.self0.eb.t}) and (\ref{eq.selfpm.eb.t}) below. The expressions for the real parts of $\rho^0$ self 
energy function are
\begin{eqnarray}
\text{Re}(\Pi^{\mu\nu}_0(q^0,q_z))_{B} &=& \text{Re}~\Pi^{\mu\nu}_{vac} + \left(\frac{g_{\rho\pi\pi}^2q_\parallel^2}{32\pi^2}\right)  \int\limits_{0}^{1}dx
\left[ \left(q_\parallel^2g_\parallel^{\mu\nu}-q_\parallel^\mu q_\parallel^\nu\right)\text{Re}
\left( 2eB\ln\Gamma\left(\frac{\Delta}{2eB}+\frac{1}{2}\right)-\Delta\ln\frac{\Delta}{2eB}+\Delta \right)
\right. \nn \\
&&\left.+~q_\parallel^2g_\perp^{\mu\nu} \text{Re}\left( \frac{\Delta}{2}\Psi\left(\frac{\Delta}{2eB}+\frac{1}{2}\right) 
+ \frac{1}{2}(\Delta+(2x-1)eB)\Psi\left(\frac{\Delta}{2eB}+\frac{1}{2}+x\right) - \Delta\ln\frac{\Delta}{2eB}\right)
 \right] \label{eq.re.pibar0.eb}
\end{eqnarray}
\begin{eqnarray}
\text{Re}(\Pi^{\mu\nu}_0(q^0,q_z))_{BT} &=& \sum\limits_{l=0}^{\infty}\sum\limits_{n=0}^{\infty}\int\limits_{-\infty}^{\infty}
\frac{dk_z}{2\pi}\frac{1}{2\omega_k^l\omega_p^n}\mathcal{P}\left[
\left(\frac{\eta^k_l\omega_p^n \mathcal{N}_{nl}^{\mu\nu}(k^0=\omega_k^l)}{(q_0-\omega_k^l)^2-(\omega_p^n)^2}\right) +
\left(\frac{\eta^k_l\omega_p^n \mathcal{N}_{nl}^{\mu\nu}(k^0=-\omega_k^l)}{(q_0+\omega_k^l)^2-(\omega_p^n)^2}\right)+\right. \nn\\ 
&&\hspace{3cm} \left.\left(\frac{\eta^p_n\omega_k^l \mathcal{N}_{nl}^{\mu\nu}(k^0=q_0-\omega_p^n)}{(q_0-\omega_p^n)^2-(\omega_k^l)^2}\right) +
\left(\frac{\eta^p_n\omega_k^l \mathcal{N}_{nl}^{\mu\nu}(k^0=q_0+\omega_p^n)}{(q_0+\omega_p^n)^2-(\omega_k^l)^2}\right) \right], \nn
\end{eqnarray} 
while the imaginary part is
\begin{eqnarray}
\text{Im}~\bar{\Pi}^{\mu\nu}_0(q^0,q_z)=&&-\pi\epsilon(q_0)\sum\limits_{l=0}^{\infty}\sum\limits_{n=0}^{\infty}
\int\limits_{-\infty}^{\infty}
\frac{dk_z}{2\pi} \frac{1}{4\omega_k^l\omega_p^n} \times \nn\\
&&\left[\mathcal{N}_{nl}^{\mu\nu}(k^0=\omega_k^l)\left\{(1+\eta^k_l+\eta^p_n)\delta(q_0-\omega_k^l-\omega_p^n)+(-\eta^k_l+\eta^p_n)
\delta(q_0-\omega_k^l+\omega_p^n)\frac{}{}\right\}\right.\nn\\ 
&&\left. +~\mathcal{N}_{nl}^{\mu\nu}(k^0=-\omega_k^l)\left\{(-1-\eta^k_l-\eta^p_n)\delta(q_0+\omega_k^l+\omega_p^n)+
(\eta^k_l-\eta^p_n)\delta(q_0+\omega_k^l-\omega_p^n)\frac{}{}\right\}\right]~.
\label{eq.im.pibar0.eb.t}
\end{eqnarray}
For the charged $\rho$ meson, the corresponding expressions are:
\begin{eqnarray}
\text{Re}(\Pi^{\mu\nu}_\pm(q^0,\vec{q}))_{B} &=& \text{Re}~\Pi^{\mu\nu}_{vac} + 
\left(\frac{g_{\rho\pi\pi}^2}{32\pi^2}\right)\int\limits_{0}^{1}\int\limits_{0}^{1}dxdz
z^{\Delta/m_\pi^2-1} \left[ 
z^{-y(1-x)q_\perp^2/m_\pi^2}\left(\frac{1}{\tilde{\zeta}}\right)\text{sech}\left(x\frac{eB}{m_\pi^2}\ln z\right) \right. \nn \\
&& \hspace{2cm} \left. \left( \frac{P^{\mu\nu}}{\ln z} + \frac{2Q^{\mu\nu}}{m_\pi^2} + \frac{R^{\mu\nu}}{\tilde{\zeta} m_\pi^2} \right) 
- z^{-x(1-x)q_\perp^2/m_\pi^2}\left(\frac{m_\pi^2q^2}{(\ln z)^2}\right)\left( q^2g^{\mu\nu}-q^\mu q^\nu \right) \right]
\label{eq.re.pibarpm.eb}
\end{eqnarray}
and 
\begin{eqnarray}
\text{Re}(\Pi^{\mu\nu}_\pm(q^0,\vec{q}))_{BT} &=& \sum\limits_{n=0}^{\infty}\int
\frac{d^3k}{(2\pi)^3}\frac{\phi_n(\alpha_p)}{2\omega_k\omega_p^n}\mathcal{P}\left[
\left(\frac{\eta^k\omega_p^n \mathcal{N}^{\mu\nu}(k^0=\omega_k)}{(q_0-\omega_k)^2-(\omega_p^n)^2}\right) +
\left(\frac{\eta^k\omega_p^n \mathcal{N}^{\mu\nu}(k^0=-\omega_k)}{(q_0+\omega_k)^2-(\omega_p^n)^2}\right)+\right. \nn\\ 
&& \hspace{3cm} \left.\left(\frac{\eta^p_n\omega_k \mathcal{N}^{\mu\nu}(k^0=q_0-\omega_p^n)}{(q_0-\omega_p^n)^2-(\omega_k)^2}\right) +
\left(\frac{\eta^p_n\omega_k \mathcal{N}^{\mu\nu}(k^0=q_0+\omega_p^n)}{(q_0+\omega_p^n)^2-(\omega_k)^2}\right) \right] \nn
\label{eq.re.pibarpm.BT}
\end{eqnarray} 
\begin{eqnarray}
\text{Im}~\bar{\Pi}^{\mu\nu}_\pm(q^0,\vec{q})&=&-\pi\epsilon(q_0)\sum\limits_{n=0}^{\infty}\int
\frac{d^3k}{(2\pi)^3} \frac{\phi_n(\alpha_p)}{4\omega_k\omega_p^n} \times \nn\\
&&\left[\mathcal{N}^{\mu\nu}(k^0=\omega_k)\left\{(1+\eta^k+\eta^p_n)\delta(q_0-\omega_k-\omega_p^n)+(-\eta^k+\eta^p_n)
\delta(q_0-\omega_k+\omega_p^n)\frac{}{}\right\} \right.\nn\\ 
&& \left.+~\mathcal{N}^{\mu\nu}(k^0=-\omega_k)\left\{(-1-\eta^k-\eta^p_n)\delta(q_0+\omega_k+\omega_p^n)+
(\eta^k-\eta^p_n)\delta(q_0+\omega_k-\omega_p^n)\frac{}{}\right\}\right]~,
\label{eq.im.pibarpm.eb.t}
\end{eqnarray}
where, $\Psi(z)$ is the digamma function, $\tilde{\zeta}=(1-x)\frac{1}{m_\pi^2}\ln z + 
\frac{1}{eB}\tanh\left(x\frac{eB}{m_\pi^2}\ln z\right)$, $y=\frac{1}{\tilde{\zeta} eB}\tanh\left(x\frac{eB}{m_\pi^2}\ln z\right)$, 
 $\omega_k^l = \sqrt{k_z^2+m_l^2}$, 
$\eta_k^l = \left[ e^{\omega_k^l/T}-1 \right]^{-1}$ and  
\begin{eqnarray}
\mathcal{N}^{\mu\nu}_{nl}(q_\parallel,k_\parallel)&=&\frac{g_{\rho\pi\pi}^2}{2}(-1)^{n+l}
\left(\frac{eB}{\pi}\right) \left[  \left\{q_\parallel^4k_\parallel^\mu k_\parallel^\nu +
(q_\parallel.k_\parallel)^2q_\parallel^\mu q_\parallel^\nu-q_\parallel^2(q_\parallel.k_\parallel)(q_\parallel^\mu
 k_\parallel^\nu+k_\parallel^\mu q_\parallel^\nu)\right\}\delta_{n,l} \right.\nn \\
&& \hspace{2cm} \left. -~\frac{eB}{4}q_\parallel^4g_\perp^{\mu\nu}\left\{ (2n+1)\delta_{n,l}-n\delta_{n-1,l}-(n+1)\delta_{n+1,l}\frac{}{} \right\}\right]. \label{eq.N_nl}
\end{eqnarray}
The expressions for $P^{\mu\nu}$, $Q^{\mu\nu}$ and $R^{\mu\nu}$ are provided in Appendix (\ref{appendix.b}).

%++++++++++++++++++++++++++++++++++++++++++++++++++++++++++++++++++++++++++++++++++++++++++++++++++++++++++++++++++++++
%
%
%
%
\section{Analytic Structure of the Imaginary Parts}\label{sec.analytic}
\begin{figure}[h]
\begin{center}
\includegraphics[angle=0, scale=0.6]{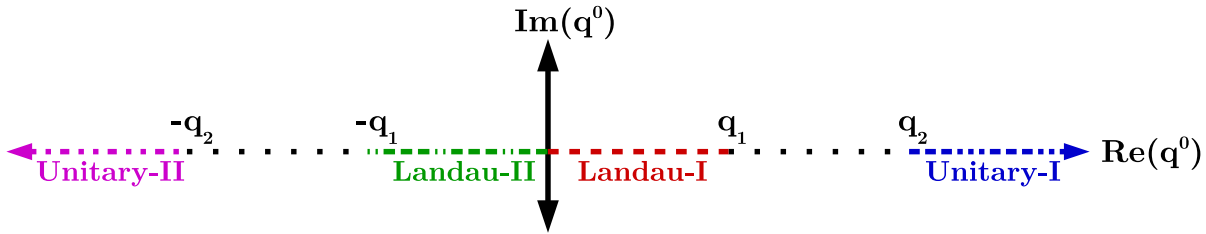}
\end{center}
\caption{Different branch cuts of the in-medium self energy function of 
	the $\rho$ at \textit{zero magnetic field} in the complex $q^0$ plane for a given $\vec{q}$. The points correspond to $q_1=|\vec{q}|$ and $q_2=\sqrt{\vec{q}^2+4m_\pi^2}$.}
\label{fig.anlytic1}
\end{figure}

Each of the imaginary parts of the self energy functions in Eqs.~(\ref{eq.im.pibar.therm}), 
(\ref{eq.im.pibar0.eb.t}) and (\ref{eq.im.pibarpm.eb.t})
contains four Dirac delta functions which will give rise to branch cuts of the self energy function in the complex $q^0$ plane, 
details of which are provided in Appendix~\ref{appendix.c}. 
Let us first discuss the analytic structure at finite temperature in absence of magnetic field. 
In Eq.~(\ref{eq.im.pibar.therm}), the first term containing 
$\delta(q^0-\omega_k-\omega_p)$ is non vanishing for $\sqrt{\vec{q}^2+4m_\pi^2}<q^0<\infty$ 
and we call this ``Unitary-I" cut. The second term containing $\delta(q^0-\omega_k+\omega_p)$ is non vanishing 
for $-|\vec{q}|<q^0<0$ and this is the ``Landau-II" cut. The third term containing $\delta(q^0+\omega_k+\omega_p)$ 
is non vanishing for $-\infty<q^0<-\sqrt{\vec{q}^2+4m_\pi^2}$ and we call this is ``Unitary-II" cut. Finally the fourth 
term containing $\delta(q^0+\omega_k-\omega_p)$ is non vanishing for $0<q^0<|\vec{q}|$ and this is the ``Landau-I" cut. 
These different cuts are shown in Fig.~\ref{fig.anlytic1} and correspond to different physical processes. 
We are interested in the physical kinematic region $q^0>0$ and $q^2>0$. In this region Unitary-I and Landau-I terms 
contribute. Unitary-I cut corresponds to the decay of a $\rho$ into two pions which can also happen in vacuum, 
whereas the Landau-I cut is purely a medium effect which corresponds to the absorption of a $\rho$ due to 
scattering with a $\pi$ producing a $\pi$ in the final state. If we take $\vec{q}=\vec{0}$, then the Landau 
contributions will be absent and we are left with only Unitary cut contributions. 

\begin{figure}[h]
\begin{center}
\includegraphics[angle=0, scale=0.6]{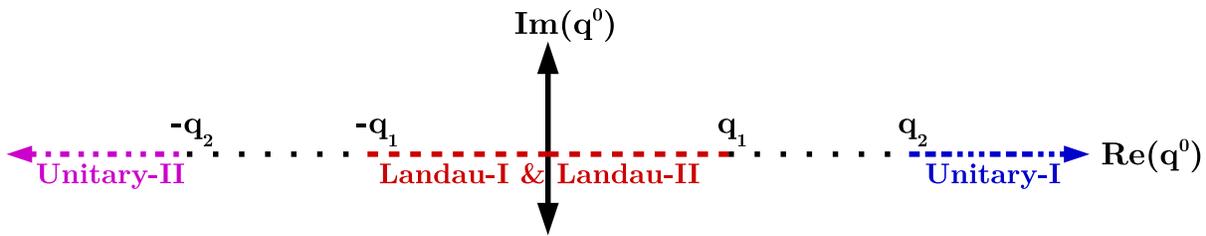}
\end{center}
\caption{Different branch cuts of the in-medium self energy function of the $\rho^0$ under external magnetic field  
in the complex $q^0$ plane for a given $\vec{q}$. The points correspond to $q_1=\sqrt{q_z^2+( \sqrt{m_\pi^2+eB}-\sqrt{m_\pi^2+3eB})^2}$ 
and $q_2=\sqrt{q_z^2+4(m_\pi^2+eB)}$.}
\label{fig.anlytic4}
\end{figure}

Let us now turn on the magnetic field. The imaginary part of $\rho^0$ self energy function in presence of the external magnetic 
field in Eq.~(\ref{eq.im.pibar0.eb.t}) is non vanishing at four different kinematic regions. Note that in this case the 
quantity $m_l$ in Eq.~(\ref{eq.ml}) for charged pions in the loop contains a contribution from the transverse momentum 
component which are quantized. As shown in Appendix~\ref{appendix.c}, the Unitary-I 
and Unitary-II cuts are defined in $\sqrt{q_z^2+4(m_\pi^2+eB)}<q^0<\infty$ and 
$-\infty<q^0<-\sqrt{q_z^2+4(m_\pi^2+eB)}$ respectively, whereas both the Landau-I and Landau-II cuts are defined 
in $|q^0|<\sqrt{q_z^2+( \sqrt{m_\pi^2+eB}-\sqrt{m_\pi^2+3eB})^2}$. These cuts are shown in Fig.~\ref{fig.anlytic4}. 
It is to be noted that, in presence of the external magnetic field, the in-medium $\rho^0$ self energy 
always possesses the Landau contribution even if $\vec{q}=\vec{0}$.

\begin{figure}[h]
\begin{center}
\includegraphics[angle=0, scale=0.6]{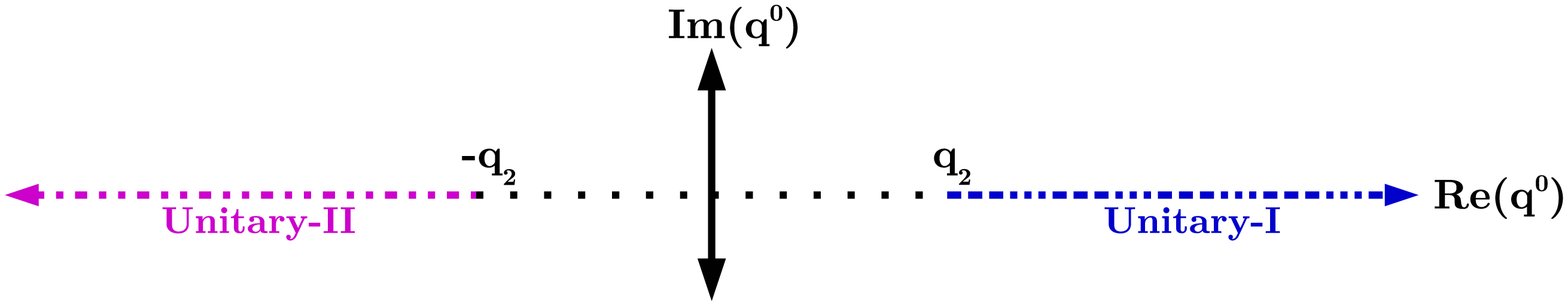} \\
\includegraphics[angle=0, scale=0.6]{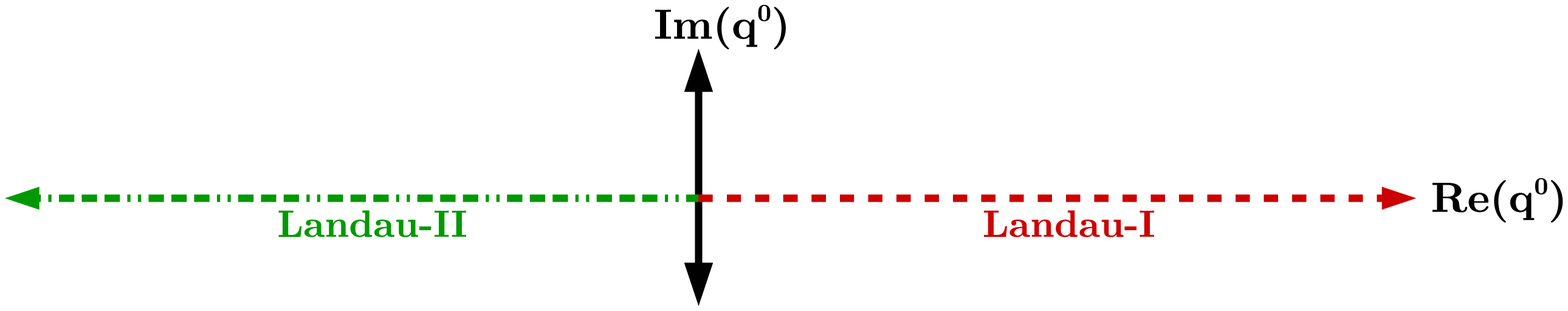}
\end{center}
\caption{Different branch cuts of the in-medium self energy functions of 
	the $\rho^\pm$ under external magnetic field in the complex $q^0$ plane for a given $\vec{q}$. Upper-Panel shows the Unitary cut 
regions with $q_2=\sqrt{q_z^2+(\sqrt{m_\pi^2+eB}+m_\pi)^2}$. Lower-Panel shows the Landau cut regions.}
\label{fig.anlytic23}
\end{figure}

In a similar way, the imaginary part of $\rho^\pm$ self energy function in presence of the external magnetic field 
in Eq.~(\ref{eq.im.pibarpm.eb.t}) has its Unitary-I and Unitary-II cuts in the kinematic domain 
$\sqrt{q_z^2+(\sqrt{m_\pi^2+eB}+m_\pi)^2}<q^0<\infty$ and $-\infty<q^0<-\sqrt{q_z^2+(\sqrt{m_\pi^2+eB}+m_\pi)^2}$ 
respectively, whereas it has its Landau-I and Landau-II cuts in the kinematic domain $0<q^0<\infty$ and $-\infty<q^0<0$ 
respectively. These cuts are displayed in Fig.~\ref{fig.anlytic23}. In this case also the in-medium $\rho^\pm$ self energy 
always has a finite Landau cut contribution even if $\vec{q}=\vec{0}$.

The imaginary parts given in Eq.~(\ref{eq.im.pibar.therm}),(\ref{eq.im.pibar0.eb.t}) and (\ref{eq.im.pibarpm.eb.t}) 
have been further simplified using the Dirac delta functions present in the integrand. For the sake of simplicity in analytic 
calculations Eq.~(\ref{eq.im.pibar.therm}) and (\ref{eq.im.pibar0.eb.t}) are simplified by taking $\vec{q}=\vec{0}$. However, for 
Eq.~(\ref{eq.im.pibarpm.eb.t}), we have taken $\vec{q}=(q_x,q_y,0)$. The simplified form of the imaginary 
parts can be obtained from Appendix~(\ref{appendix.d}),
\begin{eqnarray}
\text{Im}~\bar{\Pi}^{\mu\nu}(q^0,\vec{q}=\vec{0}) &=& \left(\frac{-\epsilon(q^0)\tilde{k}}{8\pi q^0}\right) 
\left[ U_1\left(q^0,|\vec{k}|=\tilde{k}\right)\Theta\left(q^0-2m_\pi\right) +  U_2\left(q^0,|\vec{k}|=\tilde{k}\right)\Theta\left(-q^0-2m_\pi\right) \right] \label{eq.impi.simple}
\end{eqnarray}
\begin{eqnarray}
\text{Im}~\bar{\Pi}^{\mu\nu}_0(q^0,\vec{q}=\vec{0}) &=& \left(\frac{-\epsilon(q^0)}{4|q^0|}\right)\sum\limits_{l=0}^{\infty} \sum\limits_{n=0}^{\infty}\frac{1}{\tilde{k}_z} 
 \left[ U_1^{n,l}\left( q^0,\tilde{k_z}\right)\Theta\left(q^0-m_l-m_n\right) + U_2^{n,l}\left( q^0,\tilde{k_z}\right)\Theta\left(-q^0-m_l-m_n\right) \right. \nonumber \\
&& \hspace{2cm} \left. +~L_1^{n,l}\left( q^0,\tilde{k_z}\right) \Theta\left\{-q^0-\min(m_l-m_n,0)\right\} \Theta\left\{\max(m_l-m_n,0)+q^0\right\} \nonumber  \right. \\
&& \hspace{2cm} \left. +~ L_2^{n,l}\left( q^0,\tilde{k_z}\right) \Theta\left\{q^0-\min(m_l-m_n,0)\right\} \Theta\left\{\max(m_l-m_n,0)-q^0\right\}  \frac{}{} \right] \label{eq.impi0.simple}
\end{eqnarray}
\begin{eqnarray}
\text{Im}~\bar{\Pi}^{\mu\nu}_\pm (q^0,q_x,q_y) &=& \left(\frac{-\epsilon(q^0)}{32\pi^2}\right)\sum\limits_{n=0}^{\infty}
\int\limits_{0}^{2\pi}d\phi
\left[~\int\limits_{\omega_-}^{\omega_0}\frac{d\omega_k}{|\vec{k}|\cos\theta_0}\left\{U_1^n\left(q^0,|\vec{k}|,\theta_0,\phi\right)+U_1^n\left(q^0,|\vec{k}|,-\theta_0,\phi\right)\right\}\Theta\left( q^0-m_\pi-m_n\right) \right. \nonumber \\
&& \left. + \int\limits_{-\omega_+}^{-\omega_0}\frac{d\omega_k}{|\vec{k}|\cos\theta^\prime_0} \left\{U_2^n\left(q^0,|\vec{k}|,\theta^\prime_0,\phi\right)+U_2^n\left(q^0,|\vec{k}|,-\theta^\prime_0,\phi\right)\right\}\Theta\left(-q^0-m_\pi-m_n\right) \right. \nonumber \\ 
&& \left. + \int\limits_{-\omega_0}^{-\omega_-}\frac{d\omega_k}{|\vec{k}|\cos\theta^\prime_0} \left\{L_1^n\left(q^0,|\vec{k}|,\theta^\prime_0,\phi\right)+L_1^n\left(q^0,|\vec{k}|,-\theta^\prime_0,\phi\right)\right\}\Theta\left(-q^0-m_\pi+m_n\right)\Theta(q^0) \right. \nonumber \\
&& \left.  + \int\limits_{\omega_0}^{-\omega_+}\frac{d\omega_k}{|\vec{k}|\cos\theta_0} \left\{L_2^n\left(q^0,|\vec{k}|,\theta_0,\phi\right)+L_2^n\left(q^0,|\vec{k}|,-\theta_0,\phi\right)\right\}\Theta\left(q^0-m_\pi+m_n\right)\Theta(-q^0) \right] \label{eq.impipm.simple} 
\end{eqnarray}
where, $\tilde{k}_z = \frac{1}{2q^0}\lambda^{1/2}\left(q_0^2,m_l^2,m_n^2\right)$, $\omega_\pm = (q^0\pm m_n)$ and $\omega_0 = \frac{1}{2q^0}\left(q_0^2+m_\pi^2-m_n^2\right)$. The Lorentz indices $\mu,\nu$ are
contained in $U_{1,2}$ and $L_{1,2}$ (see Appendix~\ref{appendix.d}).
%
%+++++++++++++++++++++++++++++++++++++++++++++++++++++++++++++++++++++++++++++++++++++++++++++++++++++++++++++++++++++++++++++++++

\section{Numerical Results}\label{sec.results}
We begin this section by showing the imaginary and real parts of the in-medium self energy function of $\rho$ meson 
under external magnetic field. 
We will present numerical results for the spin averaged quantity 
\begin{eqnarray}
\Pi_{0,\pm}=\frac{1}{3}g_{\mu\nu}\bar{\Pi}^{\mu\nu}_{0,\pm}~.
\label{eq.spin.averaged}
\end{eqnarray}
First, we have checked numerically that in the limit $eB\rightarrow0$, the $eB=0$ results ($T\ne0$) are exactly reproduced, i.e.
\begin{eqnarray}
\lim\limits_{eB\rightarrow0}\bar{\Pi}^{\mu\nu}_0 = \lim\limits_{eB\rightarrow0}\bar{\Pi}^{\mu\nu}_\pm = \bar{\Pi}^{\mu\nu}. \nn
\end{eqnarray}
To take $eB\rightarrow0$ limit, numerically we have taken upto 300 Landau levels for a convergent result. However, for the other 
results presented here for $eB\ge$ 0.05 GeV$^2$, the results are well convergent with 200 Landau levels.

\begin{figure}[h]
\begin{center}
\includegraphics[angle=-90, scale=0.230]{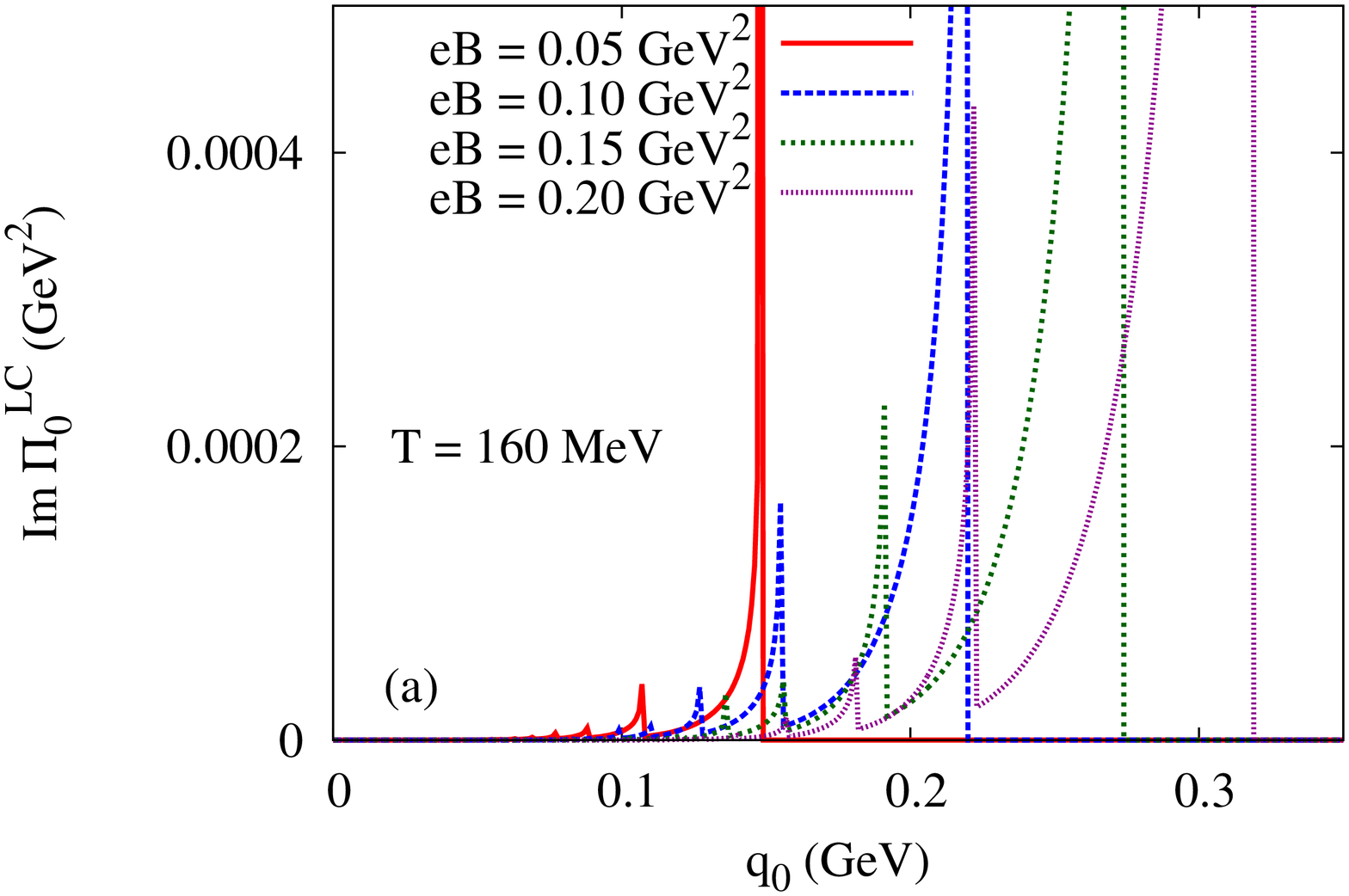} \includegraphics[angle=-90, scale=0.230]{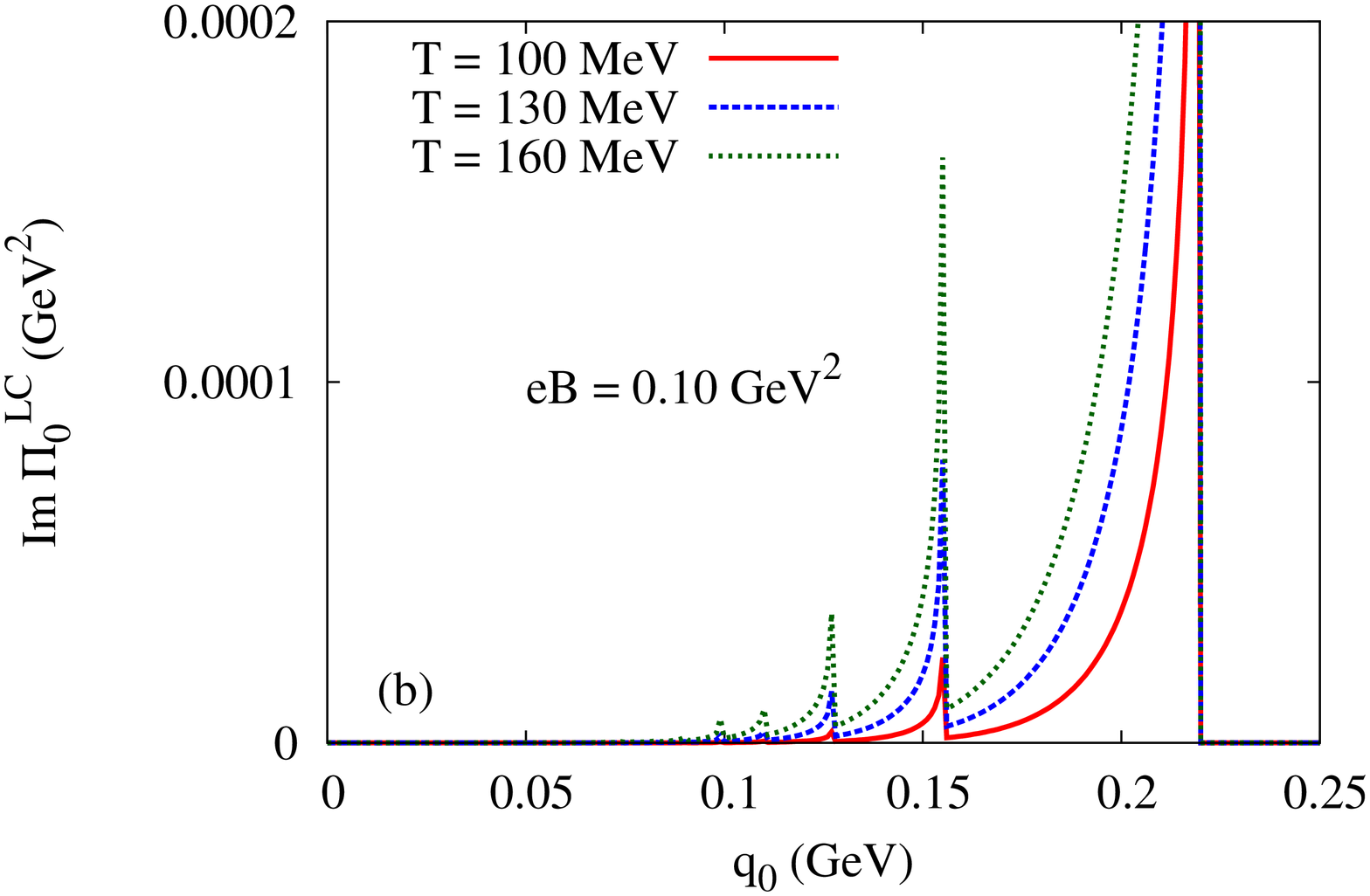}
\includegraphics[angle=-90, scale=0.230]{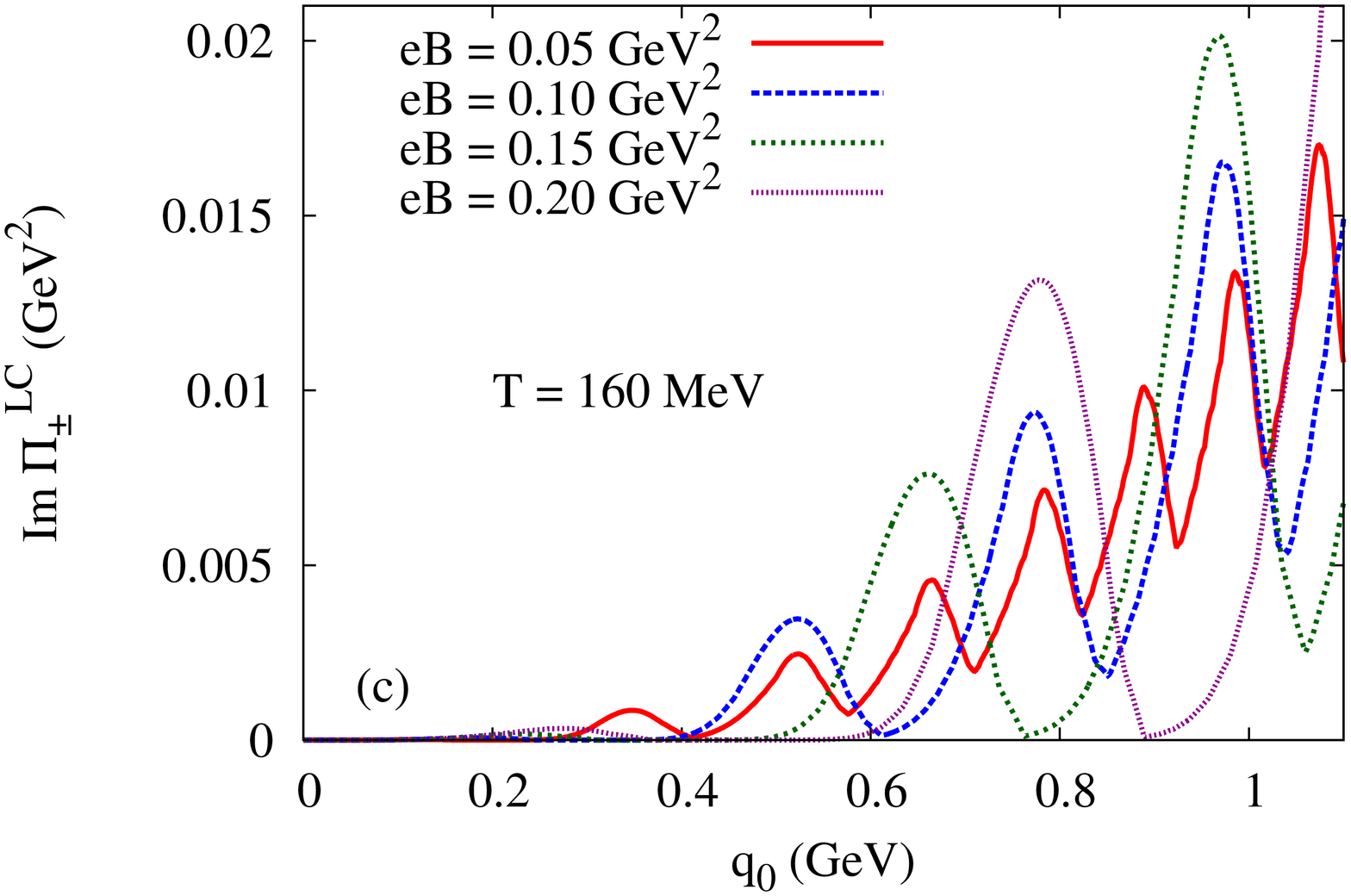} \\
\includegraphics[angle=-90, scale=0.230]{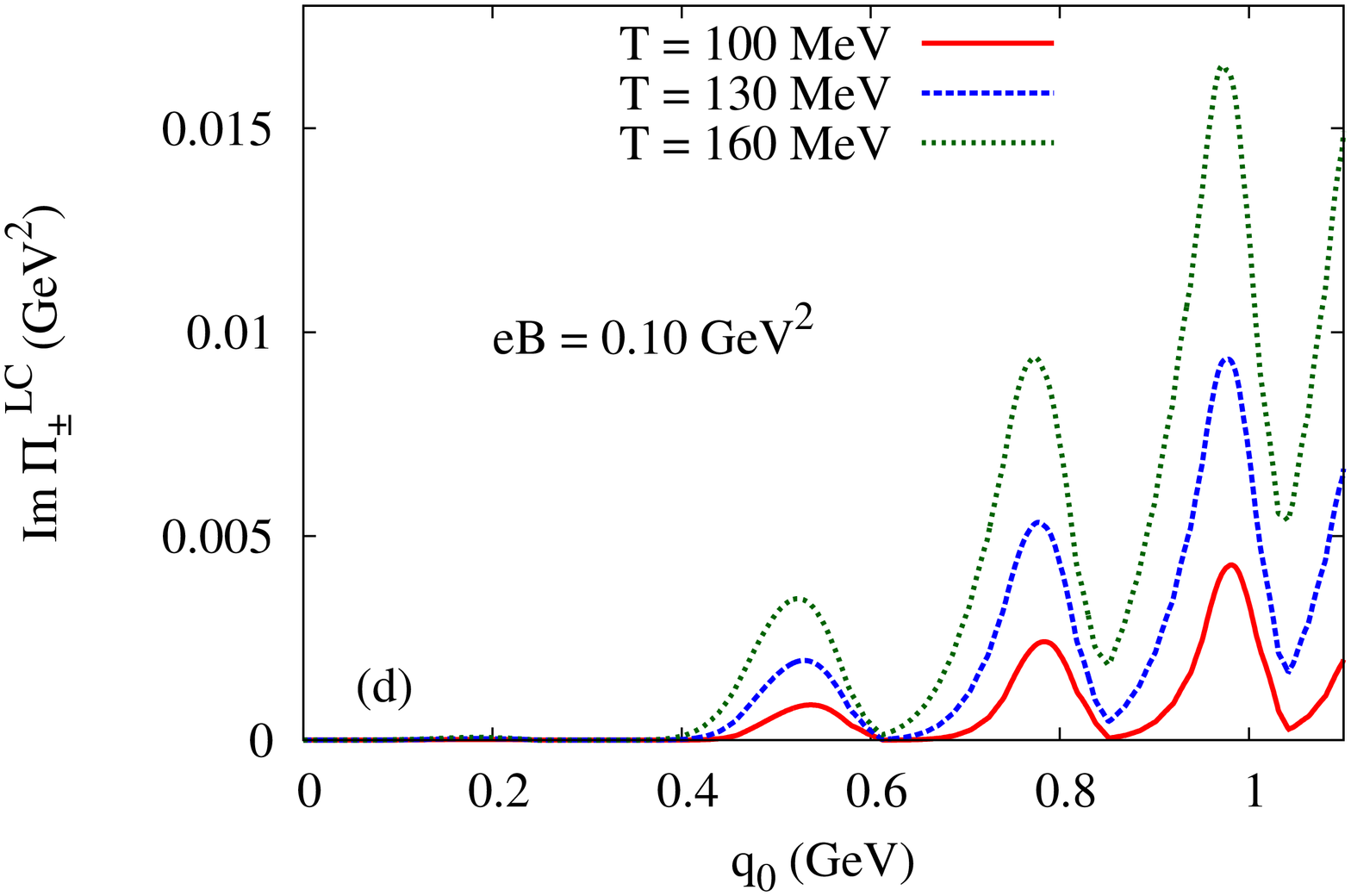} \includegraphics[angle=-90, scale=0.230]{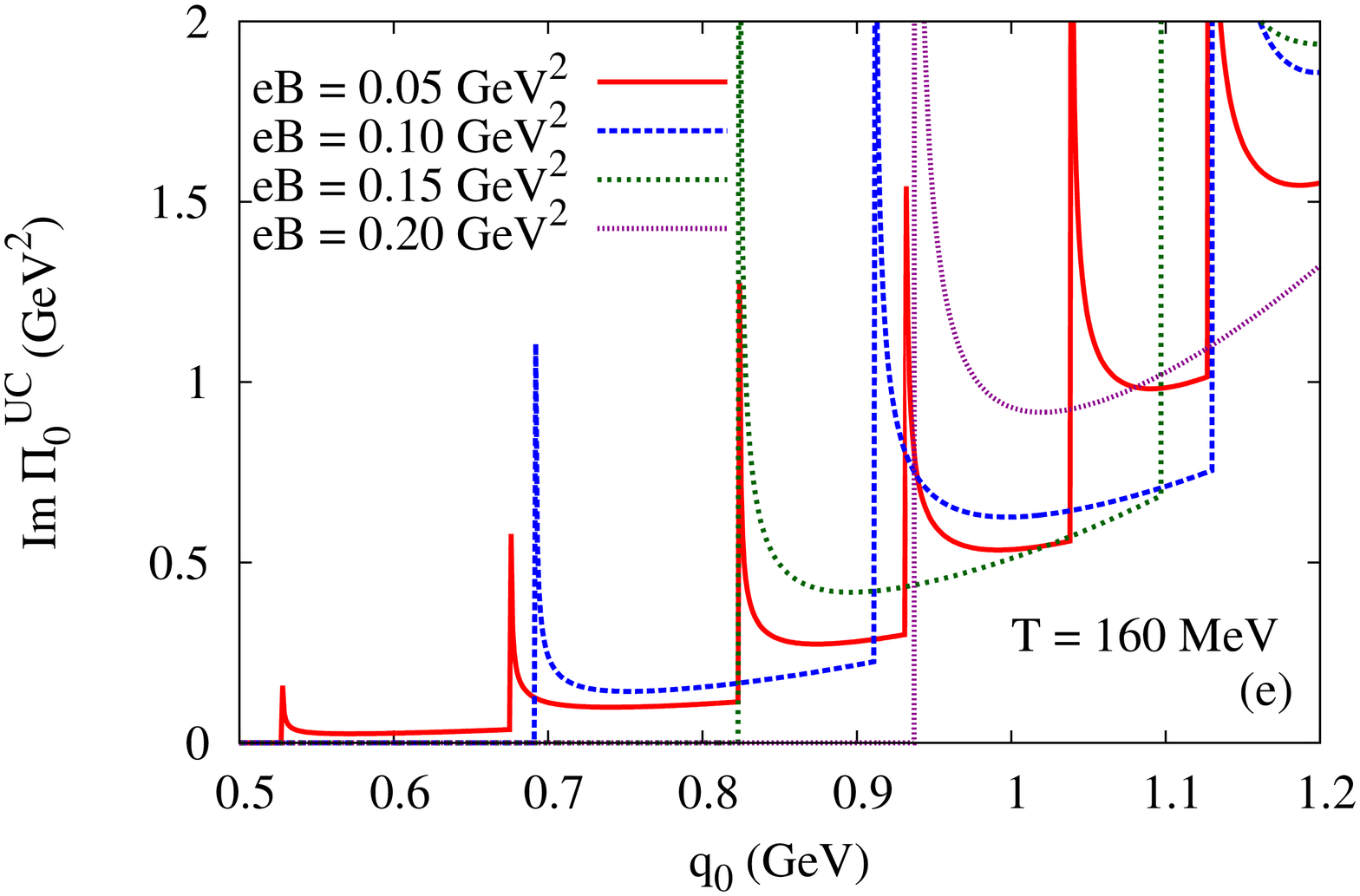} \includegraphics[angle=-90, scale=0.230]{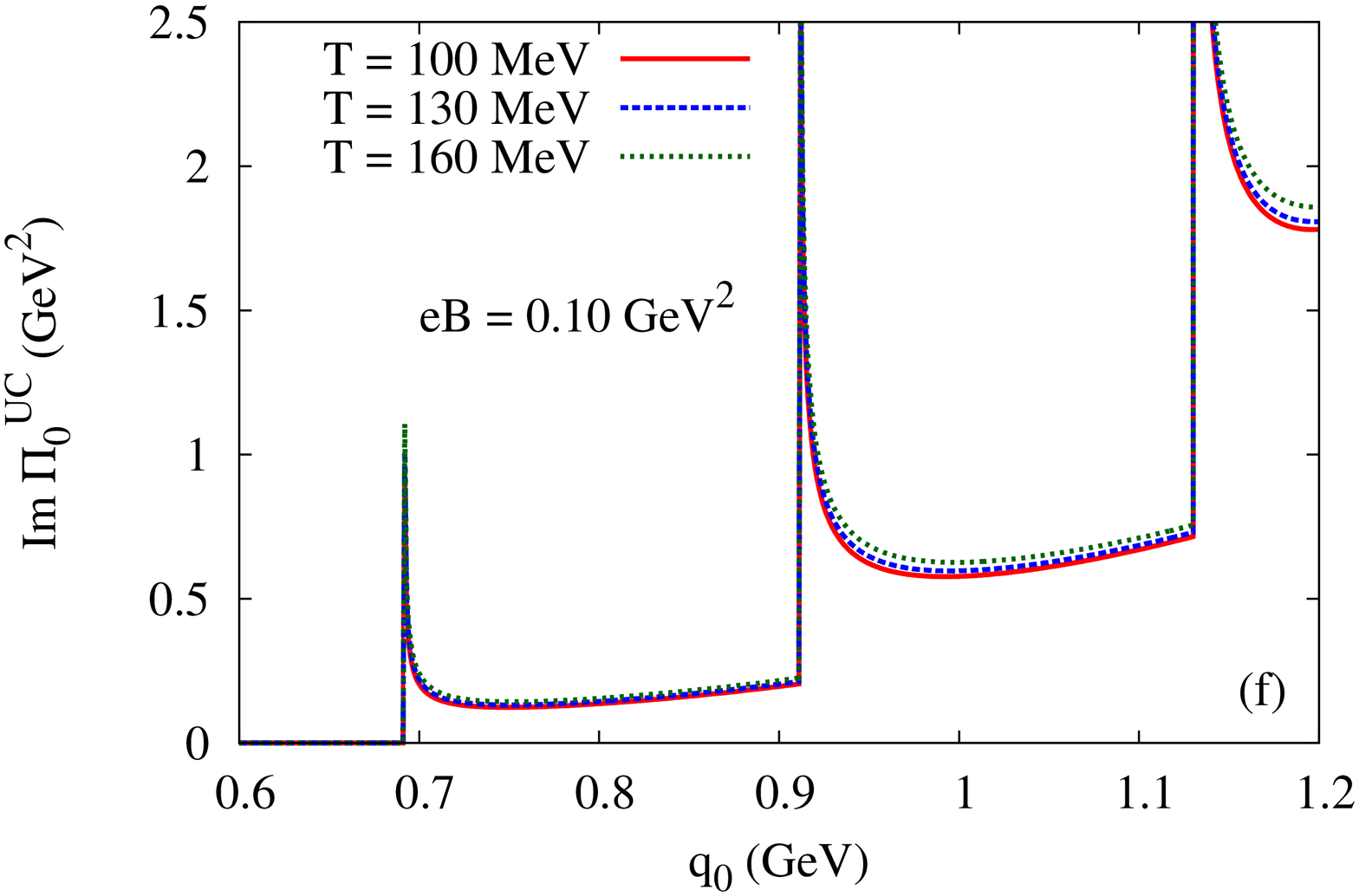} \\
\includegraphics[angle=-90, scale=0.230]{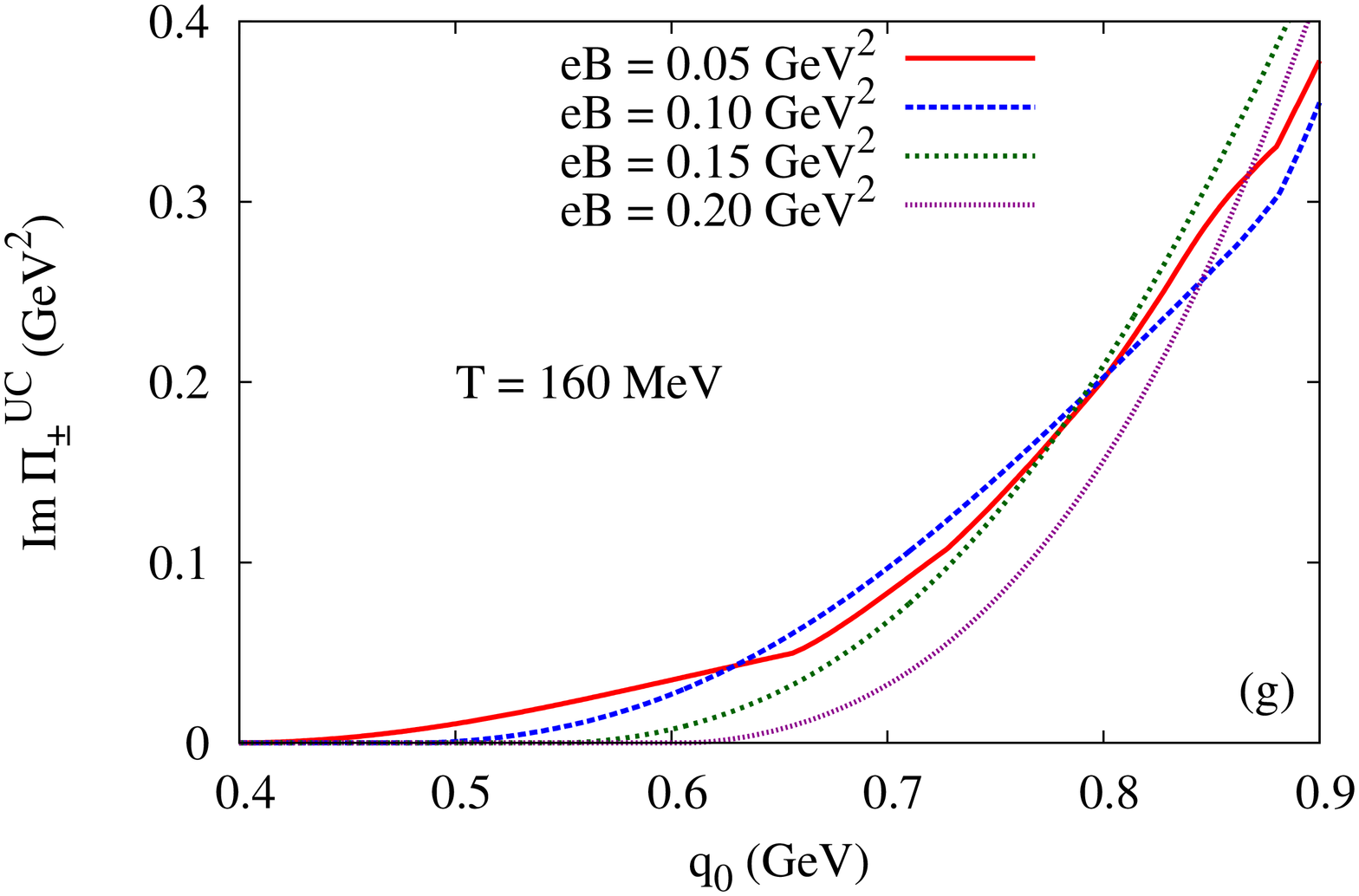} \includegraphics[angle=-90, scale=0.230]{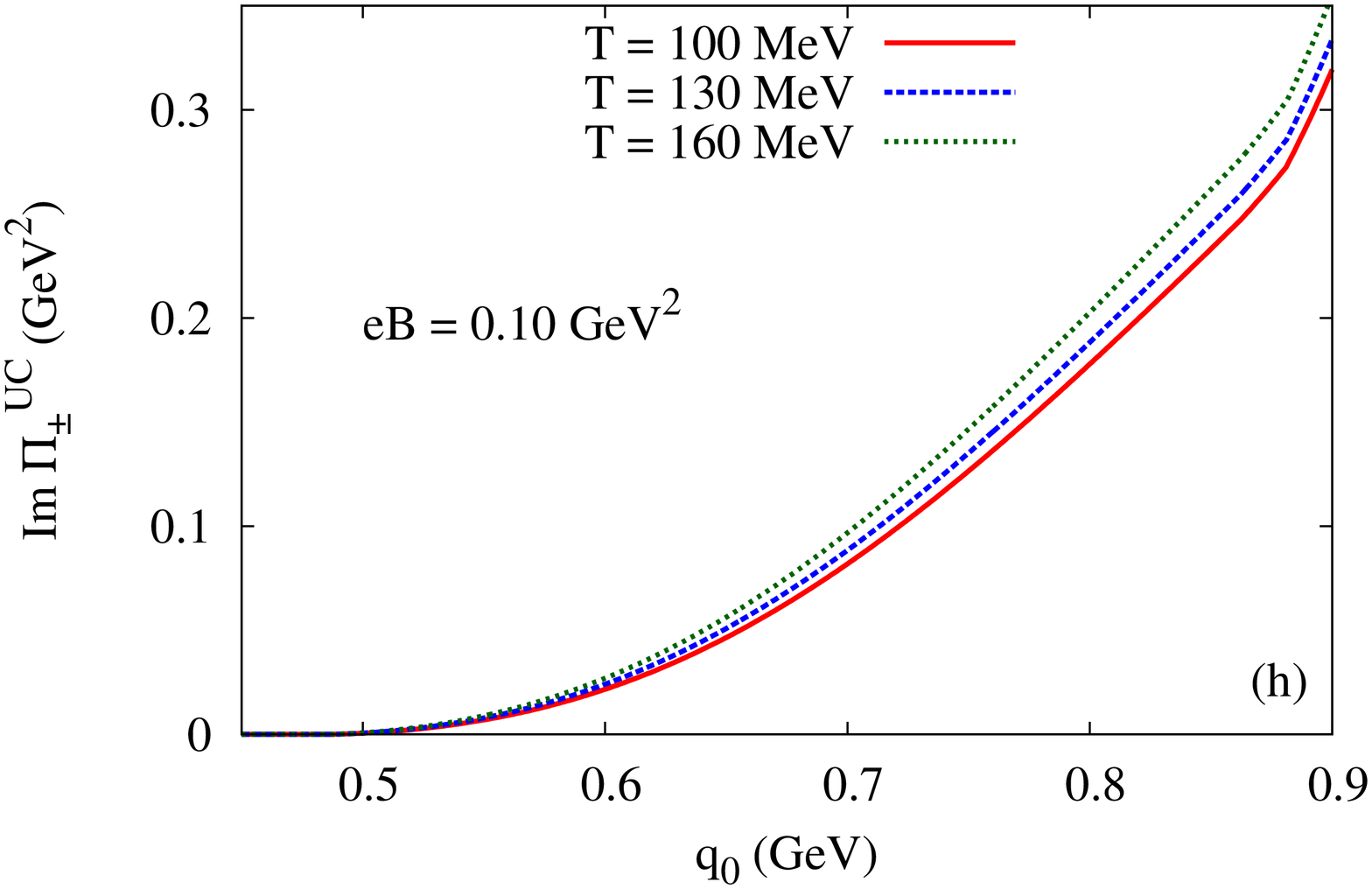}

\end{center}
\caption{ The Landau cut and Unitry cut contributions to the imaginary part of the self energy function of rho meson.  
(a) Landau cut for $\rho^0$ (c) Landau cut for $\rho^\pm$ (e) Unitary cut for $\rho^0$ (f) Unitary cut for $\rho^\pm$, are 
shown at constant temperature (160 MeV) and at different values of the magnetic field (0.05, 0.10, 0.15 and 0.20 GeV$^2$ respectively).
(b) Landau cut for $\rho^0$ (d) Landau cut for $\rho^\pm$ (f) Unitary cut for $\rho^0$ (h) Unitary cut for $\rho^\pm$, are 
shown at constant magnetic field (0.01 GeV$^2$) and at different values of the temperature (100, 130 and 160 MeV respectively).}
\label{fig.imaginary_LU}
\end{figure}
\begin{figure}[h]
\begin{center}
\includegraphics[angle=-90, scale=0.230]{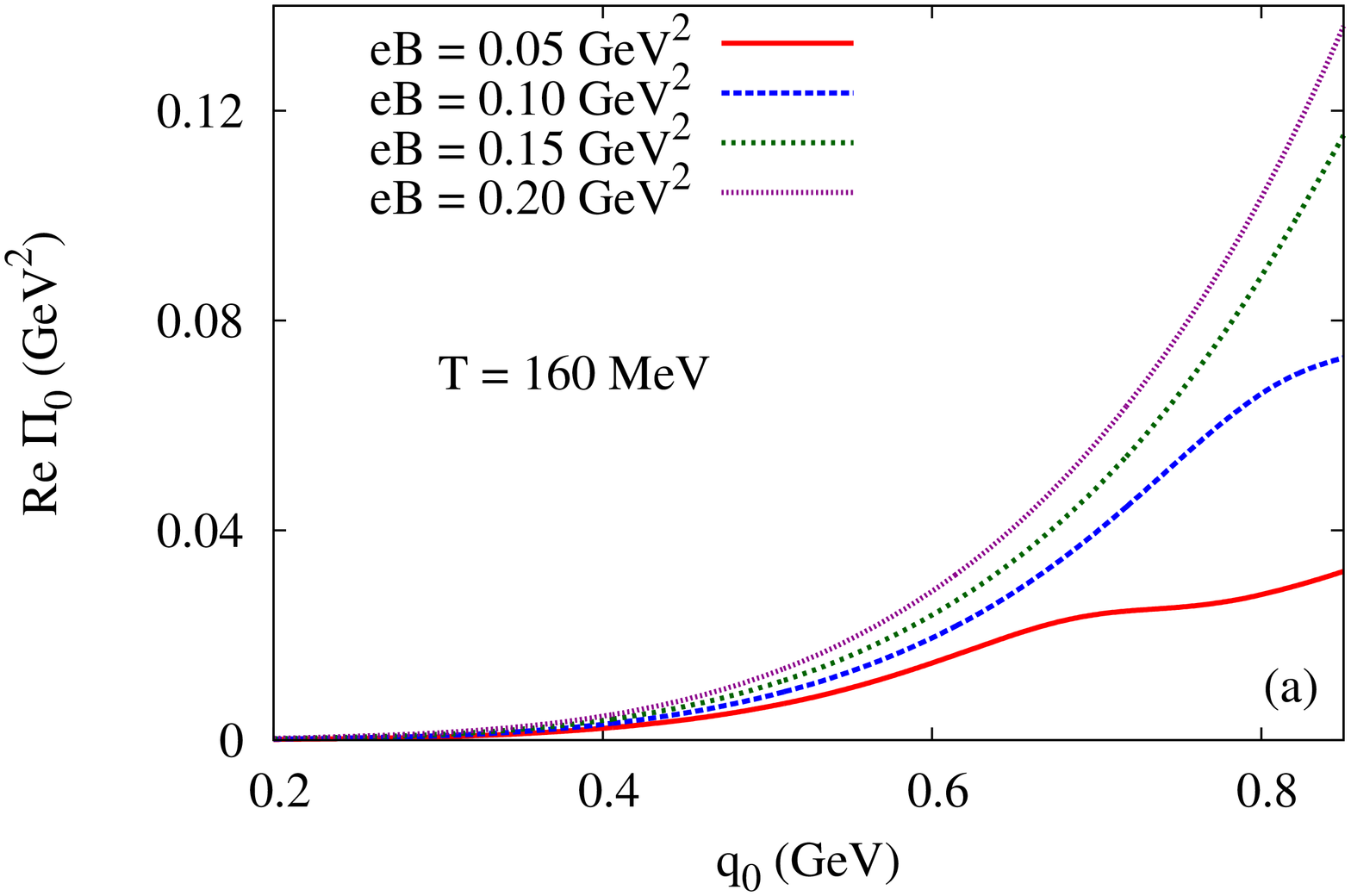} \includegraphics[angle=-90, scale=0.230]{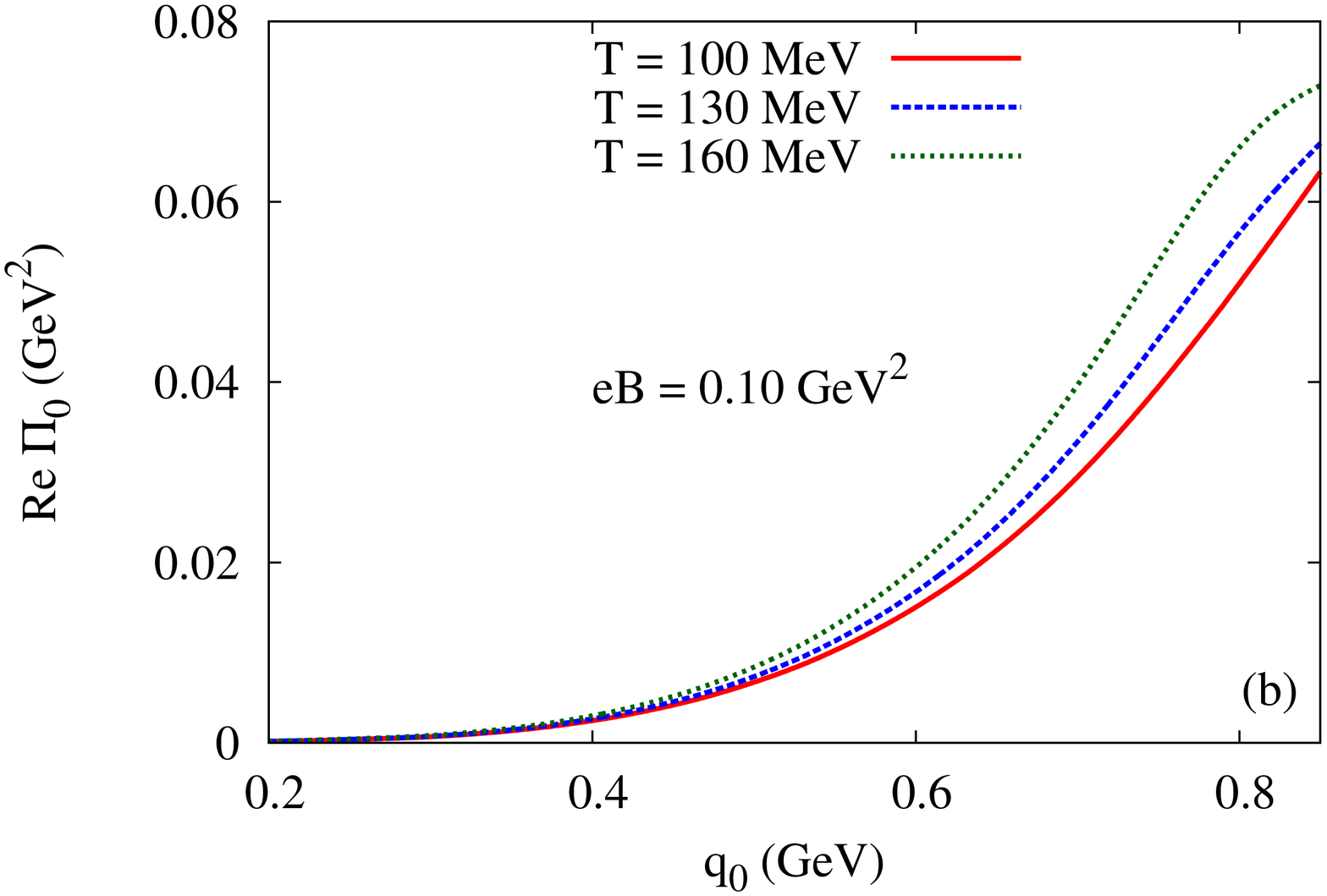} 
\includegraphics[angle=-90, scale=0.230]{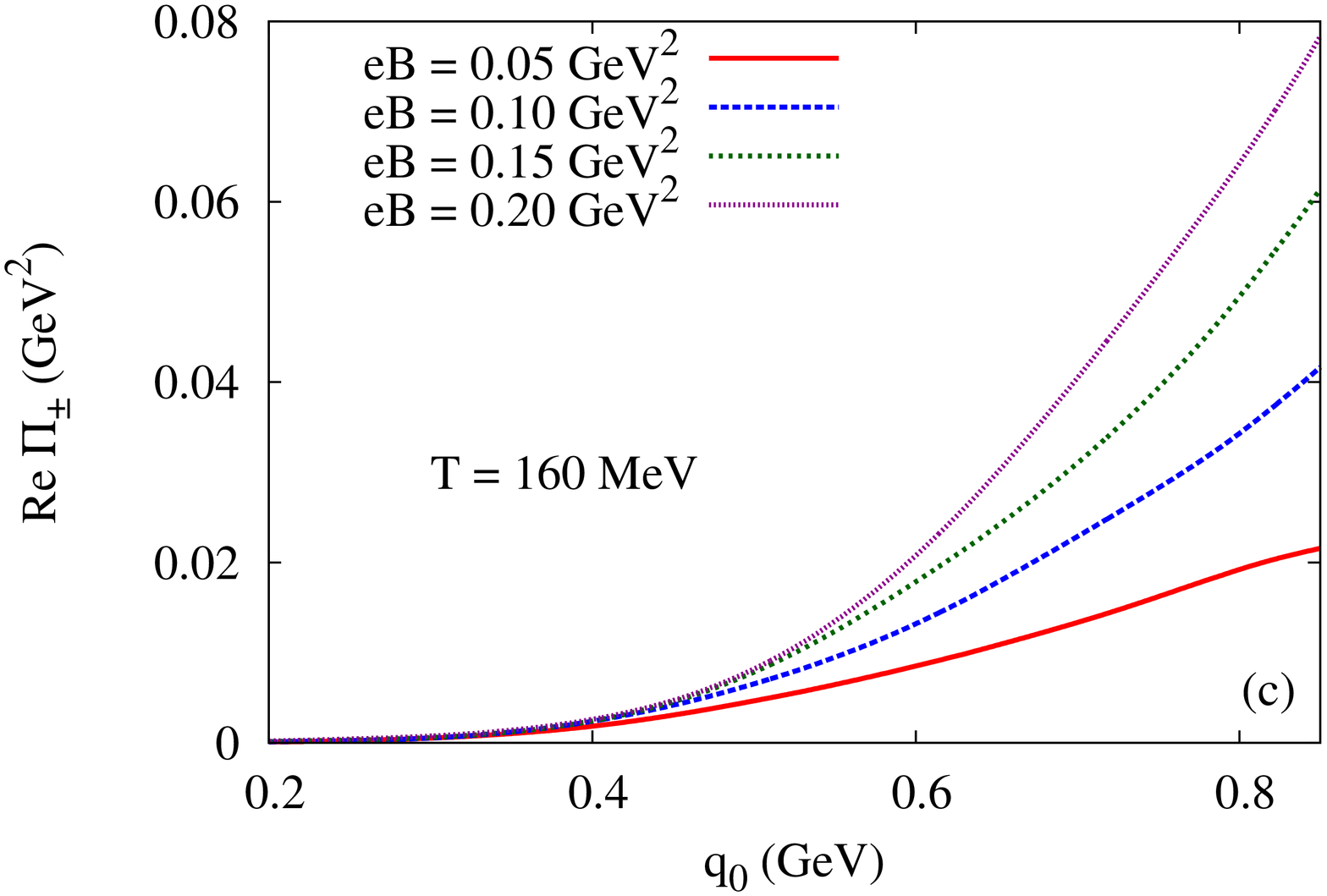} \\
\includegraphics[angle=-90, scale=0.230]{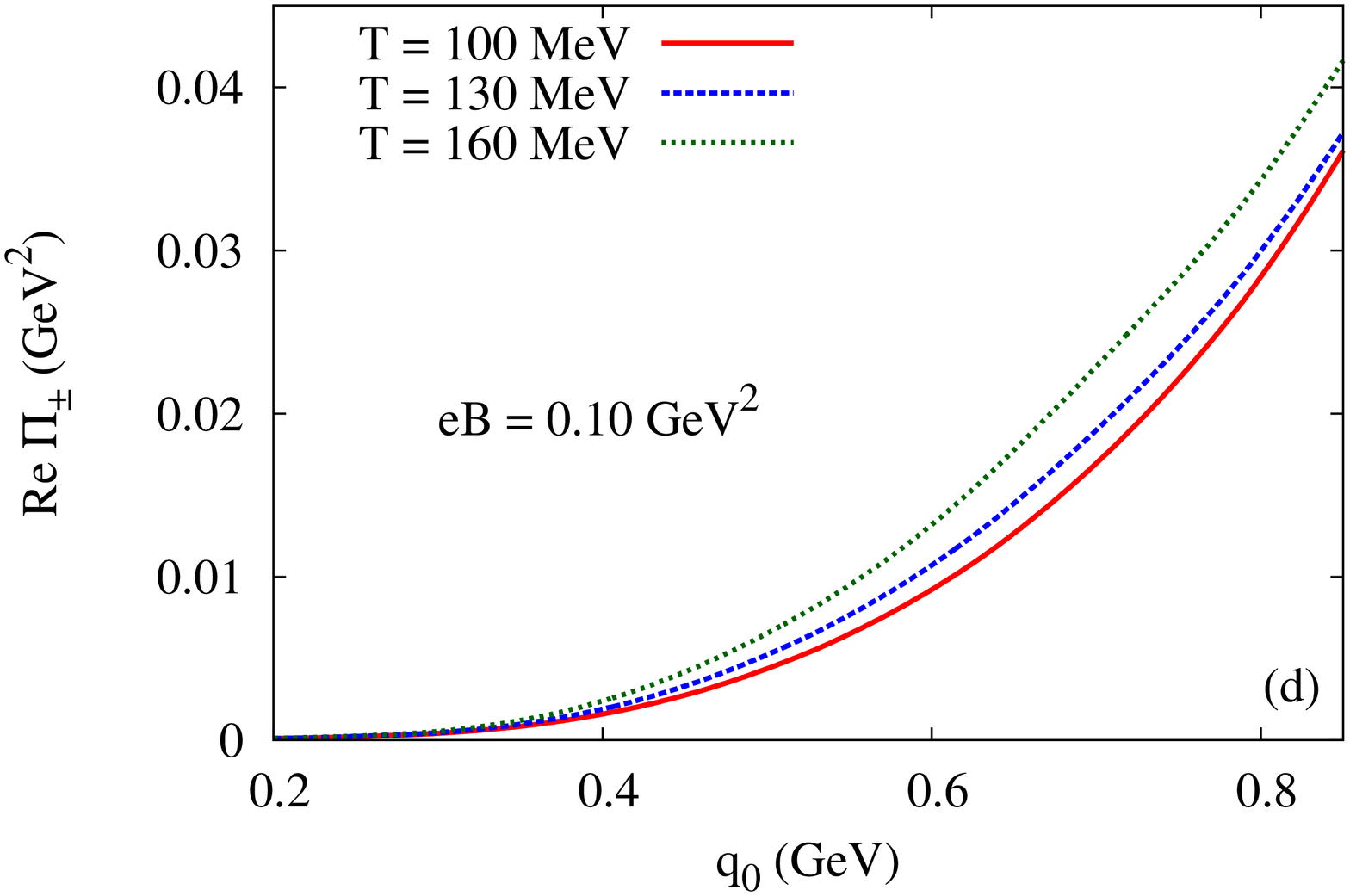} \includegraphics[angle=-90, scale=0.230]{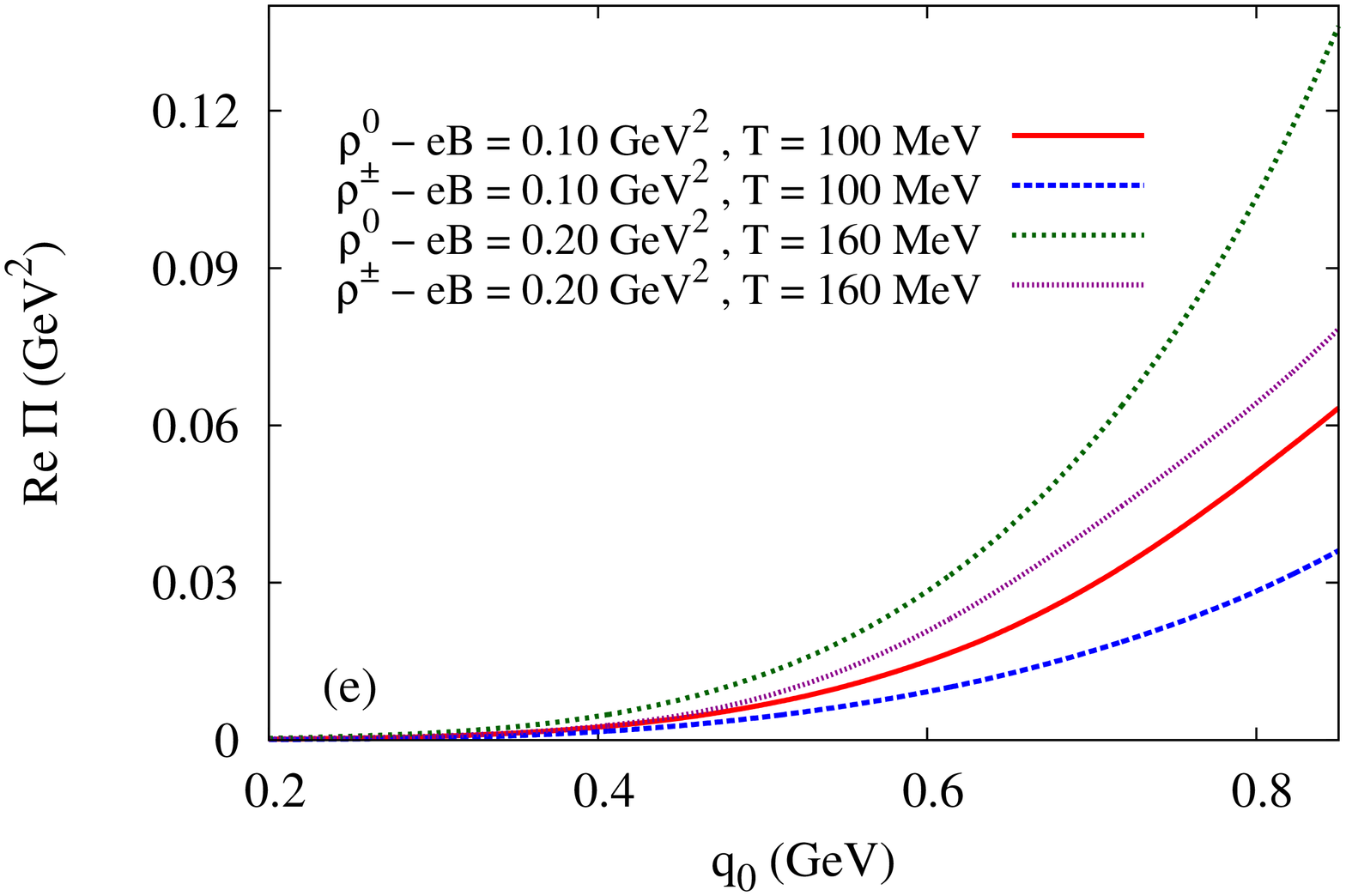}
\end{center}
\caption{The real part of the self energy function of 
(a) $\rho^0$  (c) $\rho^\pm$, at constant temperature (160 MeV) and at different values of the magnetic 
field (0.05, 0.10, 0.15 and 0.20 GeV$^2$ respectively). The real part of the self energy function of 
(b) $\rho^0$ (d) $\rho^\pm$, at constant magnetic field (0.01 GeV$^2$) and at different values of the 
temperature (100, 130 and 160 MeV) respectively. 
(e) The comparison of the real part of the self energy function between $\rho^0$ and $\rho^\pm$ at two different 
combinations of the magnetic field and temperature (eB=0.10 GeV$^2$, T=100 MeV and eB=0.20 GeV$^2$, T=160 MeV respectively).}
\label{fig.real}
\end{figure}
\begin{figure}[h]
\begin{center}
\includegraphics[angle=-90, scale=0.230]{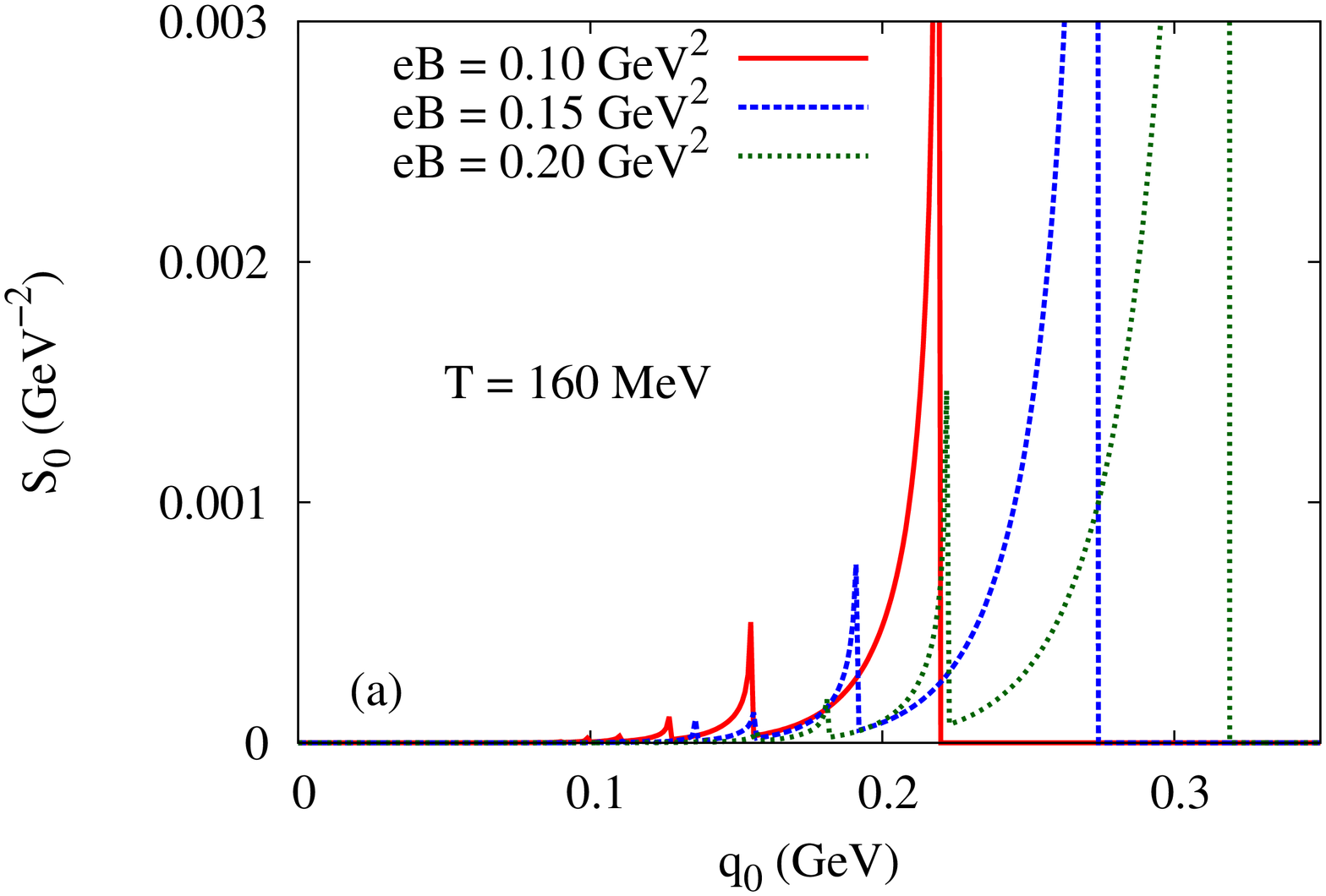} \includegraphics[angle=-90, scale=0.230]{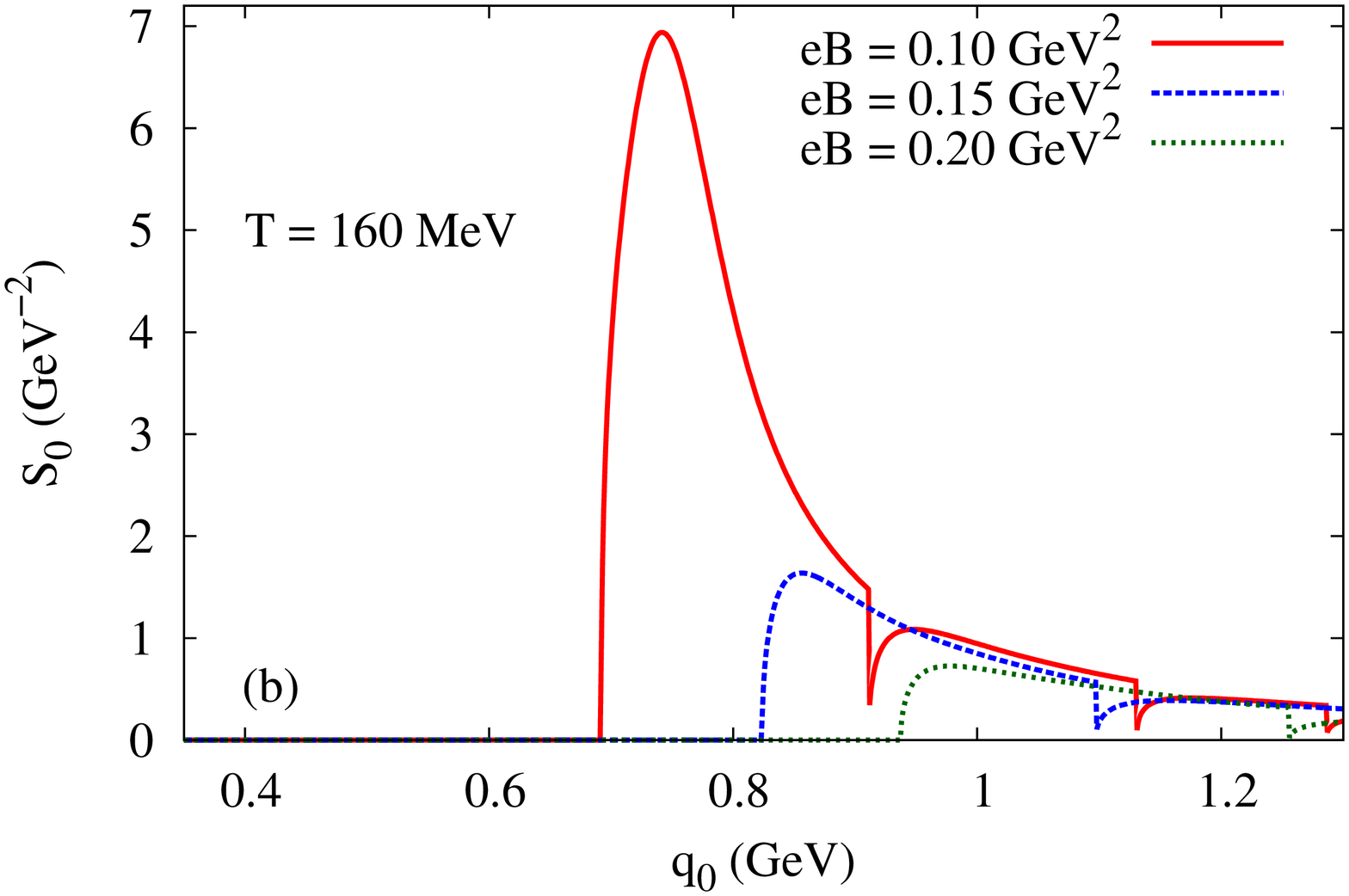}
\includegraphics[angle=-90, scale=0.230]{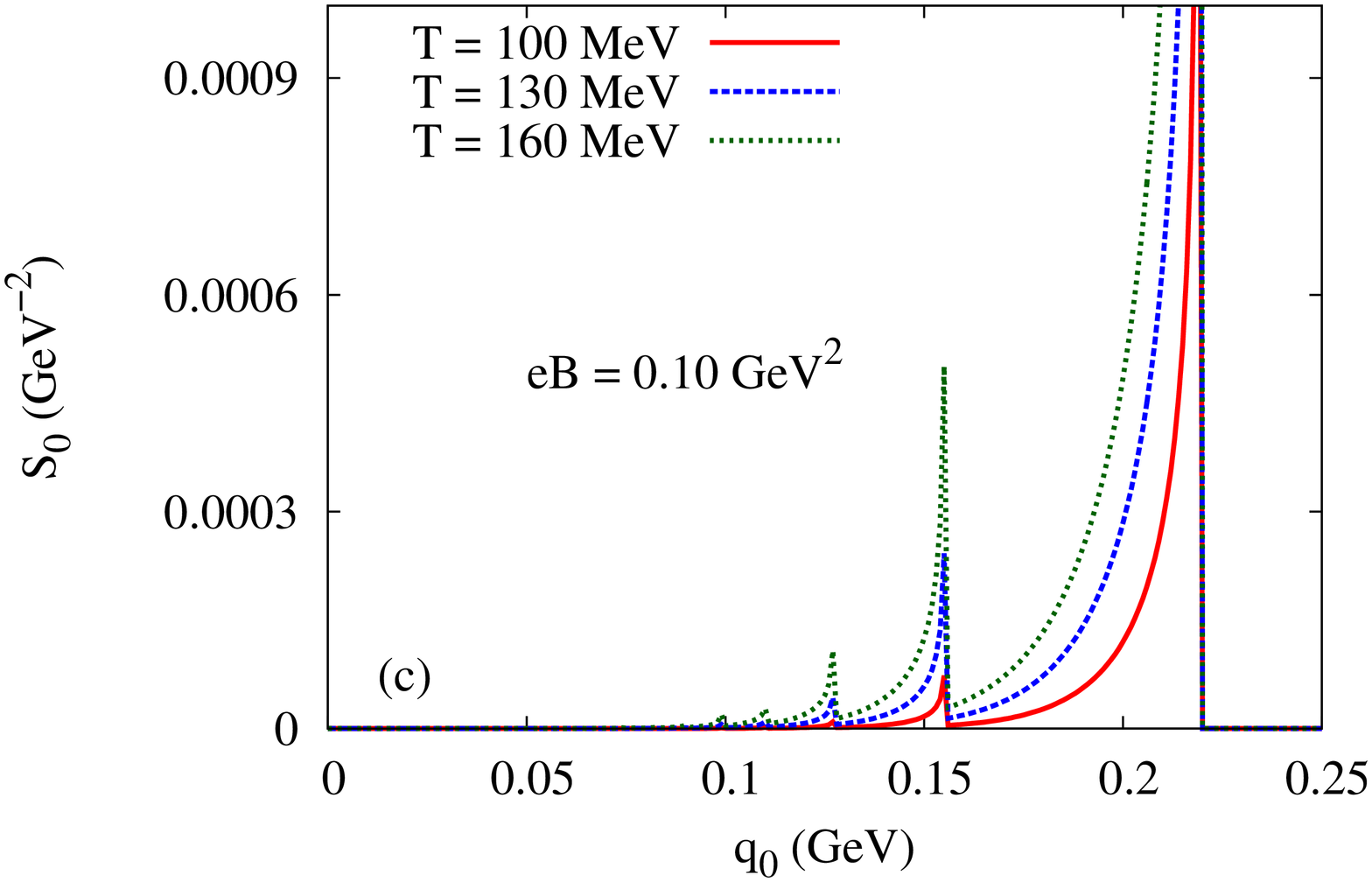} \\ \includegraphics[angle=-90, scale=0.230]{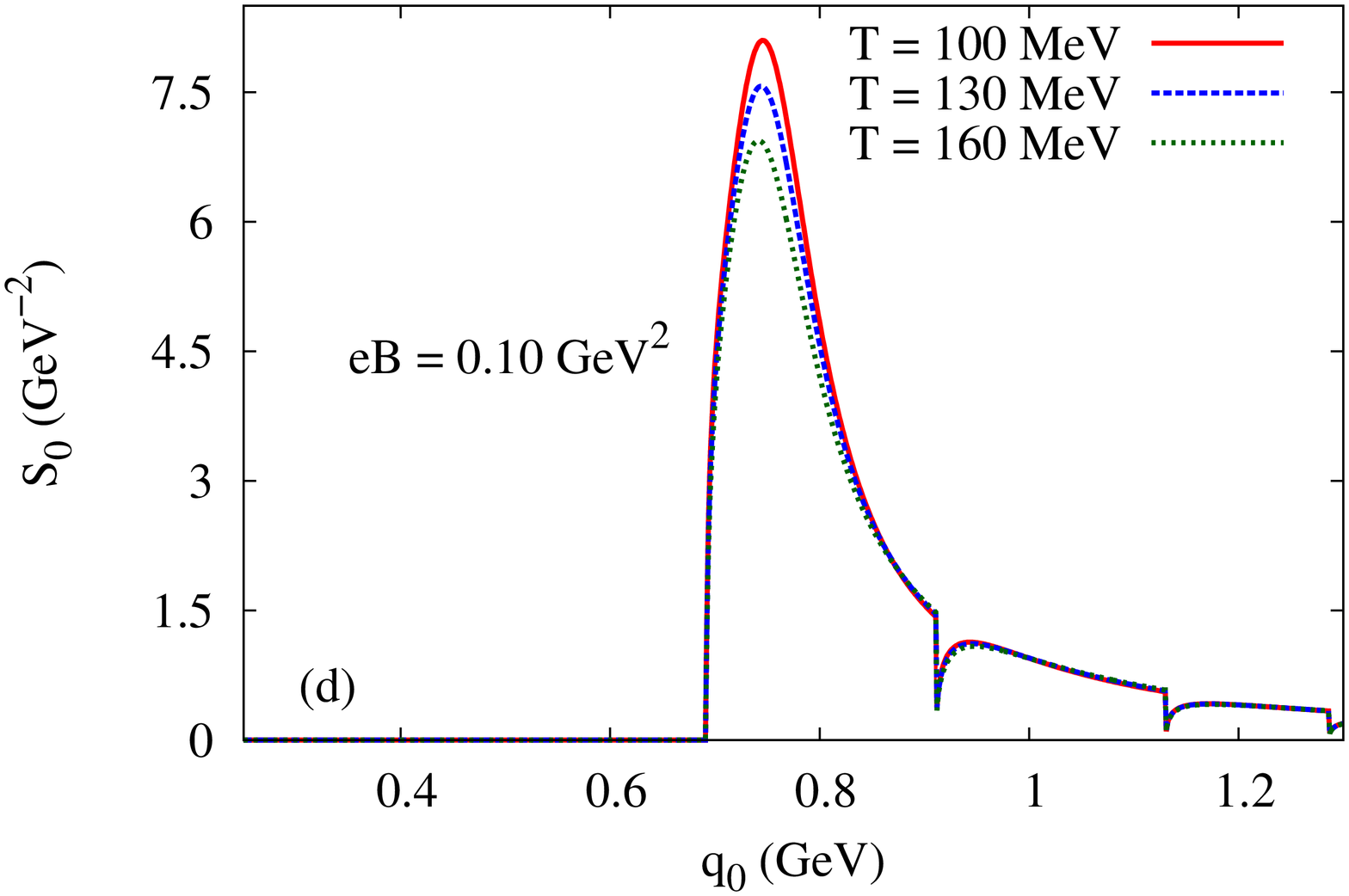}
\includegraphics[angle=-90, scale=0.230]{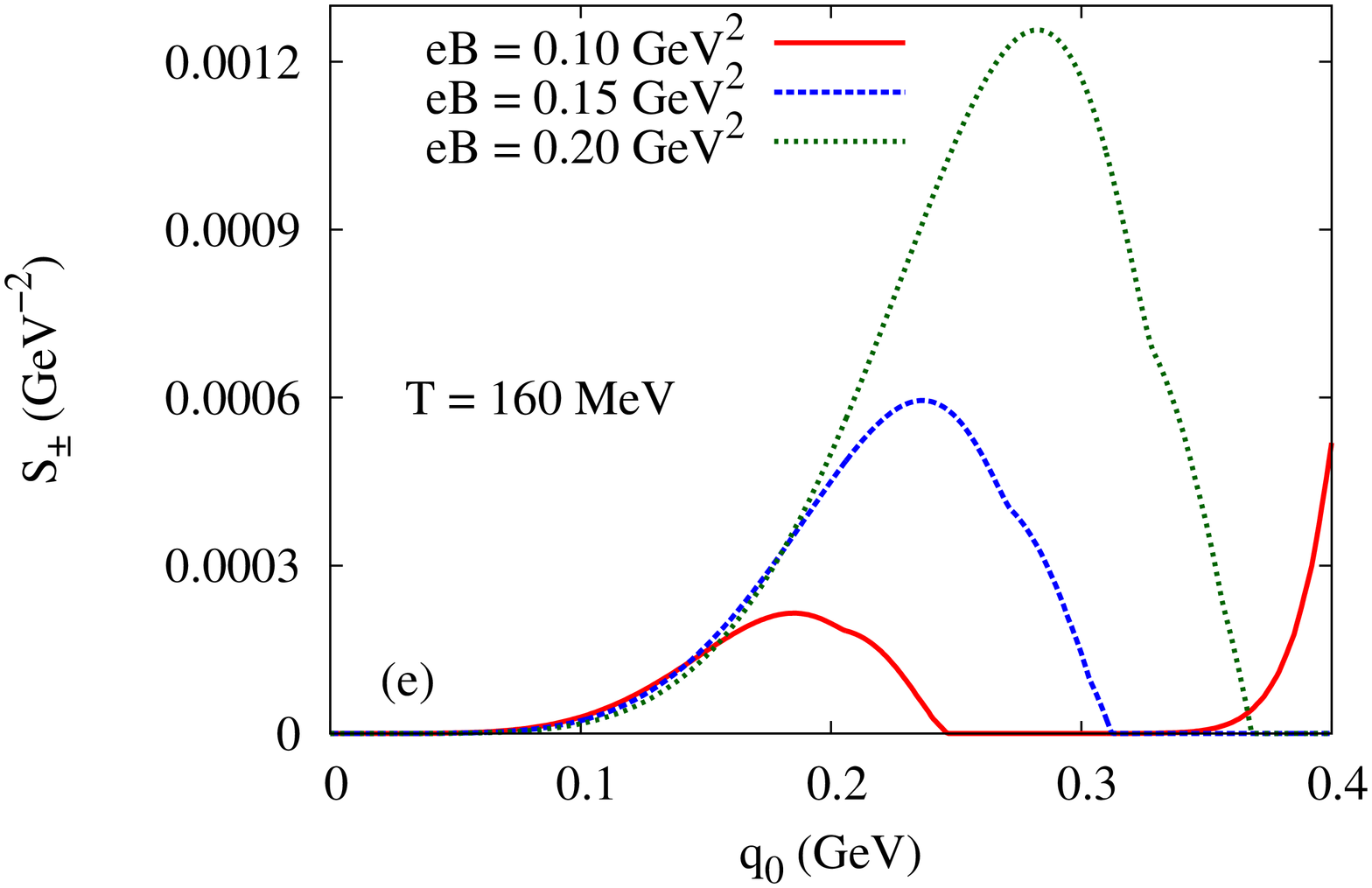} \includegraphics[angle=-90, scale=0.230]{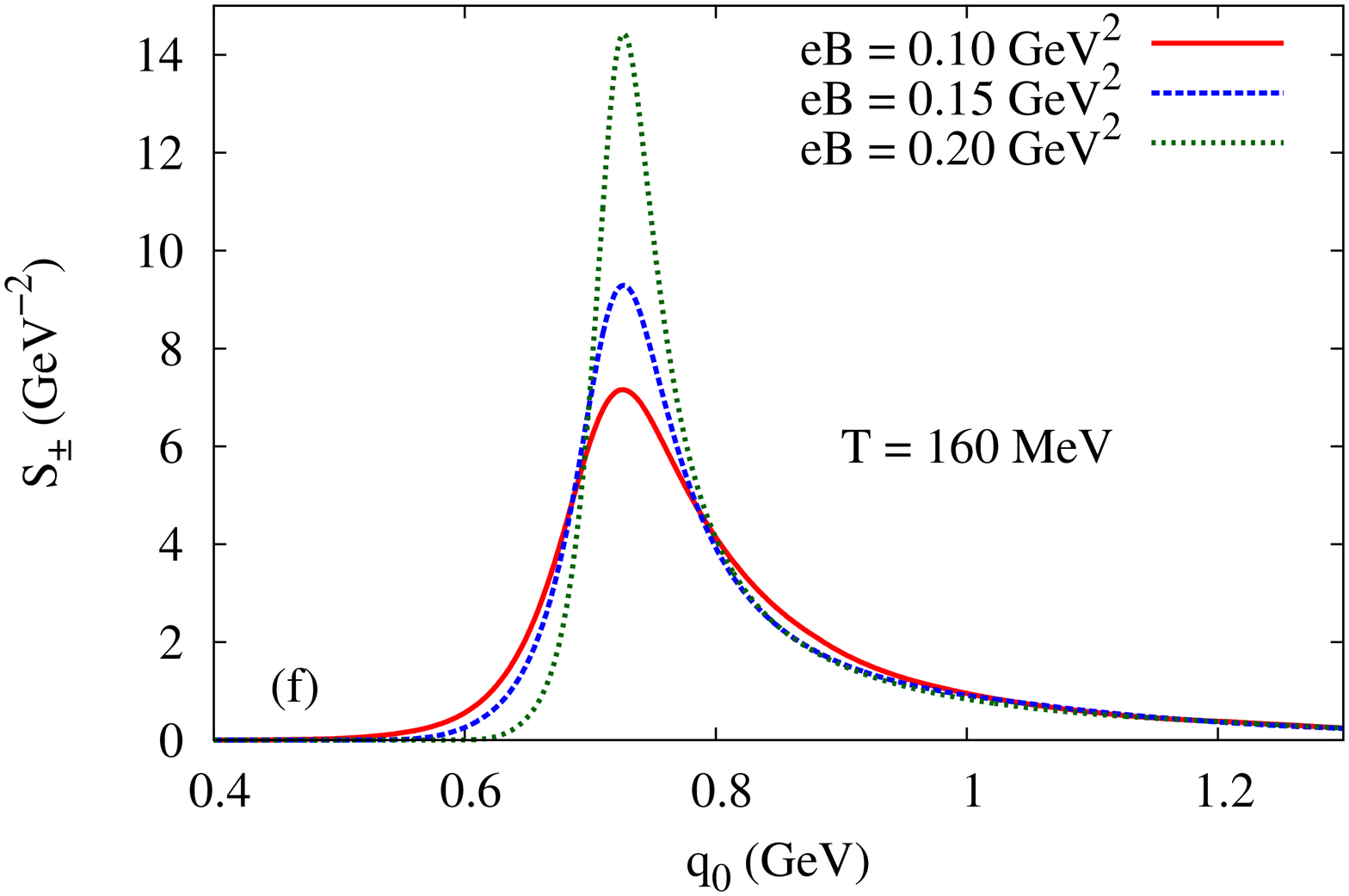} \\
\includegraphics[angle=-90, scale=0.230]{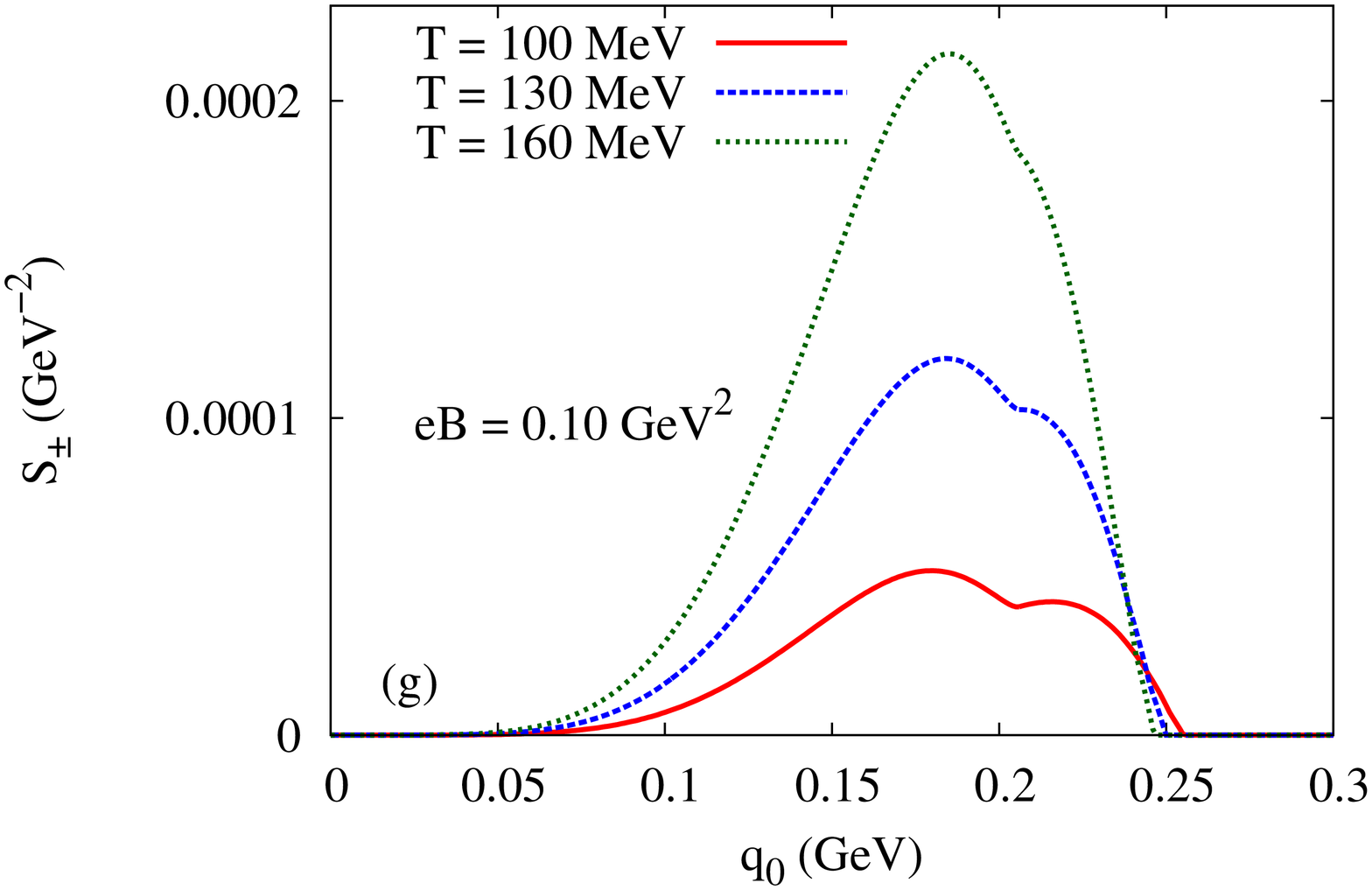} \includegraphics[angle=-90, scale=0.230]{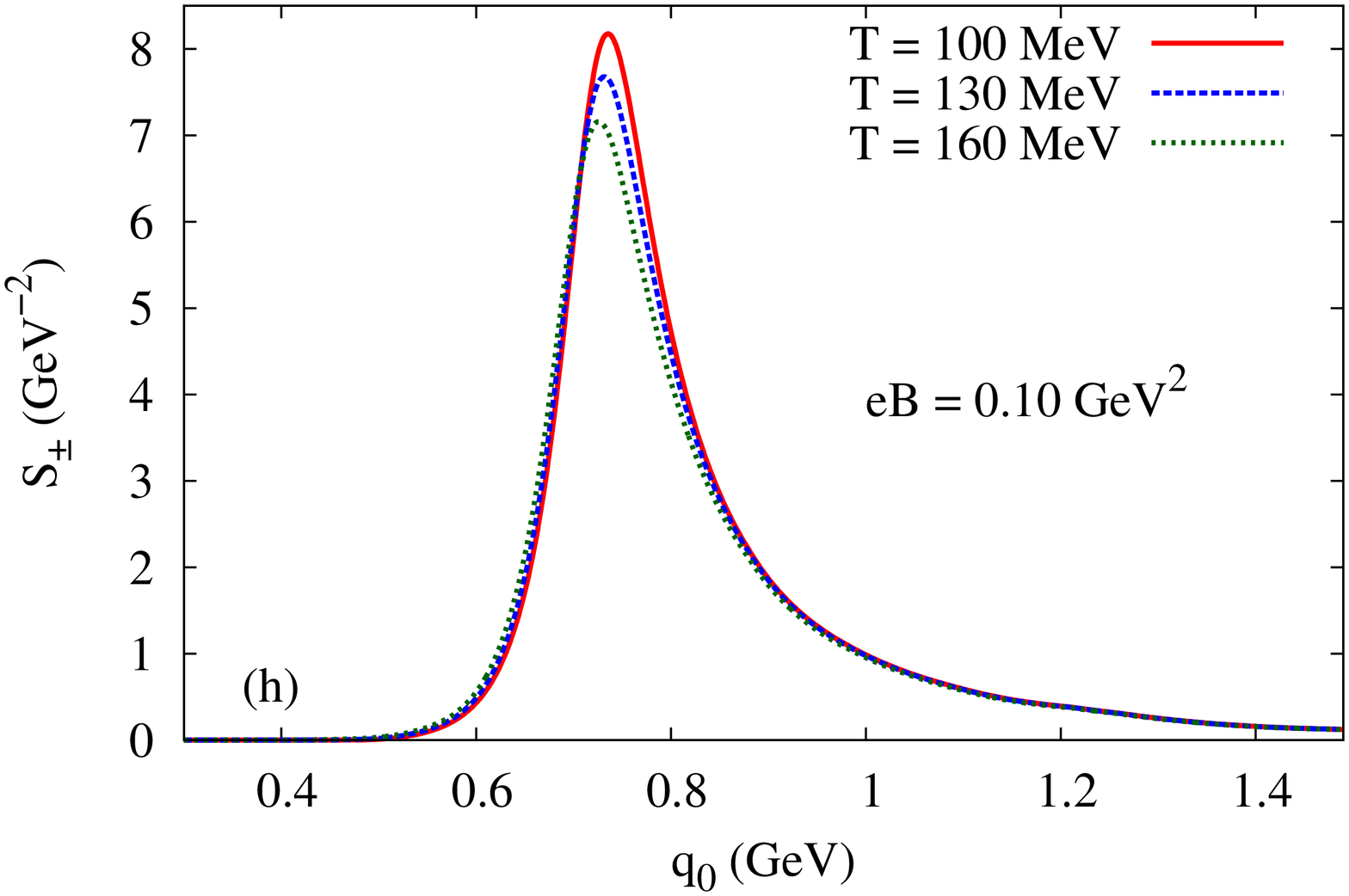}
\end{center}
\caption{The in-medium spectral functions of the $\rho^0$ 
at constant temperature (160 MeV) and at different values of the magnetic field (0.10, 0.15 and 0.20 GeV$^2$ respectively) 
(a) for $q^0$ from 0 to 0.35 GeV (b) for $q^0$ from 0.35 to 1.3 GeV. 
The same at constant magnetic field (0.10 GeV$^2$) and at different values of the temperature (100, 130 and 160 MeV respectively) 
(c) for $q^0$ from 0 to 0.25 GeV (d) for $q^0$ from 0.25 to 1.3 GeV. 
The in-medium spectral functions of the $\rho^\pm$ at constant temperature (160 MeV) 
and at different values of the magnetic field (0.10, 0.15 and 0.20 GeV$^2$ respectively) 
(e) for $q^0$ from 0 to 0.41 GeV (f) for $q^0$ from 0.41 to 1.3 GeV. 
The same at constant magnetic field (0.10 GeV$^2$) 
and at different values of the temperature (100, 130 and 160 MeV respectively) 
(g) for $q^0$ from 0 to 0.30 GeV (h) for $q^0$ from 0.30 to 1.3 GeV.}
\label{fig.spectra0_pm}
\end{figure}
\begin{figure}[h]
\begin{center}
\includegraphics[angle=-90, scale=0.30]{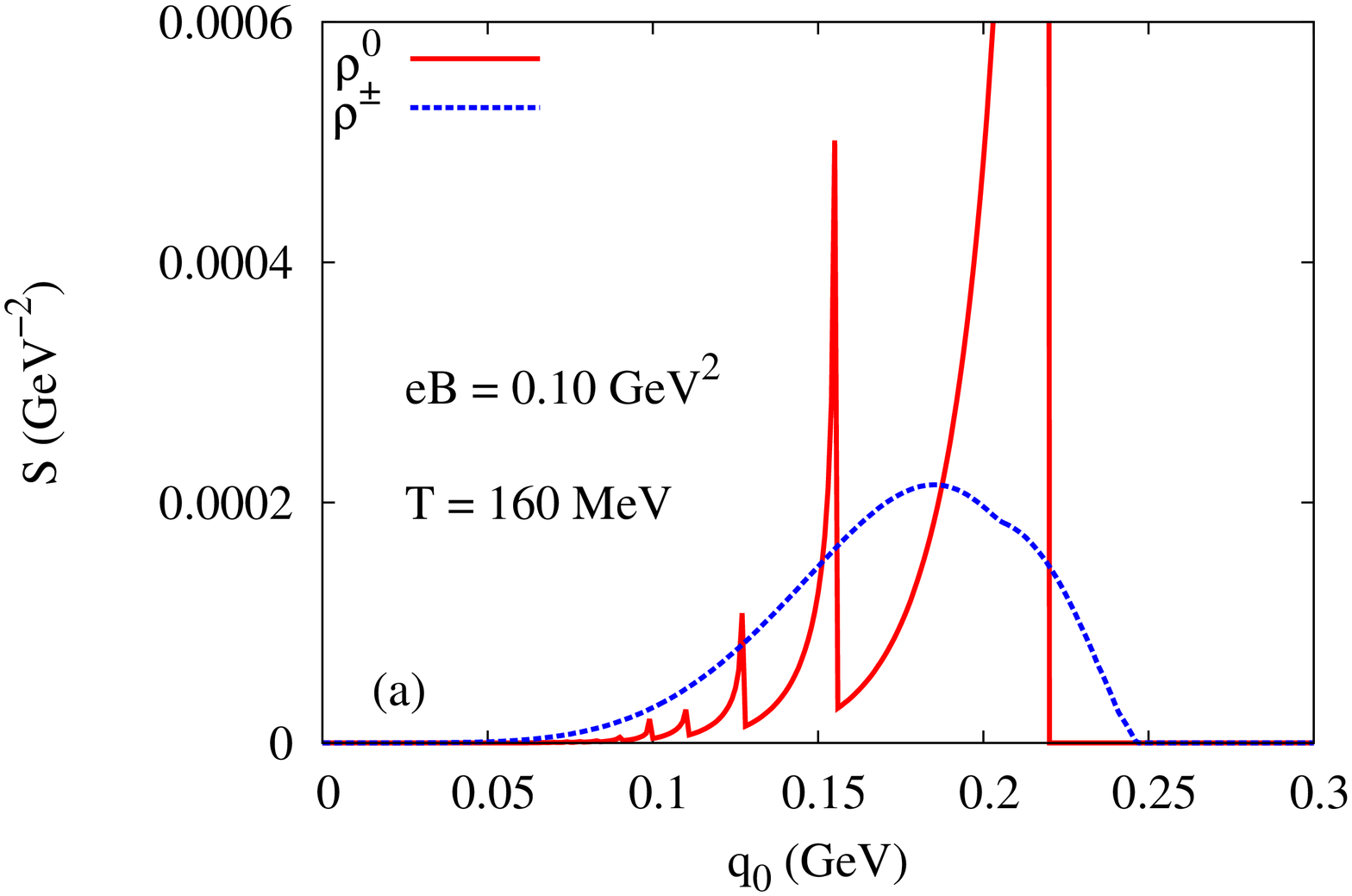} \includegraphics[angle=-90, scale=0.30]{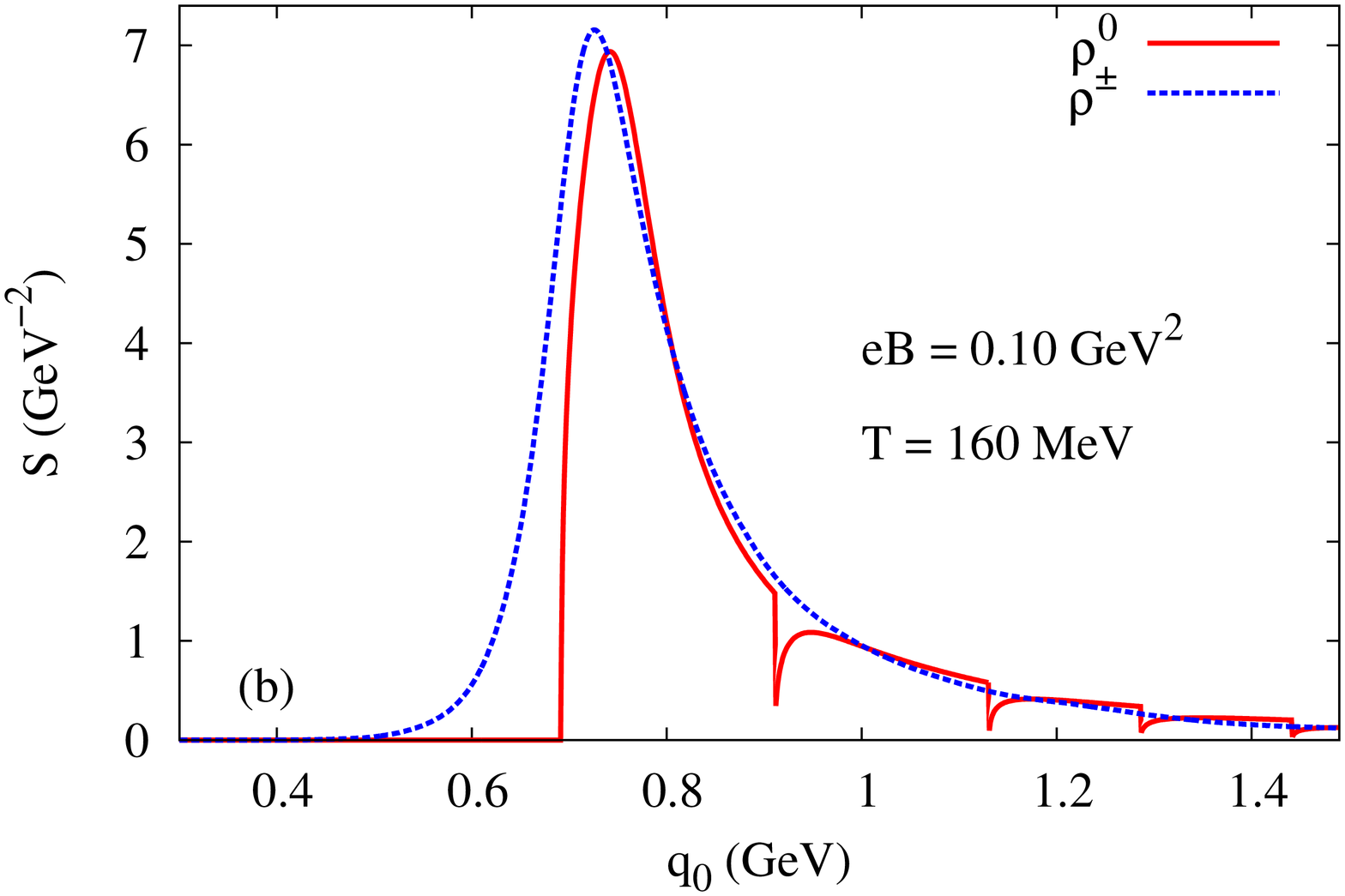}
\end{center}
\caption{The comparison of in-medium spectral functions between $\rho^0$ and $\rho^\pm$ at eB = 0.10 GeV$^2$ and T = 160 MeV 
(a) for $q^0$ from 0 to 0.30 GeV (a) for $q^0$ from 0.30 to 1.3 GeV.}
\label{fig.spectra0pm}
\end{figure}
\begin{figure}[h]
\begin{center}
\includegraphics[angle=-90, scale=0.230]{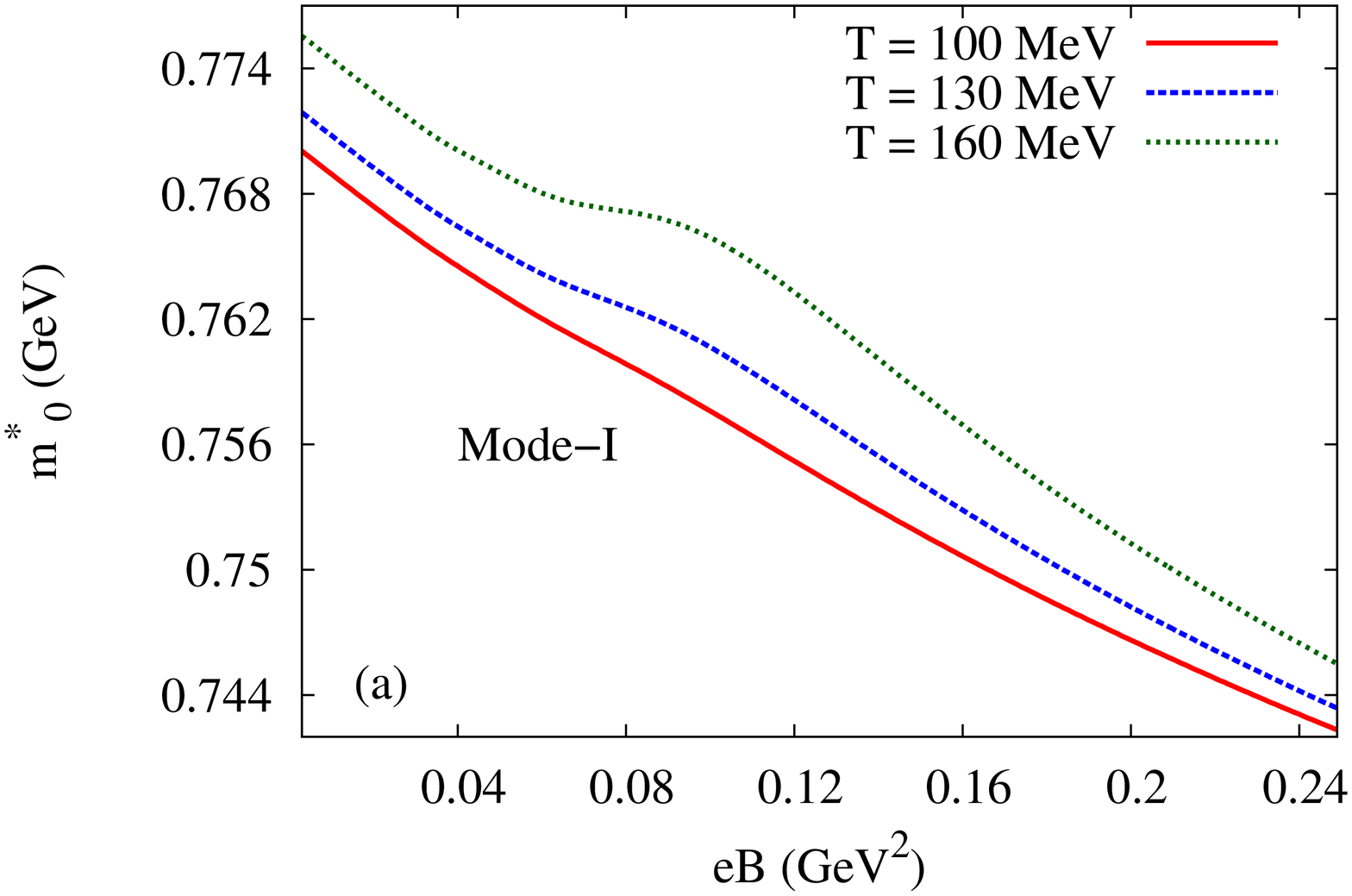} \includegraphics[angle=-90, scale=0.230]{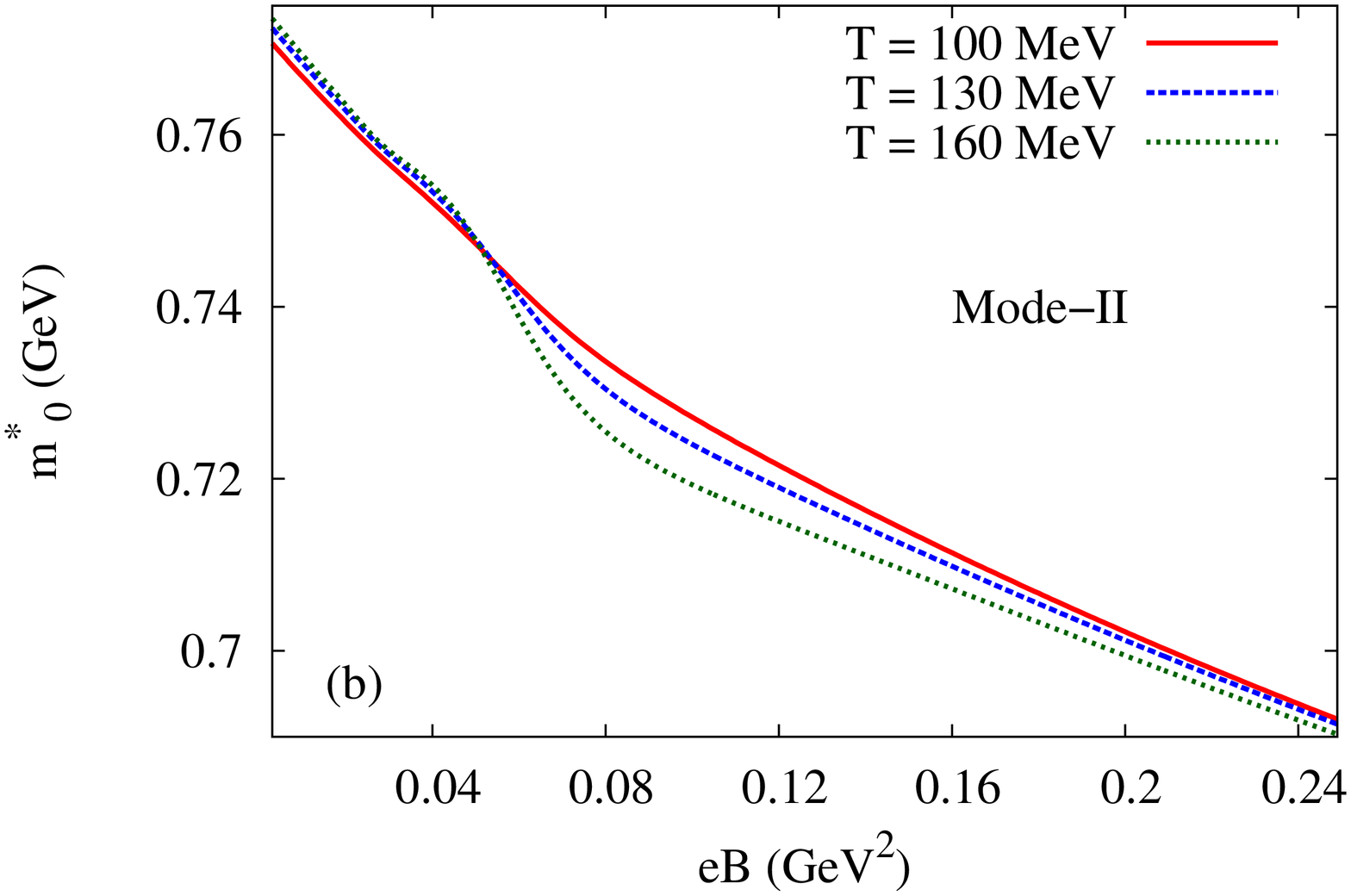} 
\includegraphics[angle=-90, scale=0.230]{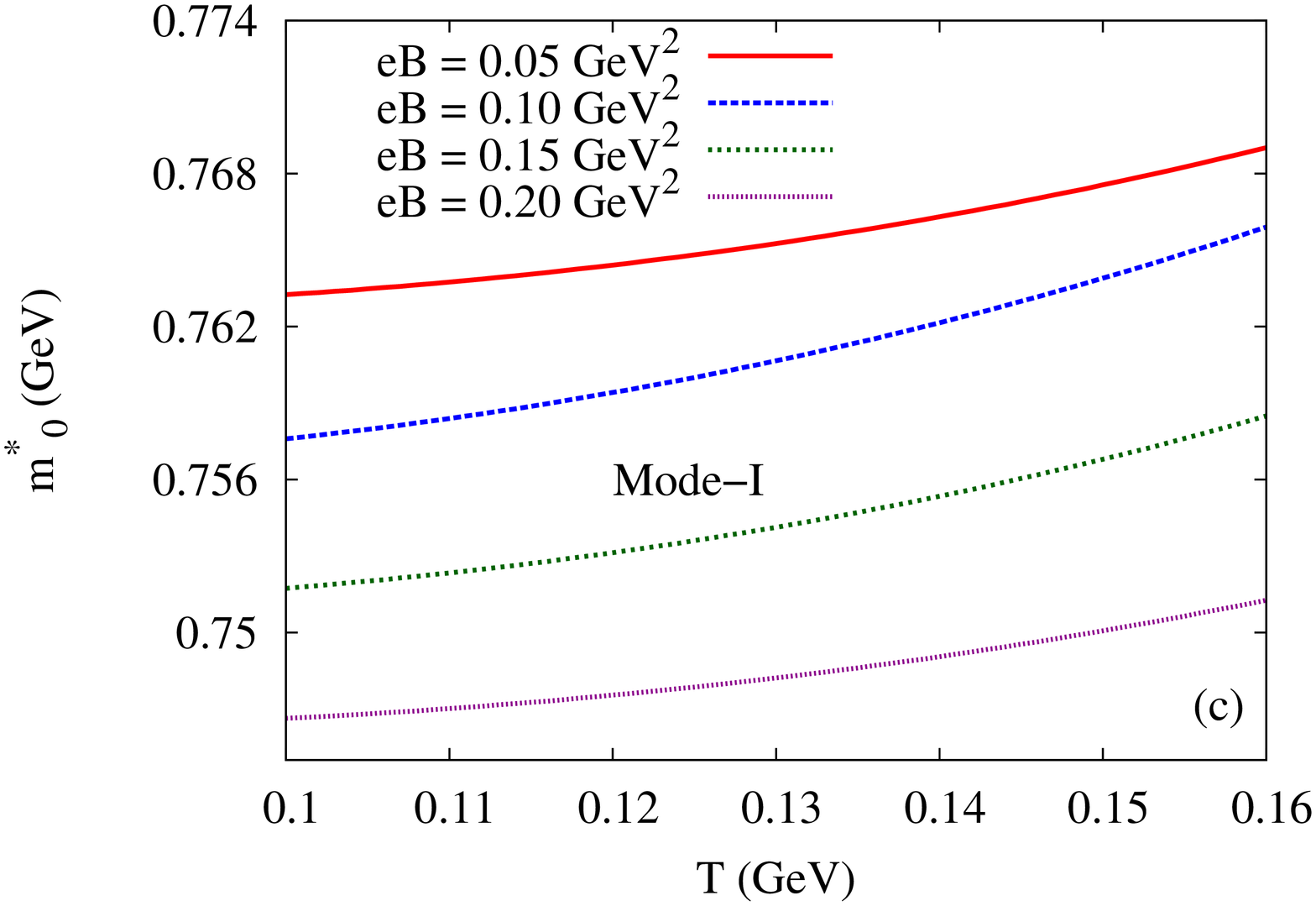} \\ \includegraphics[angle=-90, scale=0.230]{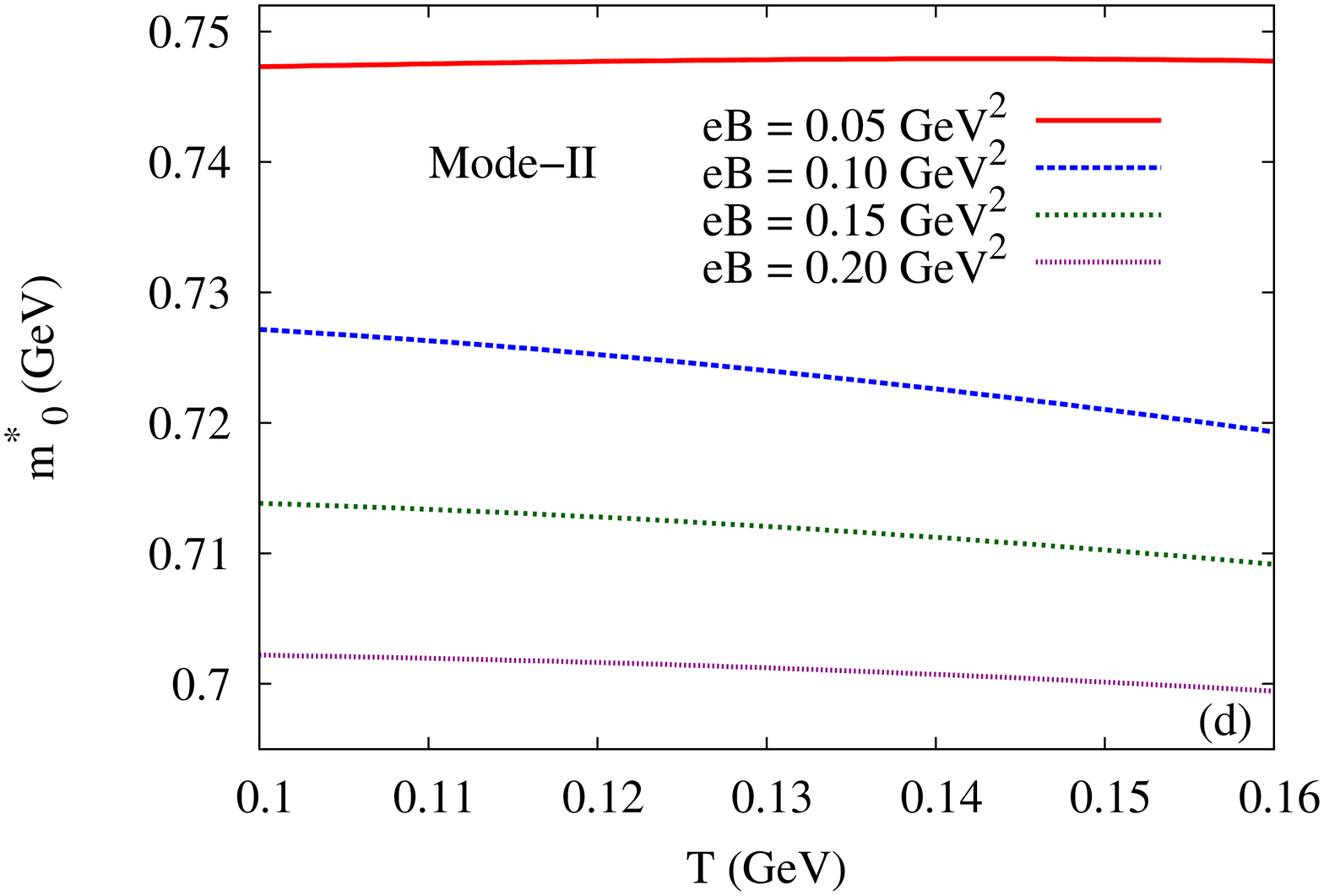}
\includegraphics[angle=-90, scale=0.230]{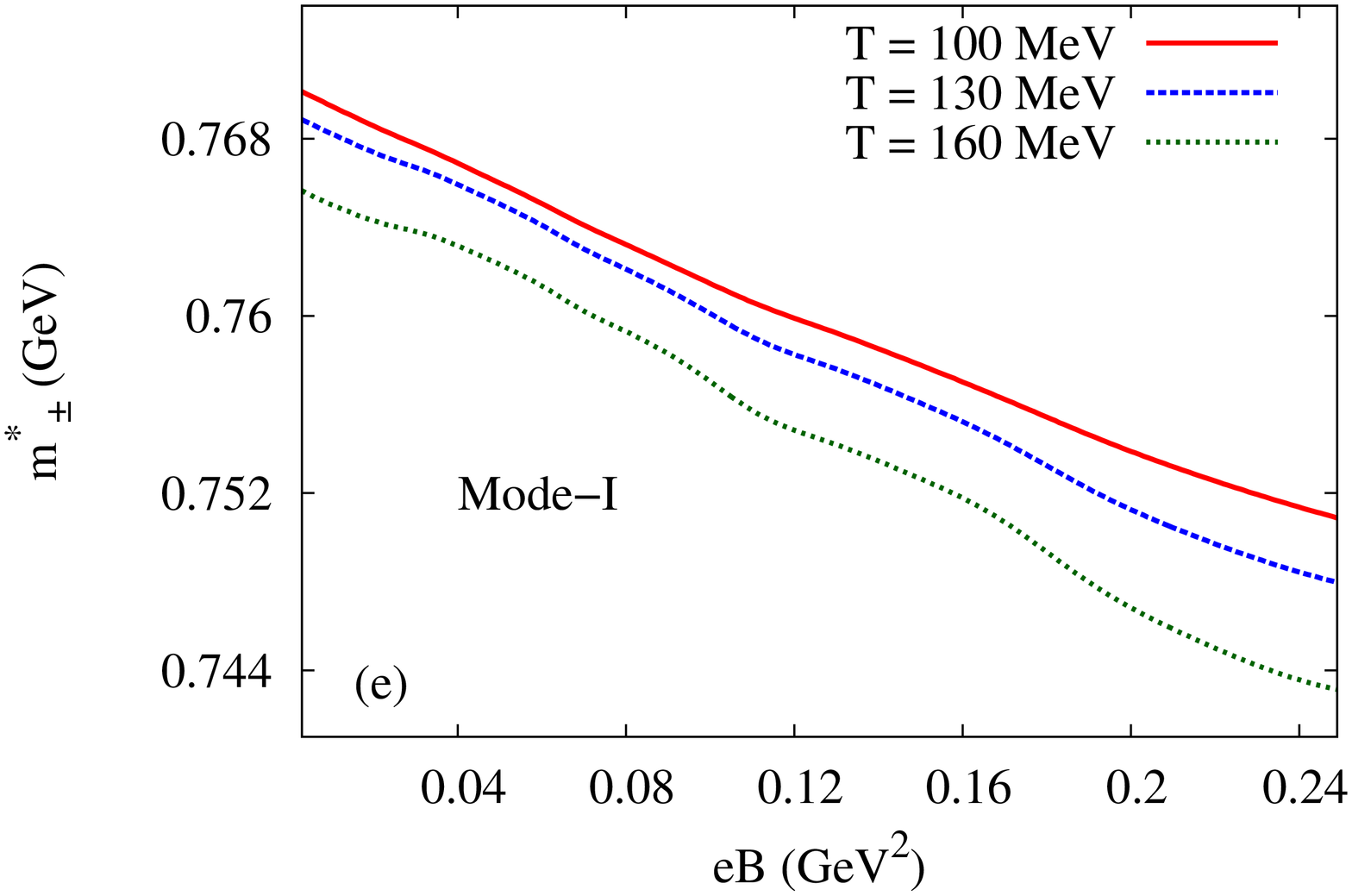} \includegraphics[angle=-90, scale=0.230]{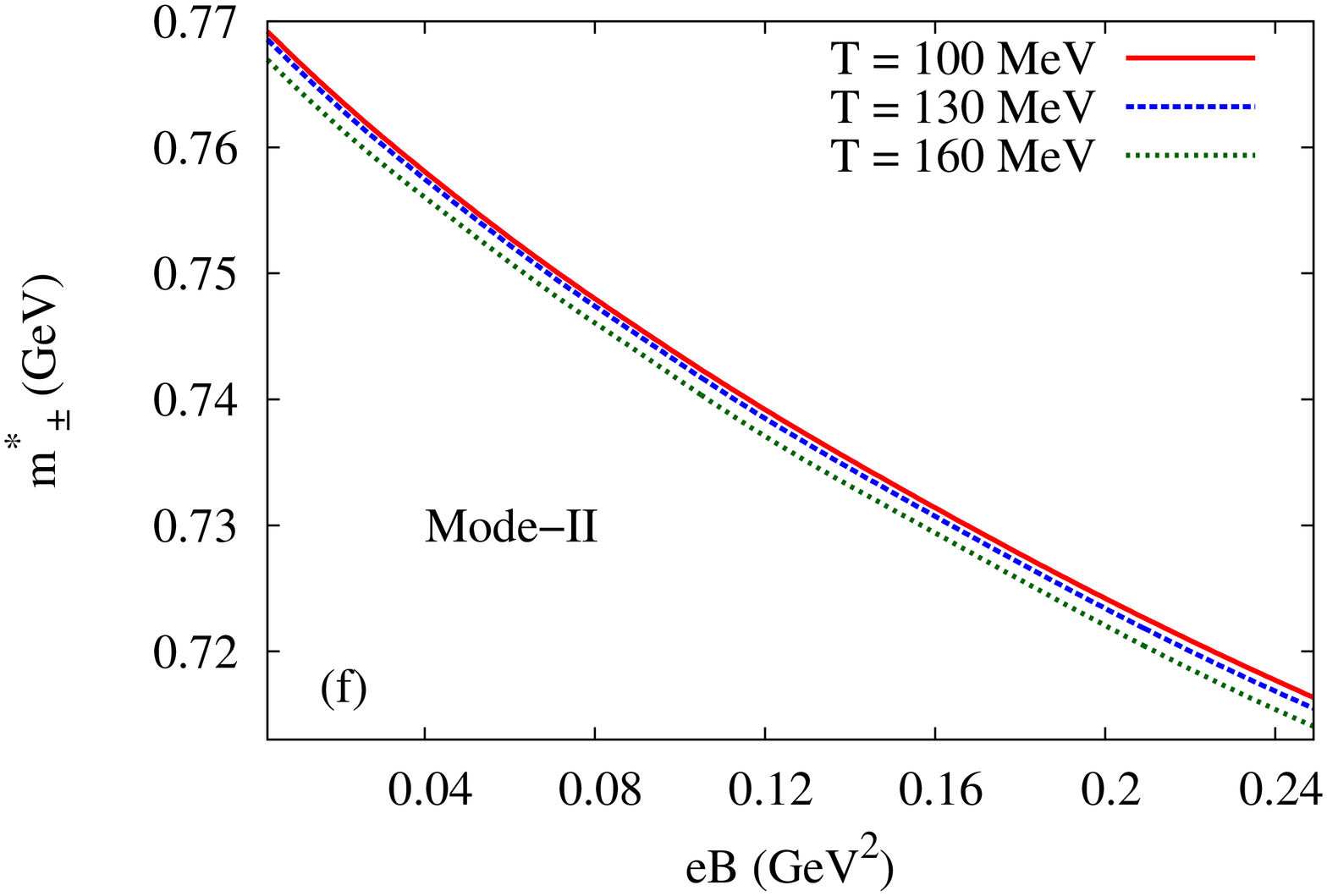} \\
\includegraphics[angle=-90, scale=0.230]{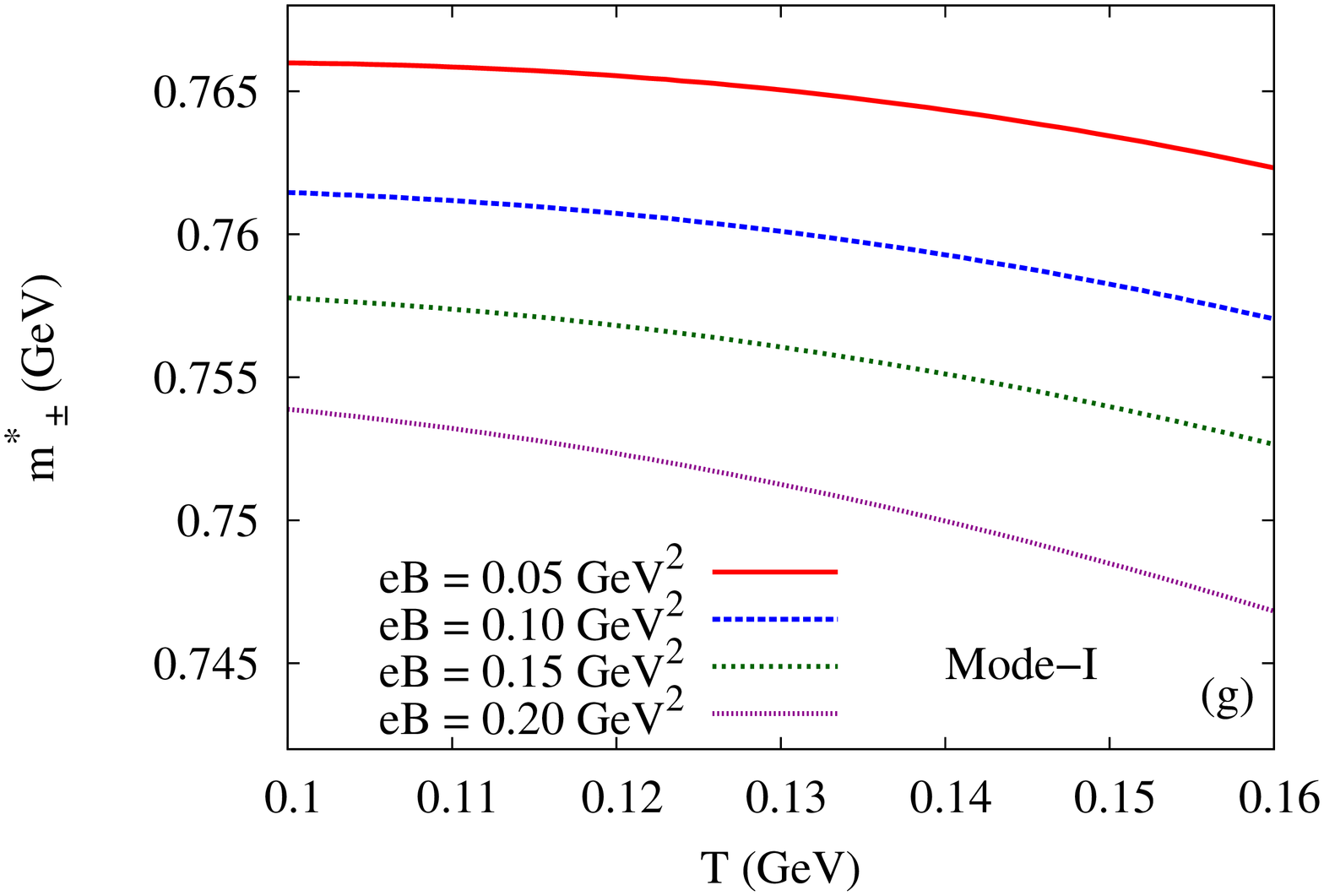} \includegraphics[angle=-90, scale=0.230]{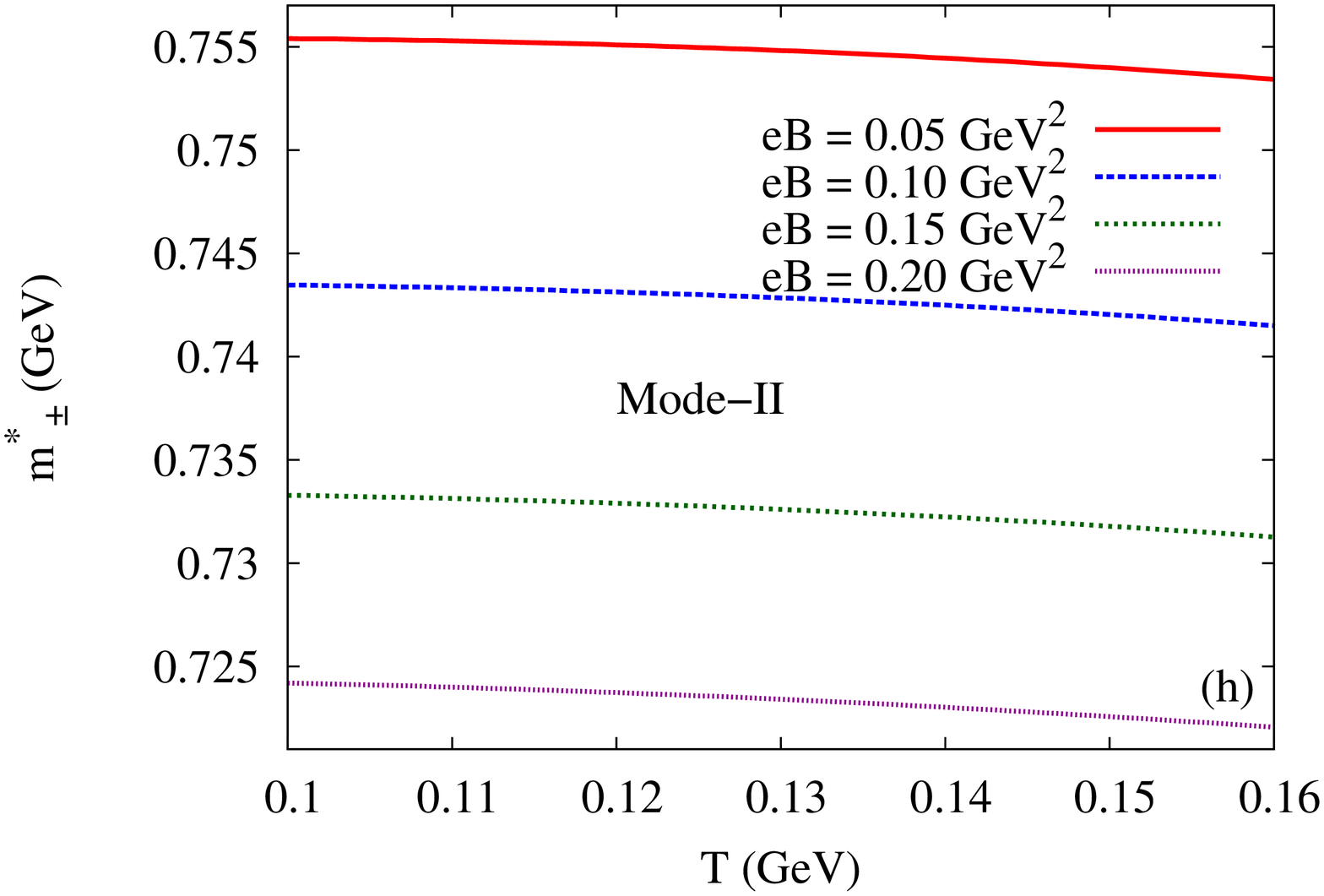}
\end{center}
\caption{The effective masses of the $\rho^0$ vs $eB$ at different values of temperatures (100, 130 and 160 MeV) for
(a) Mode-I and (b) Mode-II. 
The effective masses of the $\rho^0$ vs $T$ at different values of the magnetic field (0.05, 0.10, 0.15 and 0.20 GeV$^2$) for
(c) Mode-I and (d) Mode-II. 
The effective masses of the $\rho^\pm$ vs $eB$ at different values of temperatures (100, 130 and 160 MeV) for
(a) Mode-I and (b) Mode-II. 
The effective masses of the $\rho^\pm$ vs $T$ at different values of the magnetic field (0.05, 0.10, 0.15 and 0.20 GeV$^2$) for
(c) Mode-I and (d) Mode-II. }
\label{fig.mass_0pm}
\end{figure}
\begin{figure}[h]
\begin{center}
\includegraphics[angle=-90, scale=0.230]{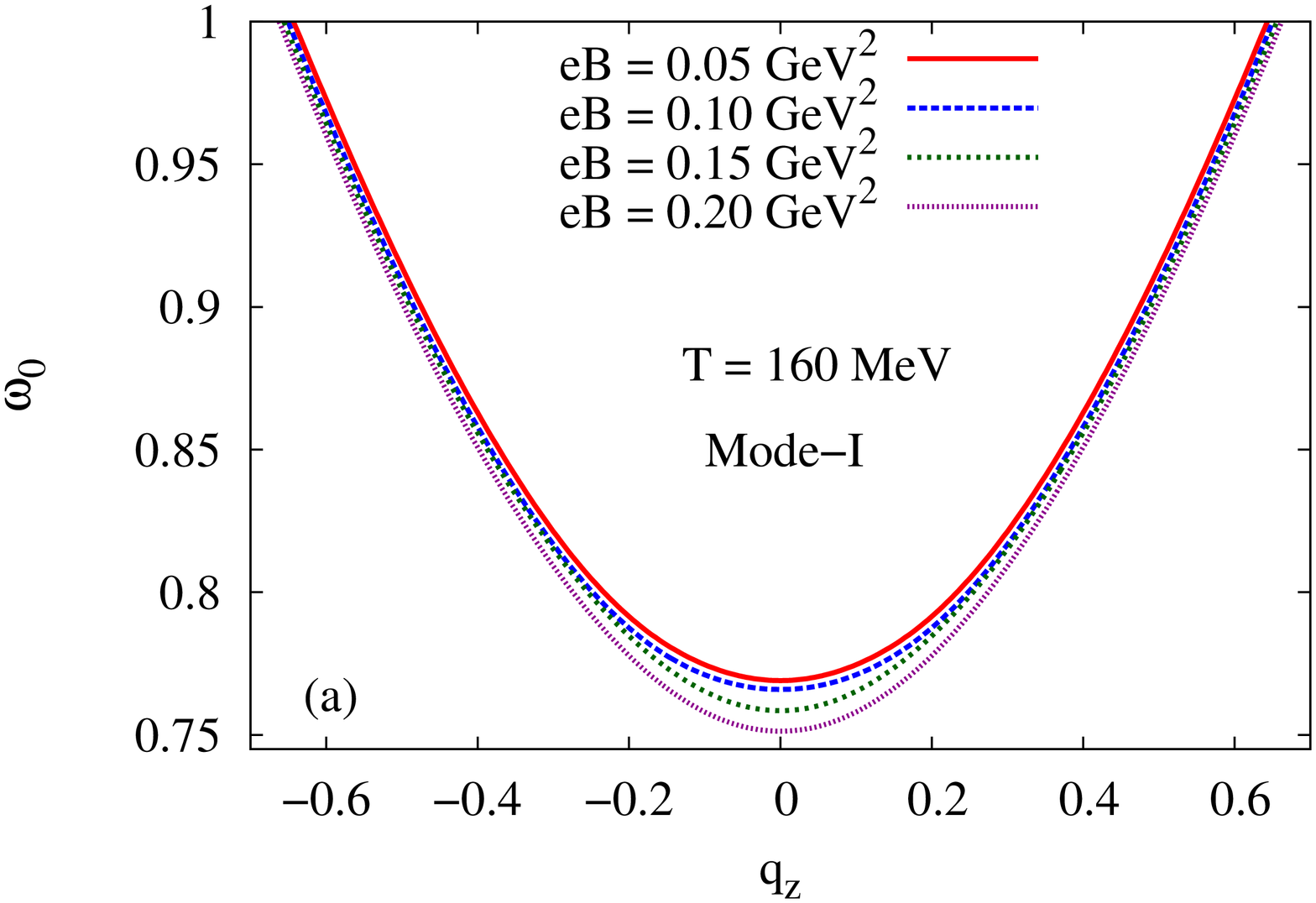} \includegraphics[angle=-90, scale=0.230]{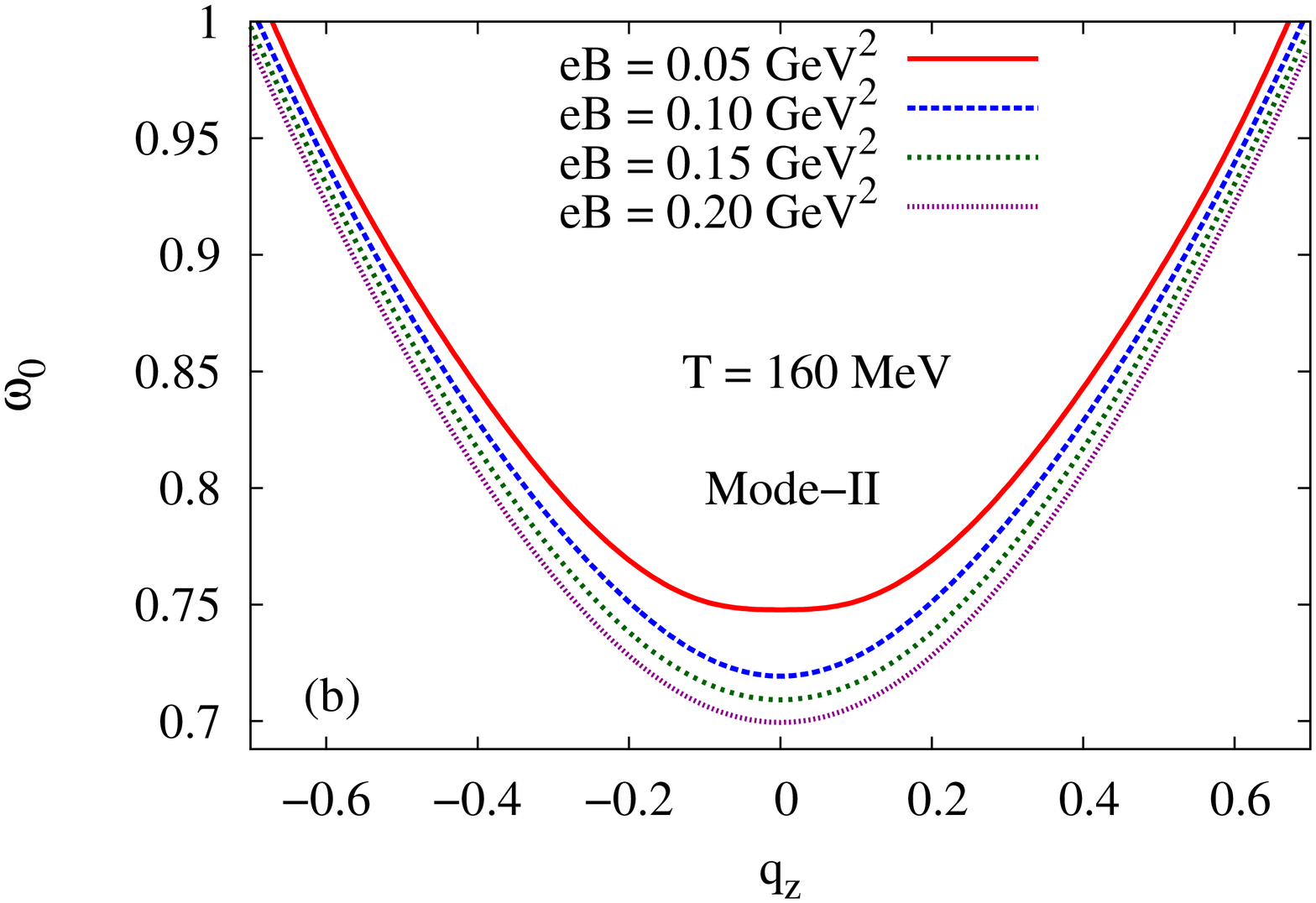}
\includegraphics[angle=-90, scale=0.230]{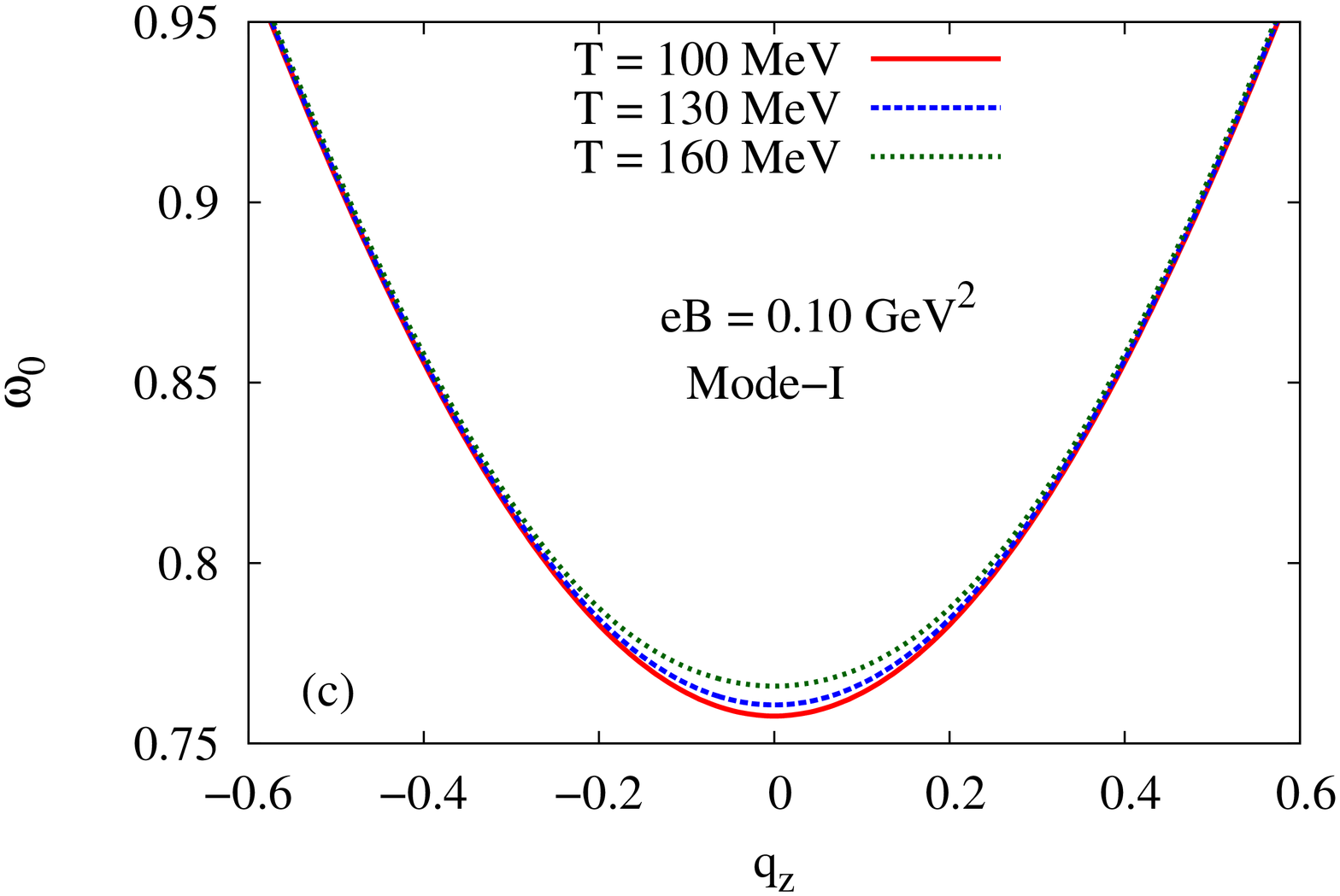} \\ \includegraphics[angle=-90, scale=0.230]{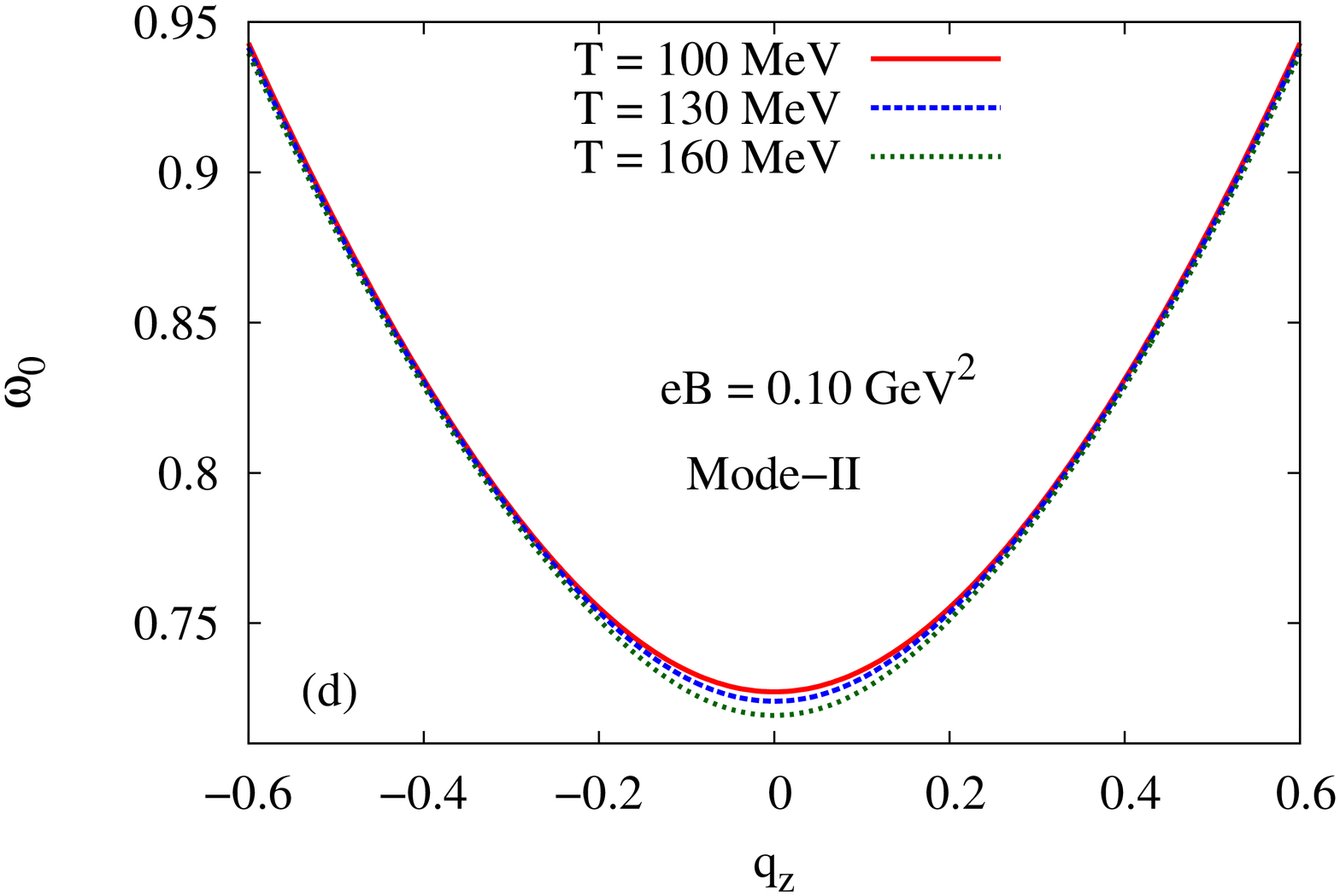}
\includegraphics[angle=-90, scale=0.230]{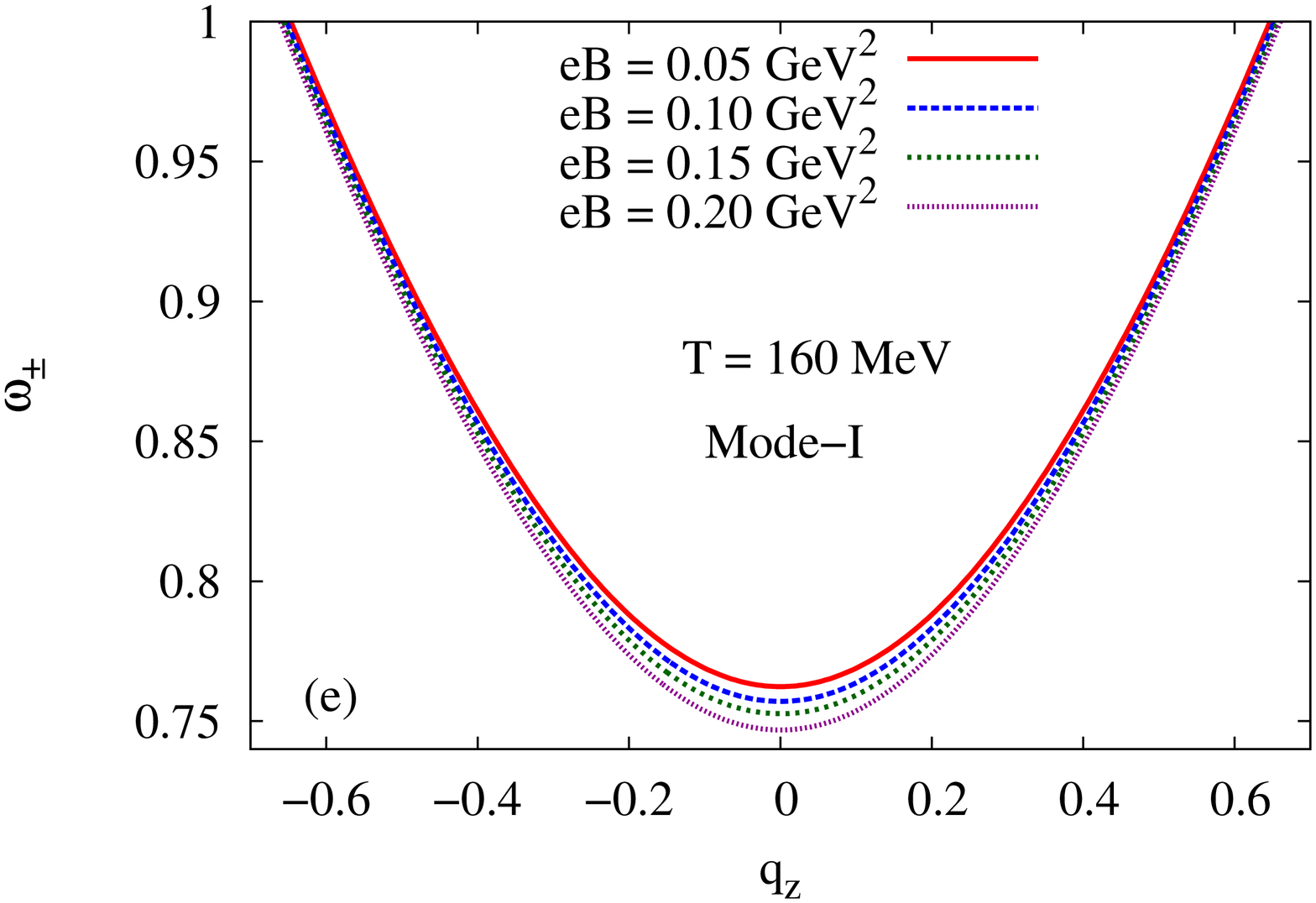} \includegraphics[angle=-90, scale=0.230]{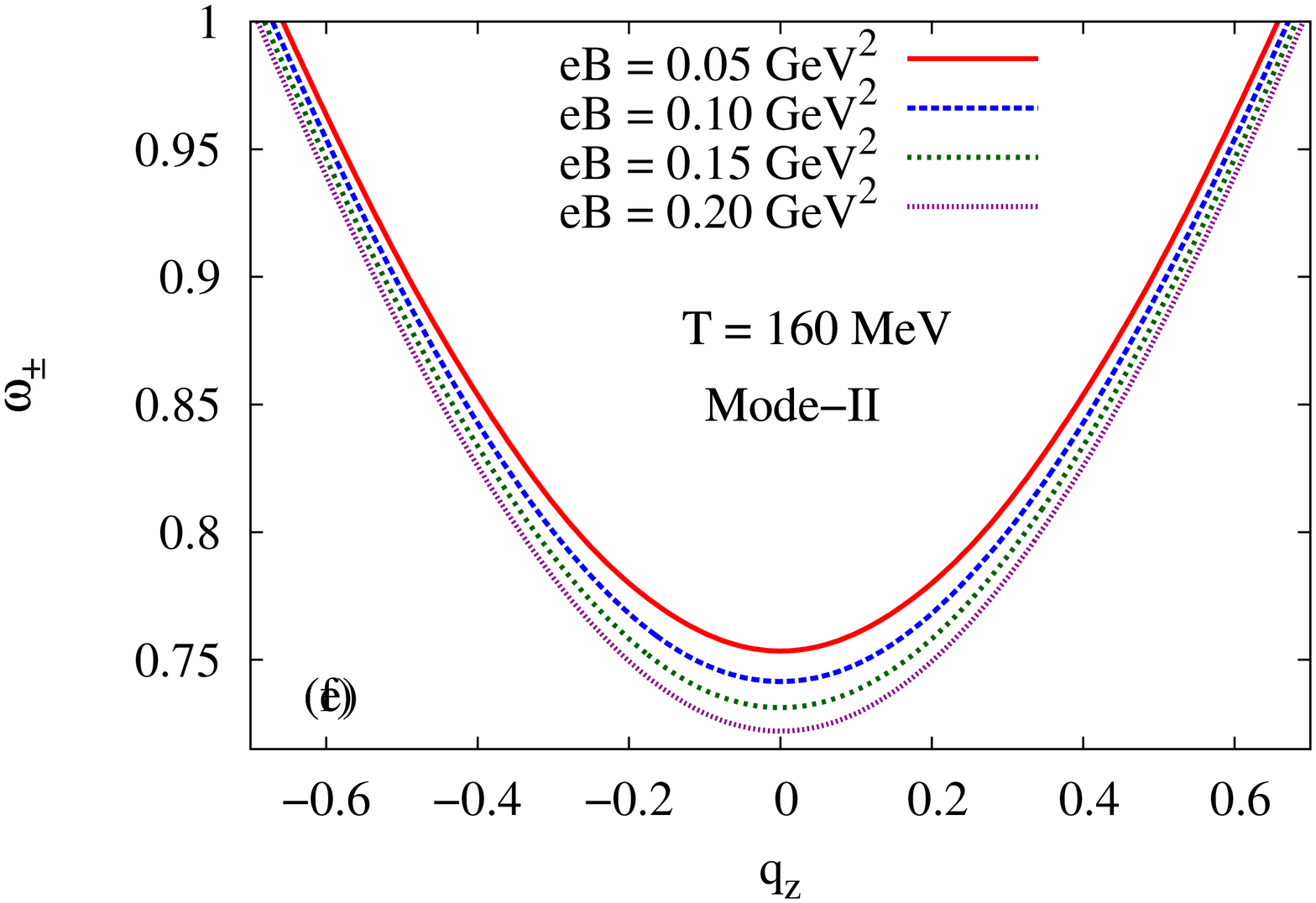}
\includegraphics[angle=-90, scale=0.230]{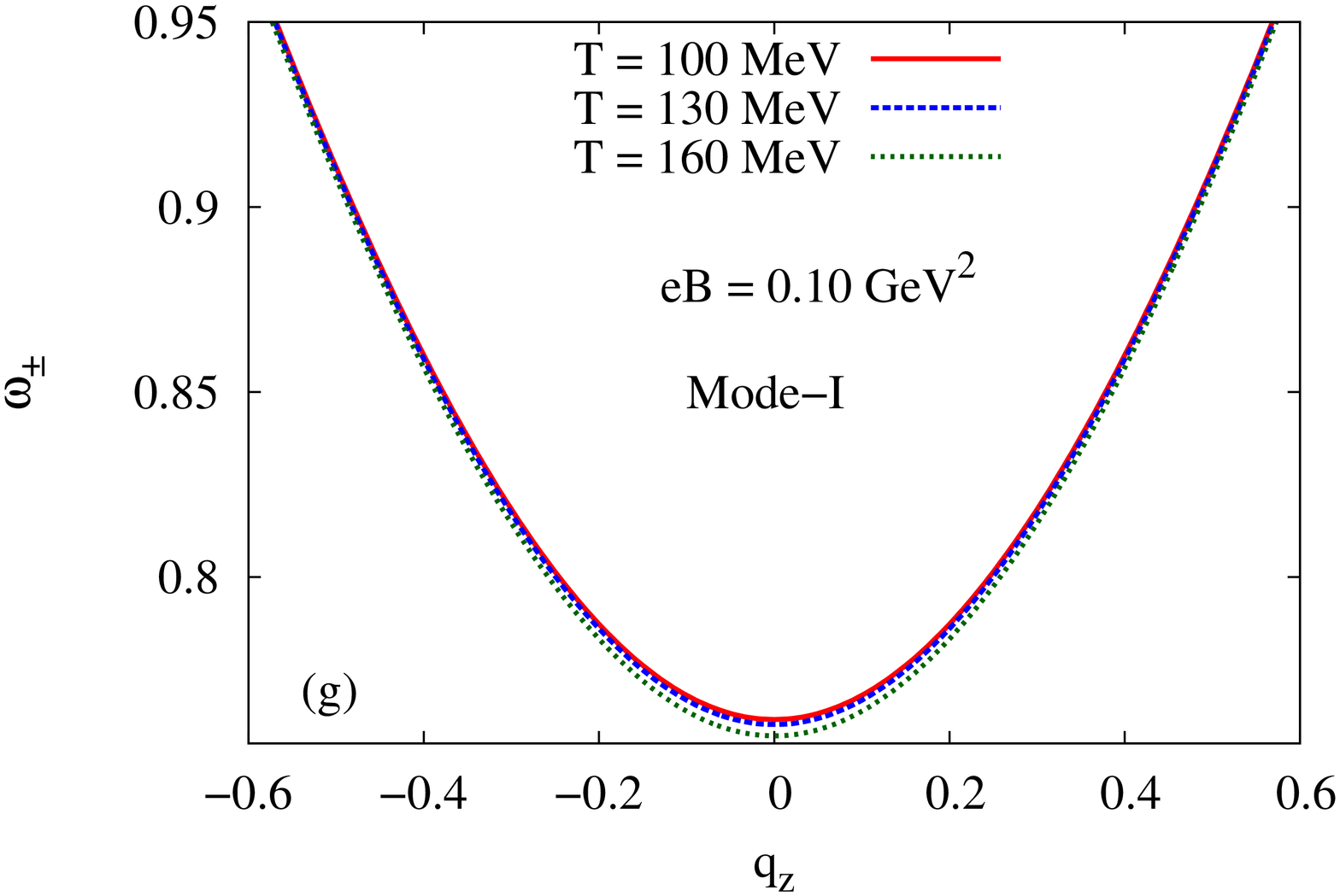} \includegraphics[angle=-90, scale=0.230]{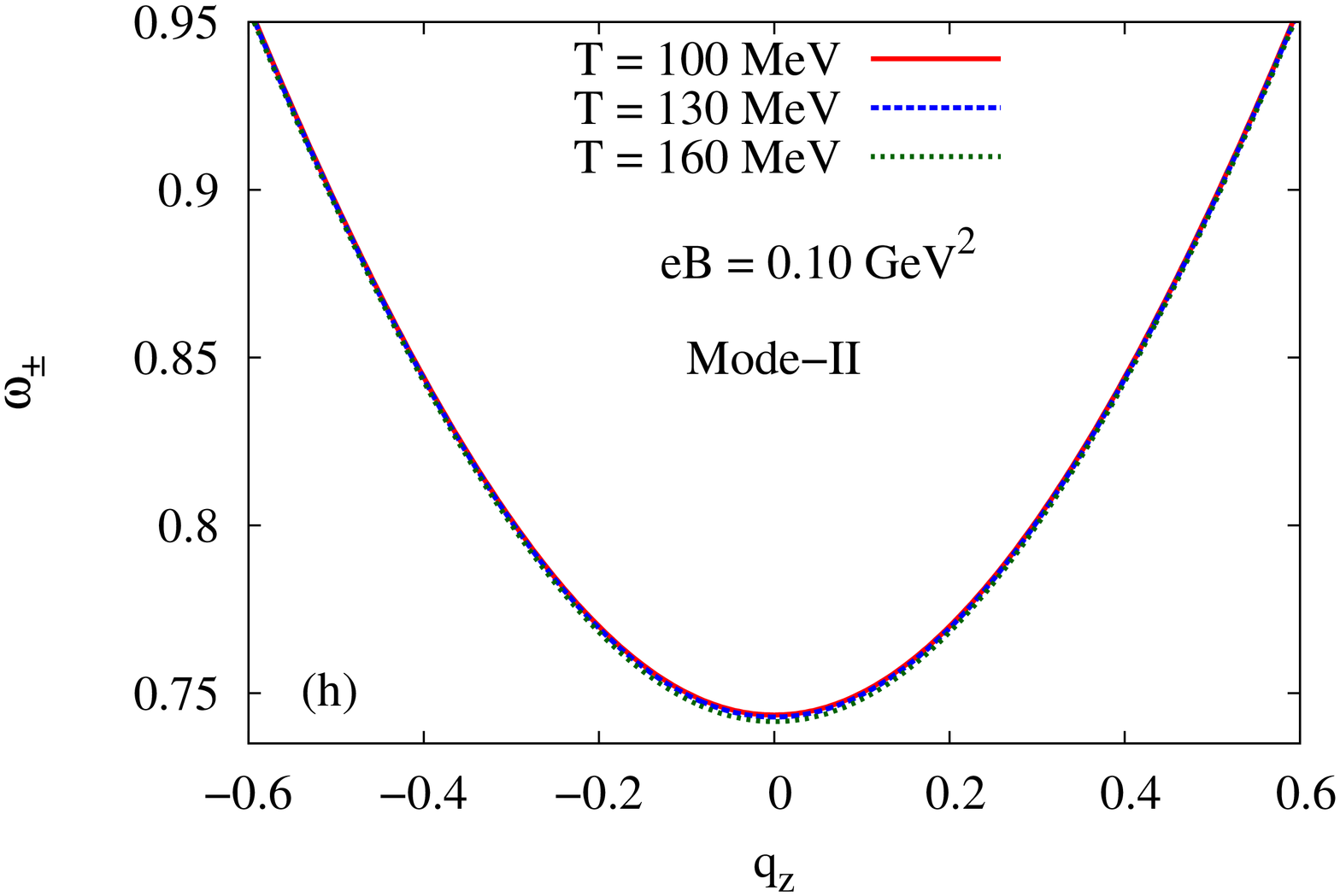}
\end{center}
\caption{The dispersion curves of the $\rho^0$ at constant temperature (160 MeV) and at different values of the magnetic 
field (0.05, 0.10, 0.15 and 0.20 GeV$^2$ respectively) for (a) Mode-I and (b) Mode-II. 
The same at constant magnetic filed (0.10 GeV$^2$) and at different values of the temperature 
(100, 130, and 160 MeV respectively) for (a) Mode-I and (b) Mode-II. 
The dispersion curves of the $\rho^\pm$ at constant temperature (160 MeV) and at different values of the magnetic 
field (0.05, 0.10, 0.15 and 0.20 GeV$^2$ respectively) for (a) Mode-I and (b) Mode-II. 
The same at constant magnetic filed (0.10 GeV$^2$) and at different values of the temperature 
(100, 130, and 160 MeV respectively) for (a) Mode-I and (b) Mode-II. }
\label{fig.disp_0pm}
\end{figure}
\begin{figure}[h]
\begin{center}
\includegraphics[angle=-90, scale=0.30]{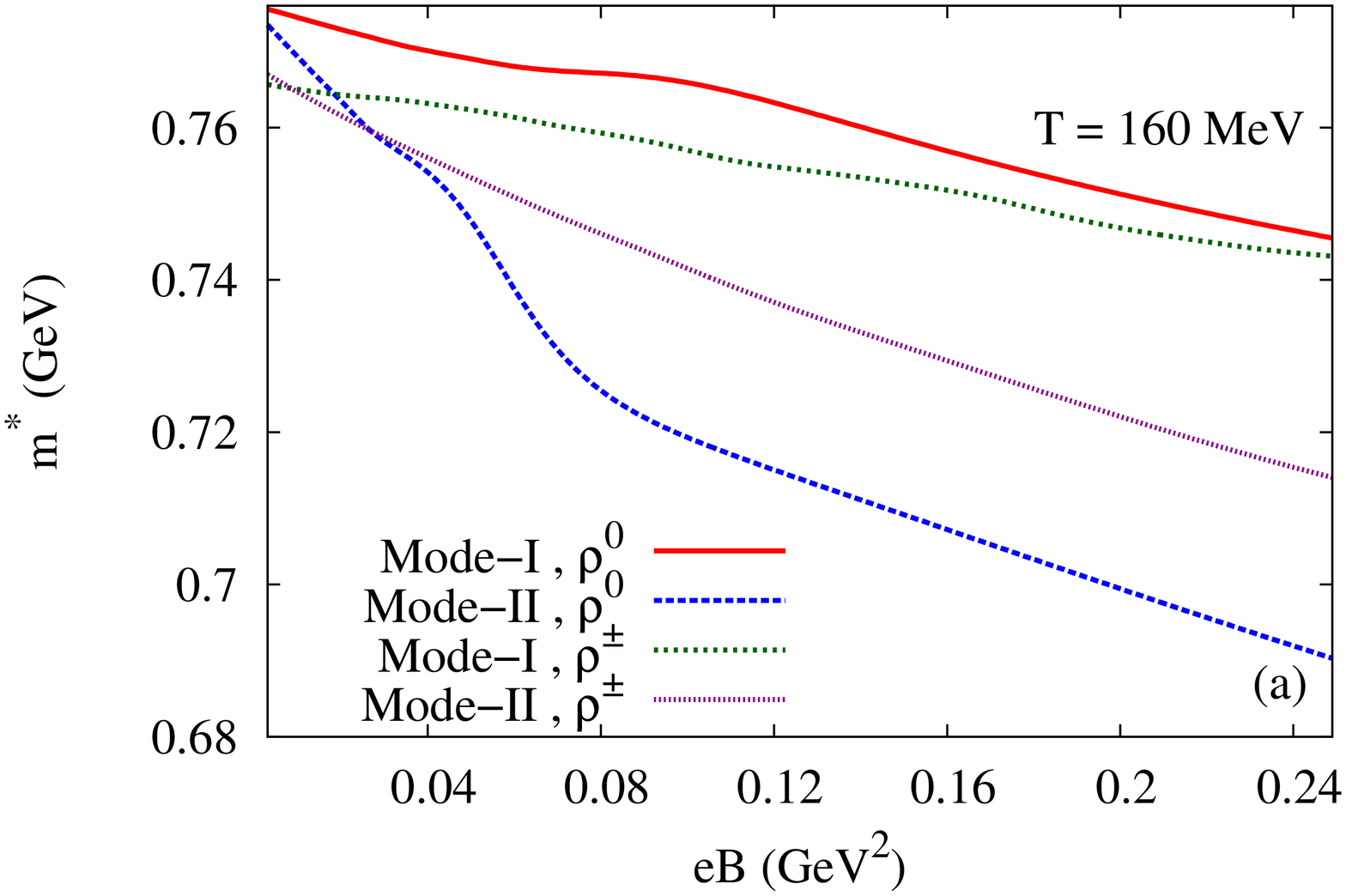} \includegraphics[angle=-90, scale=0.30]{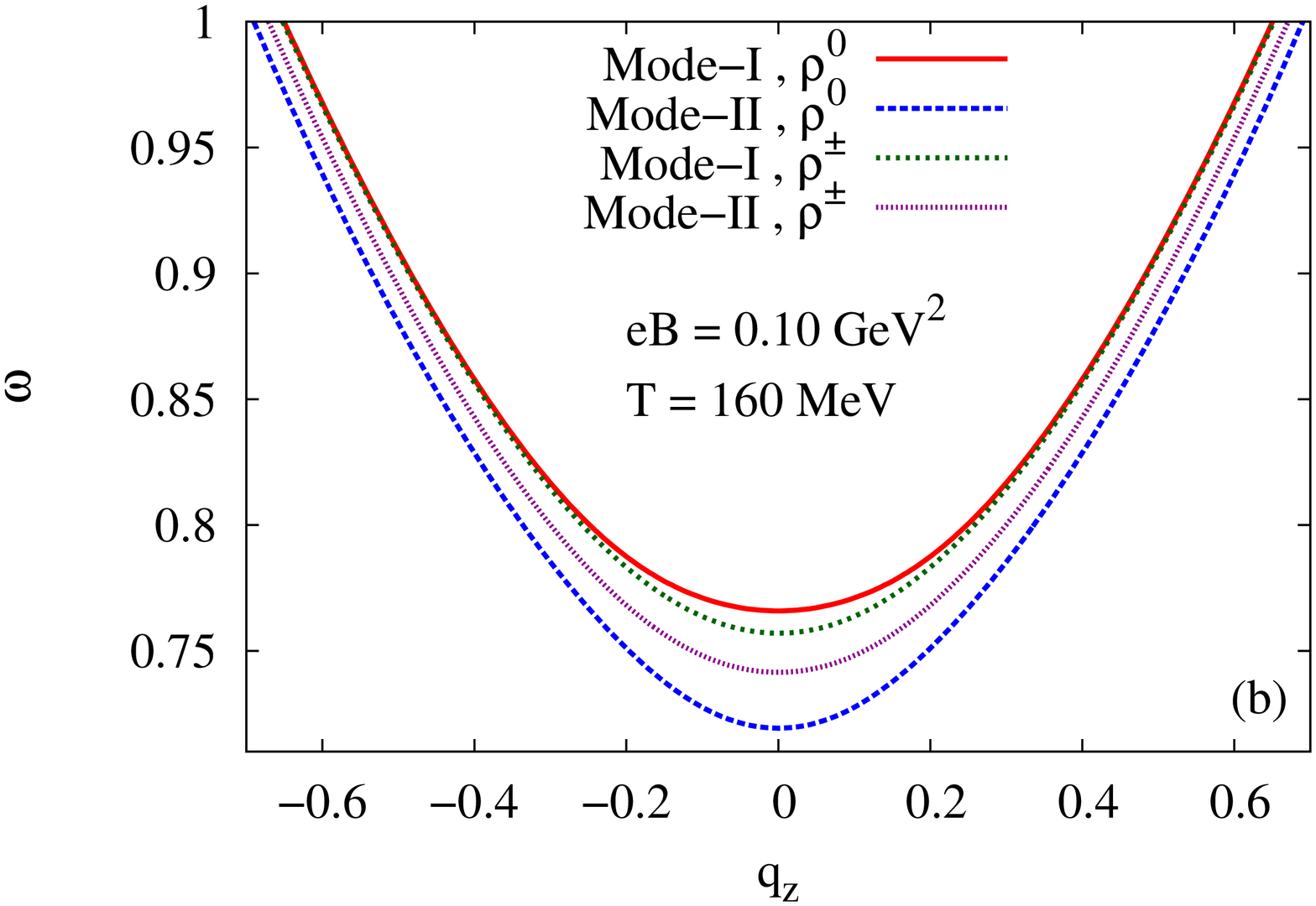}
\end{center}
\caption{The comparison of (a) effective mass (b) dispersion curve between the $\rho^0$ and $\rho^\pm$ at 
constant temperature (160 MeV) and magnetic field (0.1 GeV$^2$) for Mode-I and Mode-II.}
\label{fig.mass_disp}
\end{figure}
\begin{figure}[h]
	\begin{center}
		\includegraphics[angle=-90, scale=0.30]{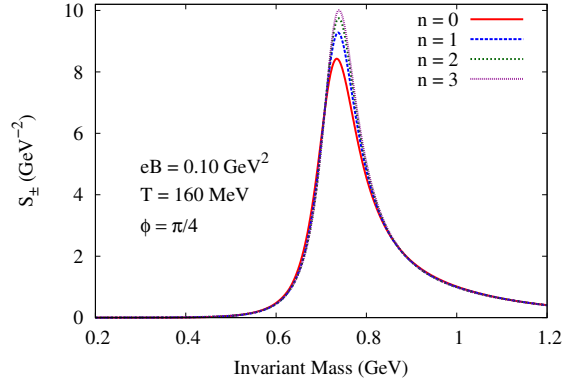}
	\end{center}
	\caption{The in-medium spectral functions of the $\rho^\pm$ at constant temperature (160 MeV), 
		magnetic field (0.10 GeV$^2$) and azimuthal angle ($\frac{\pi}{4}$) at four Landau levels ($n$ = 0, 1, 2 and 3)}
	\label{fig.spectrapm_5}
\end{figure}
\begin{figure}[h]
	\begin{center}
		\includegraphics[angle=-90, scale=0.23]{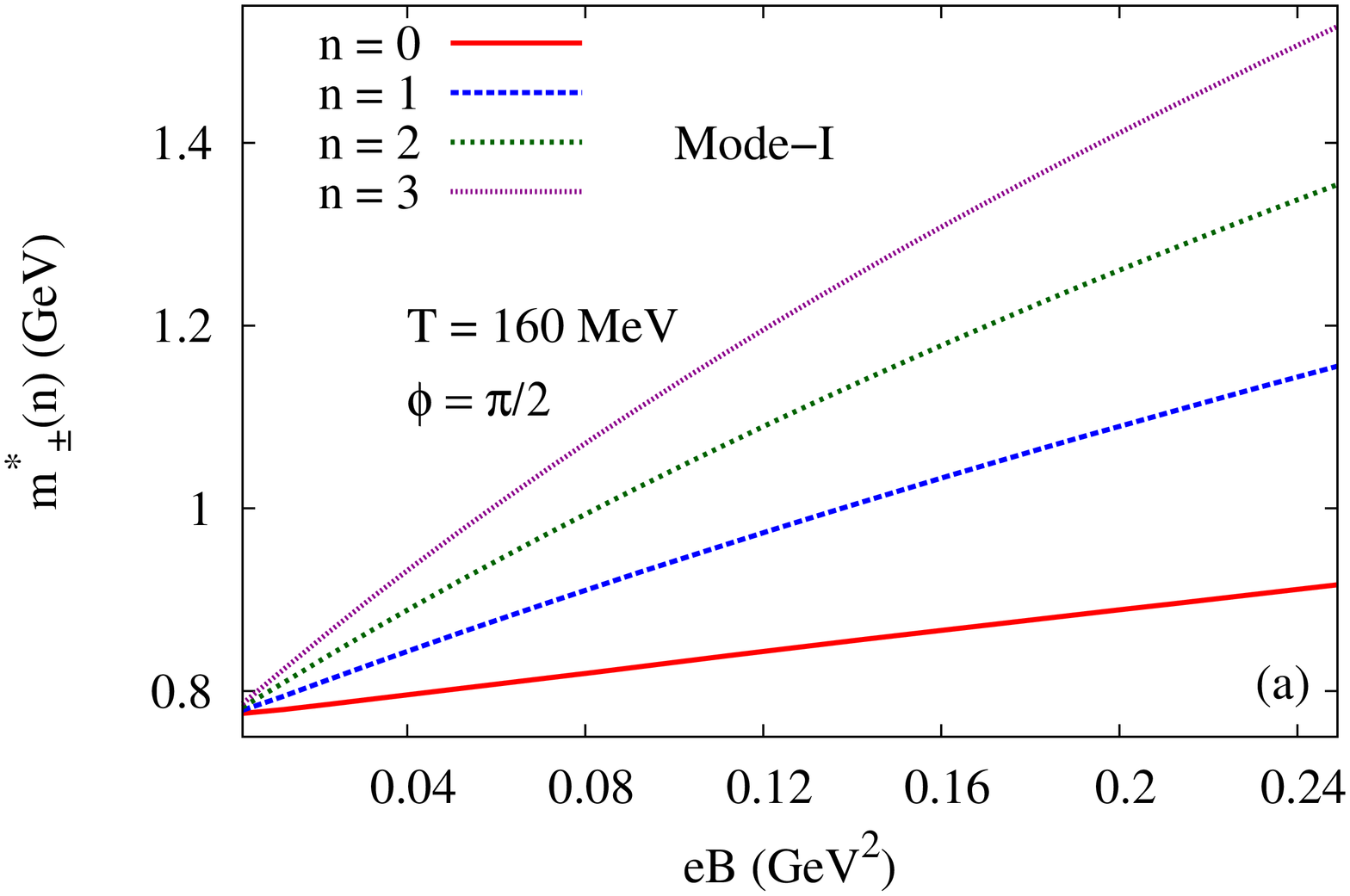} \includegraphics[angle=-90, scale=0.23]{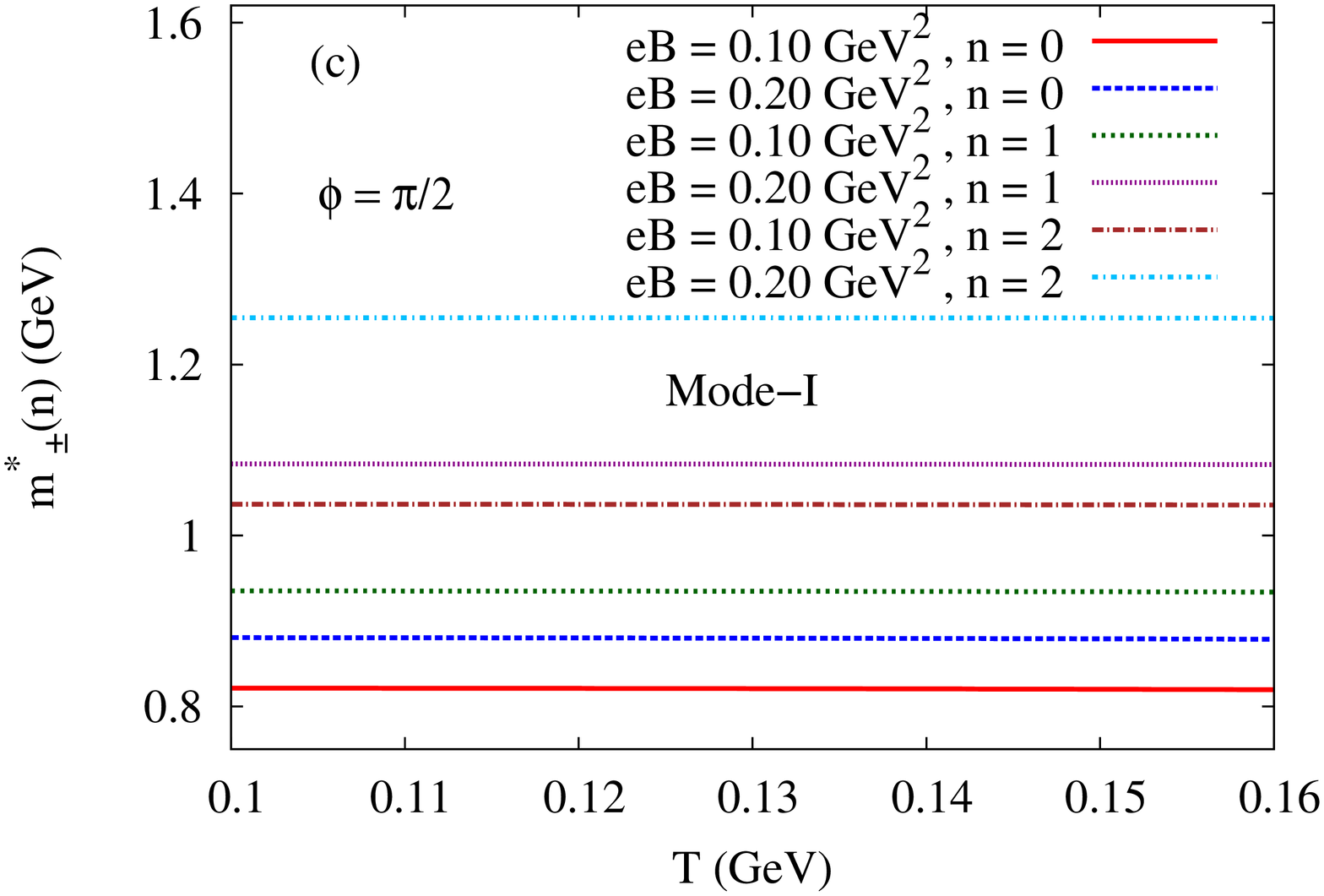}
		\includegraphics[angle=-90, scale=0.23]{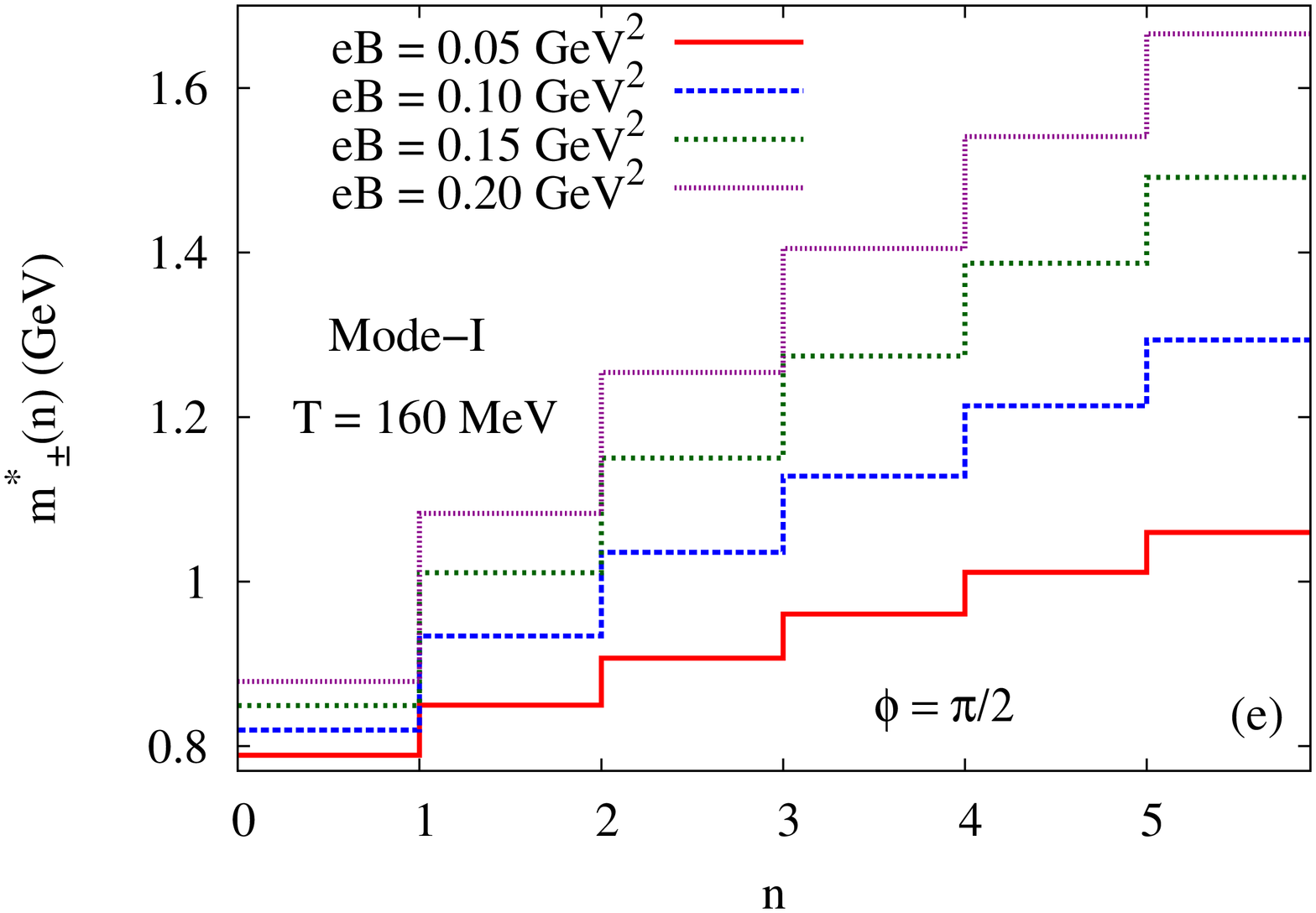} \\
		\includegraphics[angle=-90, scale=0.23]{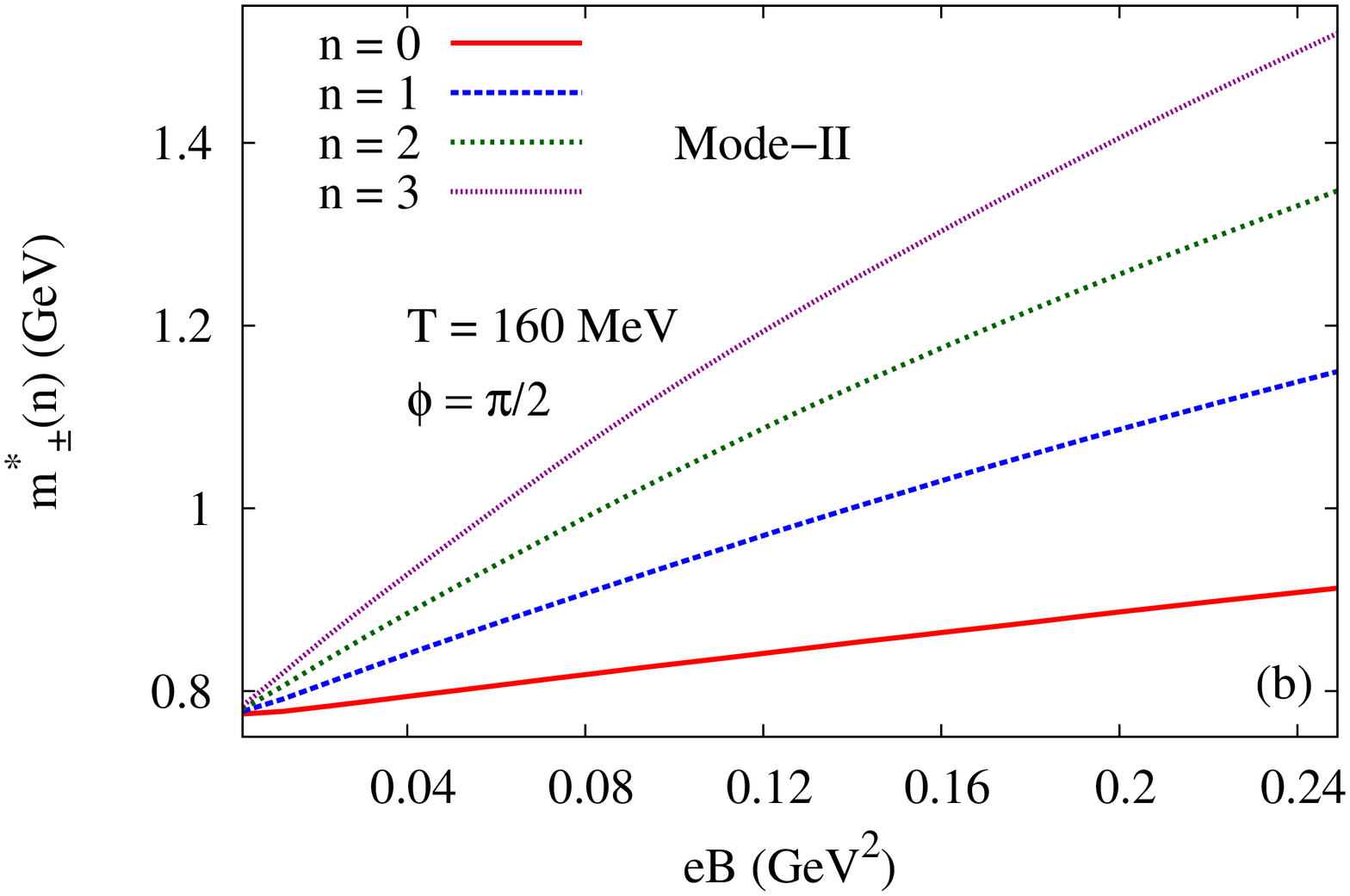}
		\includegraphics[angle=-90, scale=0.23]{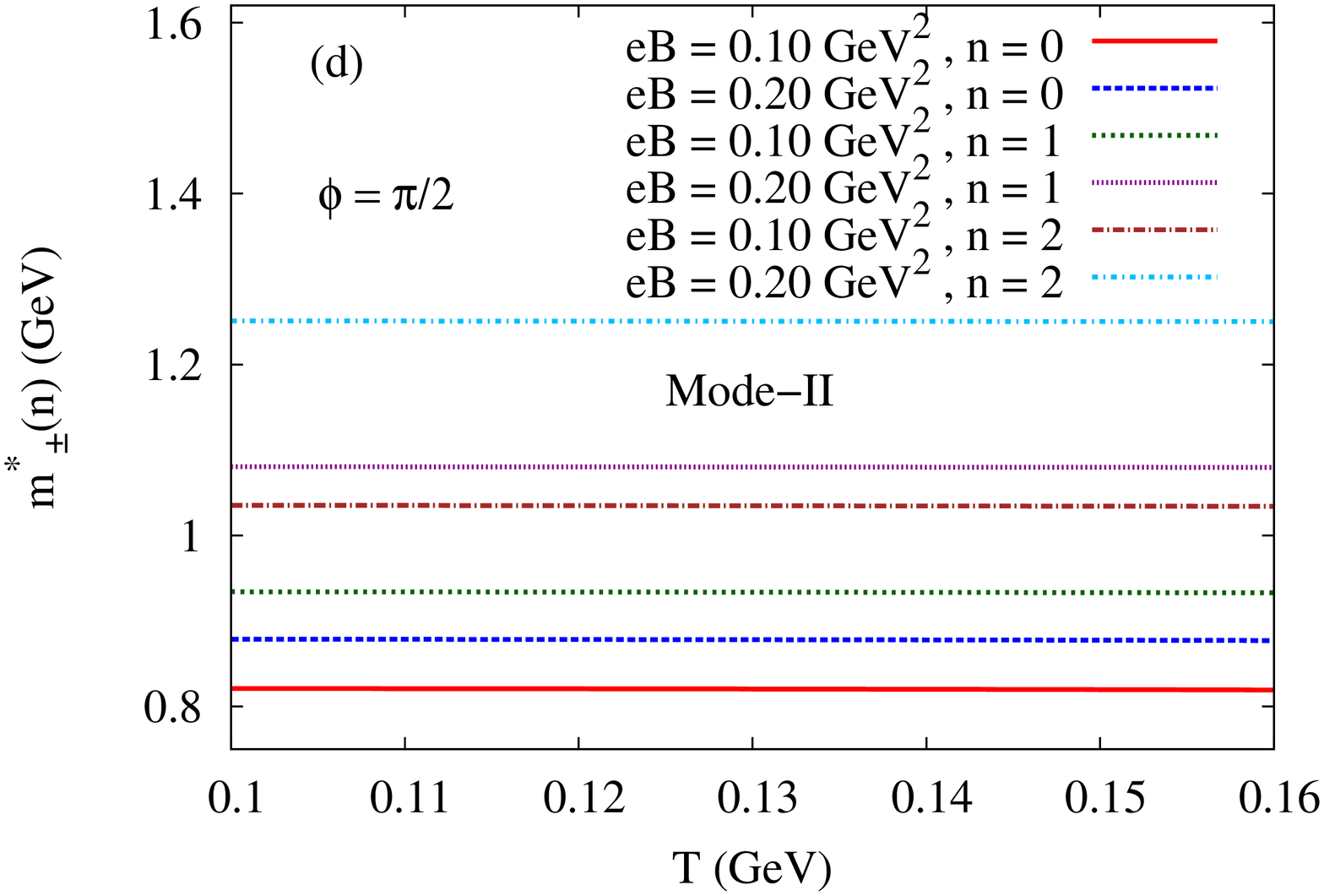} \includegraphics[angle=-90, scale=0.23]{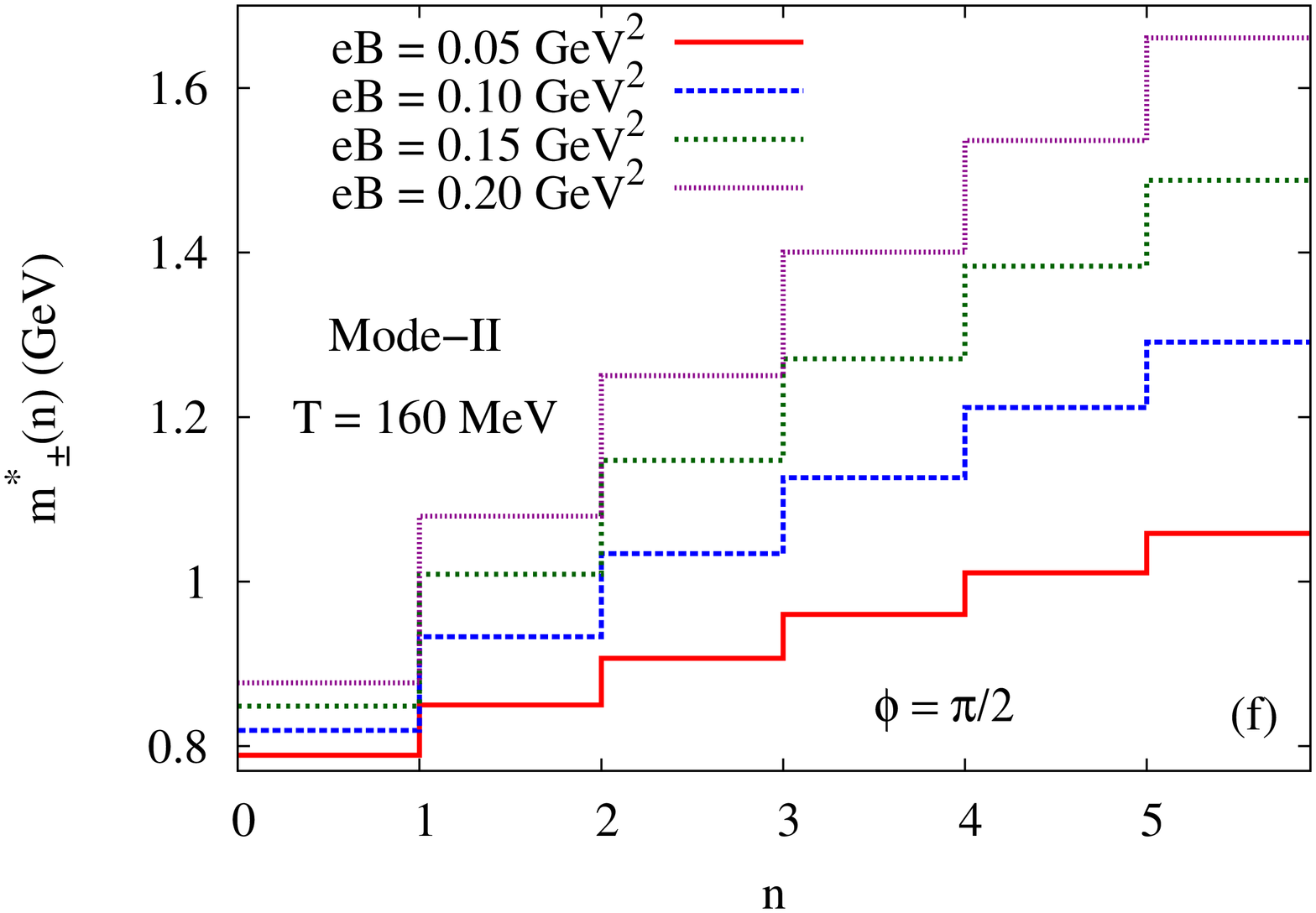}		
	\end{center}
	\caption{The Landau level ($n$) dependent effective mass of $\rho^\pm$ at $\phi=\pi/2$ (left panel) as a function of $eB$ at constant 
		temperature (160 MeV) and at four values of $n$ (middle panel) as a function of $T$ at different combinations 
		of $eB$ and $n$ (right panel) as a function of $n$ at constant temperature (160 MeV) and four different values of $eB$.
		The upper and lower panel shows results for Mode-I and Mode-II respectively.}
	\label{fig.masspm_n}
\end{figure}
\begin{figure}[h]
	\begin{center}
		\includegraphics[angle=-90, scale=0.30]{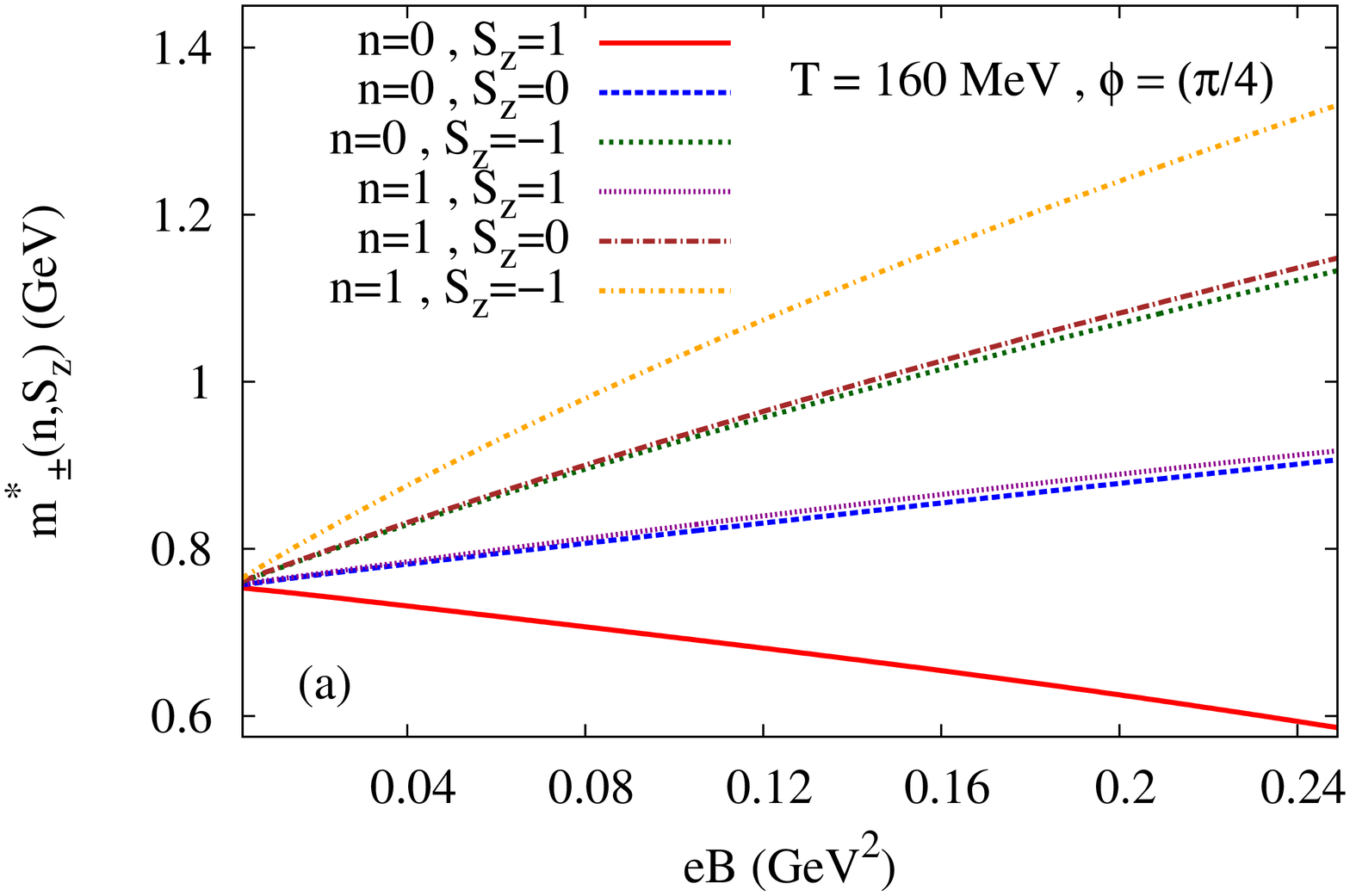}	\includegraphics[angle=-90, scale=0.30]{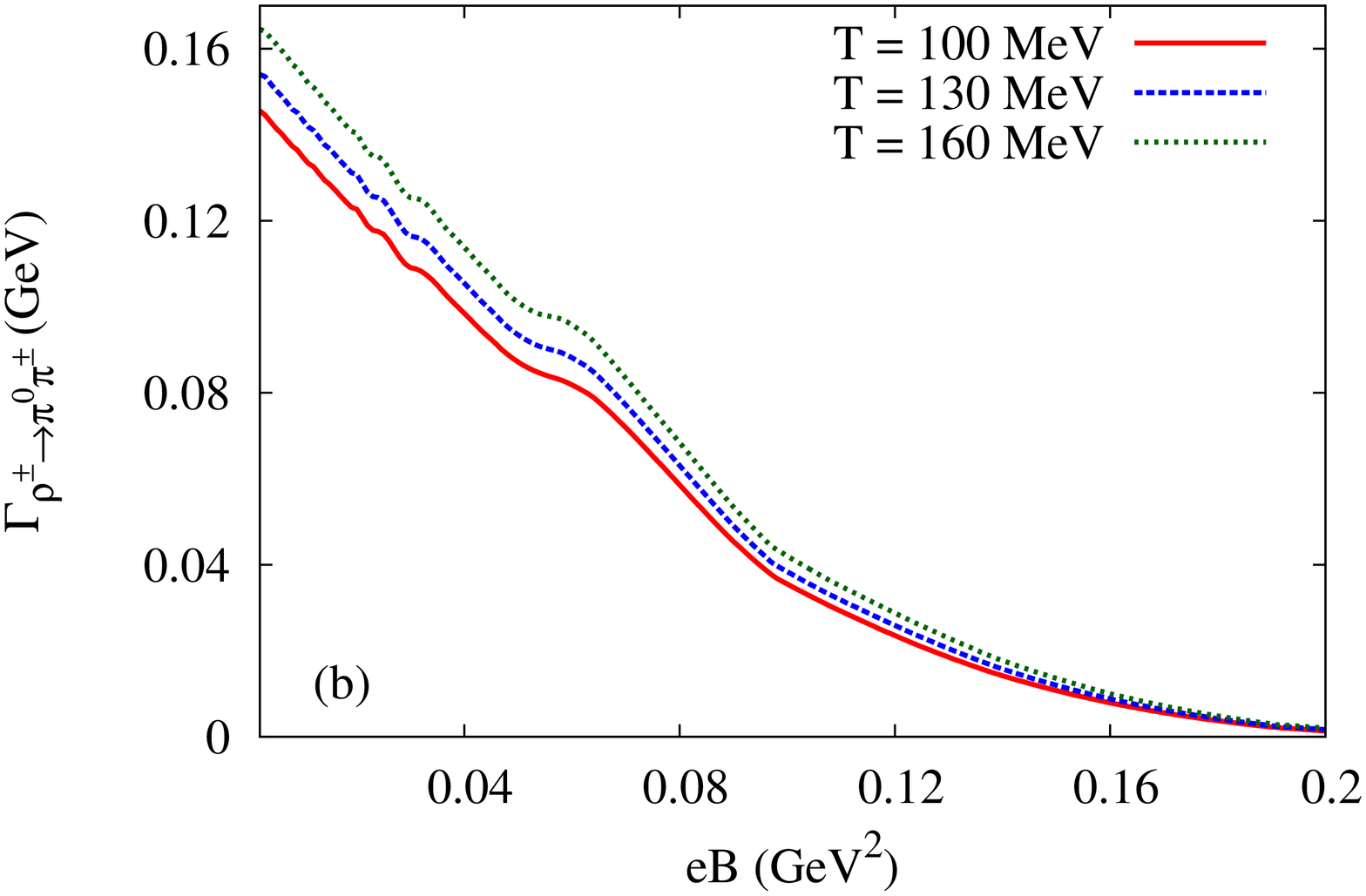}
	\end{center}
	\caption{(a) The effective mass 
		of charged rho as a function of $eB$ at $T$ = 160 MeV, $\phi = \pi/4$ and at different combinations of $n$ and $S_z$. 
		(b) The decay width of charged rho as a function of $eB$ at different temperatures.}
	\label{fig.decay_mstar_spinav}
\end{figure}

We have taken the neutral as well as charged rho meson three momenta to be zero in Figures \ref{fig.imaginary_LU}~-~\ref{fig.mass_0pm}.
In Fig.~\ref{fig.imaginary_LU}(a)-(d), the Landau cut contributions for the imaginary parts of the self energy functions are shown. 
The spikes occurring in Fig.~\ref{fig.imaginary_LU}(a) and (b) for the $\rho^0$ are due to the ``Threshold Singularity" 
for each Landau level 
as can be understood from Eq.~(\ref{eq.impi0.simple}). In this equation the $\tilde{k}_z$ present in the denominator  
has appeared due to dimensional reduction in the $\rho^0$ self energy. For a particular set of Landau levels $\{n,l\}$, we have
\begin{eqnarray}
\tilde{k}_z = \frac{1}{2q^0}\lambda^{1/2}\left(q_0^2,m_l^2,m_n^2\right) = \frac{1}{2q^0}\left(q^0+m_l+m_n\right)\left(q^0-m_l-m_n\right)\left(q^0+m_l-m_n\right)\left(q^0-m_l-m_n\right), \nn
\end{eqnarray}
which will go to zero at each threshold of Unitary and Landau cut defined in terms of the step functions in Eq.~(\ref{eq.impi0.simple}) 
which gives rise to the spike like structure in the upper panel.
Physically, the spikes correspond to the fluctuation of $\rho^0$ into two pions occurring in the transverse plane with respect to the direction of external magnetic field.
Moreover, at these threshold values of $q^0$, the $\rho^0$ will be (infinitely) unstable and will decay into pions immediately.
In Fig.~\ref{fig.imaginary_LU}(a) we have shown results for four different values of $eB$ (0.05, 0.10, 0.15 and 0.20 GeV$^2$ respectively) at a constant temperature $160$ MeV. 
As $eB$ increases, the thresholds of the Landau cuts move towards higher values of $q^0$ as evident from Eq.~(\ref{eq.impi0.simple}).
Also with the increase of $eB$, the separation among the spikes becomes larger.
In Fig.~\ref{fig.imaginary_LU}(b), results are shown at a constant $eB$ (0.10 GeV$^2$) at three different values of temperature (100, 130 and 160 MeV respectively).
In this case the threshold remains unchanged but the magnitude becomes larger which is due to the increase of the magnitude of
 thermal distribution functions with the increase of temperature.

In Fig.~\ref{fig.imaginary_LU}(c)-(d), similar results for the $\rho^\pm$ are shown. In this case, there is no threshold for the 
Landau cuts and they 
extend from $-\infty<q^0<\infty$. The oscillations are due to the presence of a Laguerre polynomial in Eq.~(\ref{eq.impipm.simple}).
It is also to be noted that, at a larger value of $eB$ the significant contributions start from a higher $q^0$. 
In Fig.~\ref{fig.imaginary_LU}(d), we see that the magnitude of the imaginary parts increases with the increase in temperature keeping the overall structure unaltered which
is again due to increase in the thermal distribution function with the increase of temperature.

In Fig.~\ref{fig.imaginary_LU}(e)-(h) the Unitary cut contributions to the imaginary part of the self energy functions are presented.
In Fig.~\ref{fig.imaginary_LU}(e), results for $\rho^0$ at a constant temperature (160 MeV) and at four different values of 
the magnetic field (0.05, 0.10, 0.15 and 0.20 GeV$^2$ respectively) are shown. Similar to the Landau cut contribution, the Unitary cut contributions of $\rho^0$
also suffer from the "Threshold Singularities". As the magnetic field increases, the thresholds of the Unitary cuts move towards higher $q^0$ and 
the separations among the spikes increase. In Fig.~\ref{fig.imaginary_LU}(f), results are shown for a constant $eB$ (0.10 GeV$^2$) and at three 
different values of the temperature (100, 130 and 160 MeV respectively). As the temperature increases, the magnitudes of the self energies increase keeping the 
thresholds unchanged similar to the Landau cut contributions. However, unlike the Landau contributions, the Unitary contributions are 
dominated by the vacuum contribution and the effect of increase of temperature is rather small. 

In Fig.~\ref{fig.imaginary_LU}(g)-(h), results are presented for the charged $\rho$. The $\rho^\pm$ has different thresholds 
for the Unitary cuts than the $\rho^0$ as given in Eq.~(\ref{eq.impipm.simple}). The displacement of the Unitary cut threshold towards higher $q^0$ with 
the increase in $eB$ is smaller compared to that of $\rho^0$ and this will have significant effect on the spectral functions. 
Small oscillations can be noticed in the graphs, which are due to
the presence of a Laguerre polynomial. However these oscillations and the effect of temperature is not of much significance as
the Unitary cut is dominated by the vacuum contributions.
Next in Fig.~\ref{fig.real}(a)-(d), the real part of the in-medium self energy functions of $\rho$ at non-zero external magnetic field are presented. 
In Fig.~\ref{fig.real}(a), results of the $\rho^0$ at constant temperature (160 MeV) with four different values of the magnetic field (0.05, 0.10, 0.15 
and 0.20 GeV$^2$ respectively) are shown. These plots are dominated by the ``$B$" terms in Eq.~(\ref{eq.self0.eb.t}). 
Small oscillatory behaviours in the plots arise due to the numerical principal value integrations. In Fig.~\ref{fig.real}(b), results are shown at a constant magnetic field 
(0.01 GeV$^2$) for three different values of the temperature (100, 130 and 160 MeV respectively). Effects of the increase in temperature 
though small, they make the real part larger. Analogous plots for $\rho^\pm$ are presented in Fig.~\ref{fig.real}(c) and (d).
Fig.~\ref{fig.real}(e) shows 
the comparison of the real part of the self energy function between $\rho^0$ and $\rho^\pm$. Results at two combinations of $eB$ and $T$ are given
($eB$=0.10 GeV$^2$, $T$=100 MeV and $eB$=0.20 GeV$^2$, $T$=160 MeV respectively). In both the cases, the real part of the $\rho^0$ self energy  has a larger magnitude than that of $\rho^\pm$. This may be due to the fact that, the $\rho^0$ contains two charged particle in the loop whereas the $\rho^\pm$ has only one. 
Having obtained the real and imaginary parts we present the in-medium spectral functions of $\rho$ which contains
both the real and imaginary parts of the self energy function. The spectral function is the imaginary part of the complete propagator
and defined as,
\begin{eqnarray}
S_{0,\pm}(q,eB,T) = \frac{\text{Im}~\Pi_{0,\pm}}{\left(q^2-m_\rho^2+\text{Re}~\Pi_{0,\pm}\right)^2+\left(\text{Im}~\Pi_{0,\pm}\right)^2} \label{eq.spectral.fn}
\end{eqnarray}
In Fig.~\ref{fig.spectra0_pm}(a)-(d), spectral functions of $\rho^0$ are shown. In Fig.~\ref{fig.spectra0_pm}(a) and (b), 
results are shown at constant 
temperature (160 MeV) and at three different values of the magnetic field (0.10, 0.15 and 0.20 GeV$^2$ respectively).
The spectral functions have the same threshold as the imaginary parts of the self energy, as is evident from Eq.~(\ref{eq.spectral.fn}). 
The spectral functions are shown in two parts, Fig.~\ref{fig.spectra0_pm}(a) shows the lower $q^0$ region dominated 
by the Landau terms
whereas Fig.~\ref{fig.spectra0_pm}(b) shows the higher $q^0$ region, dominated by Unitary terms. It is seen from Fig.~\ref{fig.spectra0_pm}(a) that, with the increase of the magnetic field, the
threshold of the spectral function shifts towards higher $q^0$. However in Fig.~\ref{fig.spectra0_pm}(b) 
because of this shift, 
the spectral function misses the $\rho$ mass pole, loses its Breit-Wigner shape, and 
becomes a continuum at large $q^0$. This physically implies the ``melting'' of the $\rho^0$ at higher 
values of the magnetic field. In Fig.~\ref{fig.spectra0_pm}(c) and (d), 
the spectral functions at constant $eB$ (0.10 GeV$^2$) for three different values of the temperature (100, 130 and 160 MeV respectively) 
are shown. The increase in temperature, increases the magnitude of the spectral function at low $q^0$ region, 
and decreases the magnitude at higher $q^0$ region without changing the threshold and shape.

In Fig.~\ref{fig.spectra0_pm}(e)-(h), similar plots of in-medium spectral function of $\rho^\pm$ are presented. In Fig.~\ref{fig.spectra0_pm}(f), it 
is observed that with the increase in $eB$, the threshold of the spectral function seems to have a displacement towards higher $q^0$.
This is due the displacement of the Unitary cut contributions to the imaginary part of $\rho^\pm$ self energy towards higher $q^0$ with the increase in $eB$, as can be seen from Fig.~\ref{fig.imaginary_LU}(g). However the Landau cut contribution is non-zero at every value of $q^0$. So even if the Unitary cut threshold crosses the $\rho$ pole at a high $eB$, the non-zero Landau cut contribution will maintain the Breit-Wigner shape of the spectral function. As the Landau cut contribution has a smaller magnitude, the spectral function will become narrow, which results in squeezing of $\rho^\pm$ width and thus making it more stable. The effect of increase of temperature is similar to that of
$\rho^0$ and is shown in Fig.~\ref{fig.spectra0_pm}(g) and (h). 
In Fig.~\ref{fig.spectra0pm}, comparison of the $\rho^0$ and $\rho^\pm$ spectral functions 
are shown, Fig.~\ref{fig.spectra0pm}(a) shows the lower $q^0$ region and Fig.~\ref{fig.spectra0pm}(b) shows the higher $q^0$ region.

Now we turn our attention to the evaluation of mass and dispersion relations.
We first consider a vanishing transverse momentum for neutral as well as charged $\rho$ mesons, i.e. 
$q^\mu\equiv(\omega,0,0,q_z)$ in Fig.~\ref{fig.mass_0pm}~-~\ref{fig.mass_disp}.
 Although the transverse momenta of $\rho^\pm$ is quantized as 
$q_\perp^2 = -(2n+1)eB$ with $n$ = 0, 1, 2 .... , we will first neglect this trivial Landau shift in order to 
show the importance of the loop corrections of $\rho^\pm$. 
From the pole of the complete $\rho$ propagator, the following dispersion relation is obtained for $q_\perp=0$:
\begin{eqnarray}
\det\left[  \left(-q_\parallel^2+m_\rho^2\right)g^{\mu\nu}+q_\parallel^\mu q_\parallel^\nu -\text{Re}~\bar{\Pi}^{\mu\nu}_{0,\pm}\left( q_\parallel,q_\perp=0 \right) \right] = 0. \label{eq.disp}
\end{eqnarray}
The effective mass($m^*$) of $\rho$ is calculated by putting its three-momentum $\vec{q}=\vec{0}$ in the dispersion relation. 
Clearly, Eq.~(\ref{eq.disp}) will admit four solutions and we should get four modes of the effective mass and dispersion relation.
Out of these modes, one is found to be unphysical and we are left with three physical modes of the effective mass and dispersion relation. These three modes corresponds to the three polarizations of the rho meson which is a spin-1 particle. 
In order to relate these modes with the physical spin state propagations, one needs to decompose the self energy tensor as well 
as the complete propagator in spin projection basis. In our approach, in order to avoid the complicated Lorentz structure of 
the self energy function at finite temperature and external magnetic field, we solve Eq.~(\ref{eq.disp}) instead. 
Thus in our case, although we have three modes, the correspondence with definite spin states is not obvious.
%Therefore, the modes presented in this paper 
%are mixture of the individual spin states propagation of $\rho$ meson.      
Since, we have taken $\vec{q}=(0,0,q_z)$, two modes are identical and 
we are left with two distinct modes. We refer to these modes as Mode-I and Mode-II respectively.

In Fig.~\ref{fig.mass_0pm}(a)-(d), the effective mass($m_0^*$) of $\rho^0$ is shown. In the Fig.~\ref{fig.mass_0pm}(a) and (b),  $m_0^*$ is plotted as a function of 
$eB$ at three different temperatures (100, 130 and 160 MeV) for Mode-I and Mode-II respectively. We find significant decrease of $\rho^0$ mass with the increase of $eB$ in both modes which is due to a 
strong positive contribution coming from the real part of the self energy. This decrease of $m_0^*$ is more in Mode-I as compared to Mode-II. In Fig.~\ref{fig.mass_0pm}(c)-(d), $m_0^*$ is plotted as a function of the $T$ at four different values of $eB$ (0.05, 0.10, 0.15 and 0.20 GeV$^2$) for Mode-I and Mode-II. The variation of $m_0^*$ is slow as the contribution of the ``$BT$" terms are small compared to the ``$B$" terms as seen from Eqs.~(\ref{eq.self0.eb.t}) and (\ref{eq.selfpm.eb.t}). We find small increase (decrease) of $m_0^*$ with the increase of $T$ in Mode-I (Mode-II).

In Fig.~\ref{fig.mass_0pm}(e)-(h), corresponding plots are shown for $\rho^\pm$. The effective mass ($m_\pm^*$) of $\rho^\pm$ shows similar 
behaviour as the $\rho^0$. 
Due to the presence of one Laguerre polynomial in the expression of the real part of thermal self energy in Eq.~(\ref{eq.re.pibarpm.BT}), 
small oscillatory behaviour is present in the Fig.~\ref{fig.mass_0pm}-(a).
In the lower panel, the $m_\pm^*$ is decreasing with the increase in temperature in both the modes.

Having obtained the effective masses, we now proceed to present the dispersion relations of rho meson. 
In Fig.~\ref{fig.disp_0pm}(a)-(d), 
dispersion curves of $\rho^0$ are shown. We have plotted the energy($\omega_0$) of $\rho^0$ as a function of $q_z$ taking $q_x=q_y=0$. 
In Fig.~\ref{fig.disp_0pm}(a) and (b), dispersion curves are shown at constant temperature (160 MeV) and at four different values of $eB$ (0.05, 0.10, 0.15 and 0.20 GeV$^2$) for Mode-I and Mode-II respectively. 
In all the cases, the dispersion curves are well separated at $q_z=0$ (which is actually the effective mass). As the momentum 
increases, the kinetic energy term dominates over the self energy corrections and the dispersion relation becomes light-like.  
In Fig.~\ref{fig.disp_0pm}(c) and (d), dispersion curves are shown at constant $eB$ (0.10 GeV$^2$) and at three different values of the temperature 
(100, 130 and 160 MeV) for Mode-I and Mode-II. The separation among the curves are very small due to smallness of 
the ``$BT$" terms as compared to ``$B$" terms in the real part of self energy.
In Fig.~\ref{fig.disp_0pm}(e)-(h), corresponding plots are shown for $\rho^\pm$. In this case we find results similar to the $\rho^0$ 
except the fact that the separation among the dispersion curves becomes less. This is due to the fact that the real part of 
the $\rho^\pm$ self energy has a lower magnitude than that of $\rho^0$.

In Fig.~\ref{fig.mass_disp}, we have shown comparison of effective masses and dispersion curves between $\rho^0$ and $\rho^\pm$ at 
constant temperature (160 MeV). From Fig.~\ref{fig.mass_disp}(a), it is observed that, 
when $eB \approx 0$ both the $\rho^0$ and $\rho^\pm$ have identical effective masses. This is because of the fact that the 
$\rho^0$ and $\rho^\pm$ self energies at $eB$=0 are equal. With the increase in $eB$, the splitting between them increases. Moreover 
the effect of the external magnetic field is more on $\rho^0$ than that on $\rho^\pm$.  
Fig.~\ref{fig.mass_disp}(b) shows comparison plots of dispersion curves at contant magnetic field (0.10 GeV$^2$). 

%
%
% ++++++++++++++++++++++++++++++++++++++++++++++++++++++++++++++++++
Now we will introduce non-zero three momenta for the charged rho mesons. Keeping in mind the Landau quantization of the transverse 
momenta of $\rho^\pm$, we parametrize the $\rho^\pm$ four momenta as,
\begin{eqnarray}
q^\mu \equiv \left( q^0,\sqrt{(2n+1)eB}\cos\phi,\sqrt{(2n+1)eB}\sin\phi,0 \right)~~~~~; ~~ n=0,1,2,3..... , \label{eq.rhopm.4.momenta}
\end{eqnarray} 
where $\phi$ is the azimuthal angle. In Fig.~(\ref{fig.spectrapm_5}), we have shown the $\rho^\pm$ spectral function as a function of the invariant mass ($\sqrt{q^2}=\sqrt{q_0^2-\vec{q}^2}$) of rho meson. Results are shown at a constant magnetic field (0.10 GeV$^2$), temperature (160 MeV) and azimuthal angle ($\frac{\pi}{4}$) with four values of $n$ (0, 1, 2 and 3). The peak of the spectral function increases with increase of $n$ with very small shifts towards higher $\sqrt{q^2}$ making the $\rho^\pm$ more stable at higher transverse momenta. We have not shown the azimuthal angle ($\phi$) dependence of the spectral function since it is insignificantly small.
The disperion relation for $\rho^\pm$ will follow from the pole of the complete propagator. For simplicity we have taken the azimuthal 
angle $\phi=\left(\pi/2\right)$, for which $\vec{q}\equiv(0,\sqrt{(2n+1)eB},0)$ and the corresponding dispersion relation is, 
\begin{eqnarray}
\det\left[  \left(-q_0^2+(2n+1)eB+m_\rho^2\right)g^{\mu\nu}+q^\mu q^\nu -\text{Re}~\bar{\Pi}^{\mu\nu}_{0,\pm}\left( q^0,q_x=0,q_y=\sqrt{(2n+1)eB},q_z=0 \right) \right] = 0. \label{eq.disp2}
\end{eqnarray}
Clearly, the effective mass obtained from Eq.~(\ref{eq.disp2}) will be function of the Landau level $n$ in addition to temperature ($T$) and 
magnetic field ($eB$). As before, we have obtained two distinct physical modes for the $n$-dependent effective mass ($m^*_\pm(n)$) of $\rho^\pm$. We will refer these modes as Mode-I and Mode-II respectively. 
In Fig.~\ref{fig.masspm_n}, $m^*_\pm(n)$ is plotted as a function of $eB$, $T$ and $n$ in the left, middle and right panels respectively 
at azimuthal angle ($\phi=\pi/2$).
Fig.~\ref{fig.masspm_n}(a) shows $m^*_\pm(n)$ for Mode-I at constant temperature (160 MeV) and at four values of $n$. 
It is clear that the trivial Landau term completely dominates over the self-energy correction and effective mass monotonically 
increases with the increase of $eB$. Fig.~\ref{fig.masspm_n}(b) shows corresponding plot for Mode-II.
In Fig.~\ref{fig.masspm_n}(c), $m^*_\pm(n)$ for Mode-I is shown as a function of the temperature at six different combinations of 
$eB$ and $n$. Since the variation of $T$ affects only the self energy part which is subleading compared to the dominant trivial Landau term, 
we see no significant variation of effective mass with temperature. Corresponding plot for Mode-II is shown in Fig.~\ref{fig.masspm_n}(d).
Finally, we have shown $m^*_\pm(n)$ vs $n$ at constant temperature (160 MeV) and at four different values of the magnetic field (0.05, 0.10, 0.15 and 0.20 GeV$^2$) for Mode-I and Mode-II in Fig.~\ref{fig.masspm_n}(e) and Fig.~\ref{fig.masspm_n}(f) respectively. Here also 
the self energy correction have insignificant effect with respect to the Landau level term ($(2n+1)eB$) so that the separation among them increases with the increase of $n$.

% ++++++++++++++++++++++++++++++++++++++++++++++++++++++++++++++++++++++++++++++++++++++++++++++++++++++++++

Until now, we have neglected the trivial coupling between the magnetic moment of $\rho^\pm$ and the external magnetic field. 
Following Ref.~\cite{Chernodub:2010qx}, the dispersion relation of $\rho^\pm$ (without loop correction) under external magnetic field is given by,
\begin{eqnarray}
q_0^2 = q_z^2 + \left(2n+1-2S_z\right)eB + m_\rho^2~,
\end{eqnarray}
where, $S_z \in \left(1,0,-1\right)$ and $n\geq0$ are the spin projections on the field axis and Landau level index respectively. 
It is evident from the above equation, that the effective mass of $\rho^\pm$ will be dominated by the leading order trivial 
contribution coming from the second term on the R.H.S. which is a linear term in $eB$. Corresponding corrections in the 
effective mass due to the self-energy contribution are subleading compared to this trivial term and this can be noticed in 
Fig.~\ref{fig.mass_0pm}(g)-(h) and Fig.~\ref{fig.masspm_n}(c)-(d). In view of this, let us approximate the contribution 
of self energy in different modes (which corresponds to different spin projections) by the spin-averaged one. So the 
dispersion relation of $\rho^\pm$ becomes,
\begin{eqnarray}
q_0^2 = q_z^2 + \left(2n+1-2S_z\right)eB + m_\rho^2 - \text{Re}~\Pi_\pm\left(q^0,q_x=\sqrt{(2n+1)eB}\cos\phi,q_y=\sqrt{(2n+1)eB}\sin\phi,q_z\right)~,
\label{eq.disp.1}
\end{eqnarray}
where, $\Pi_\pm$ is defined in Eq.~(\ref{eq.spin.averaged}). The effective mass $m^*_\pm\left(n,S_z\right)$ 
obtained from the above equation by putting $q_z=0$, will 
depend on $n$ and $S_z$ in addition to $eB$, $T$ and $\phi$. 
In Fig.~\ref{fig.decay_mstar_spinav}(a), we have plotted $m^*_\pm\left(n,S_z\right)$ as a function of $eB$ at 
$T$ = 160 MeV and $\phi=(\pi/4)$ for the two Landau levels with different spin projections. The only case where the 
effective mass decreases rapidly with the increase of $eB$ is for $n=0$ and $S_z=1$ as evident from the R.H.S. 
of Eq.~(\ref{eq.disp.1}). This is due to the fact that, with the increase of $eB$, both the terms $(2n+1-2S_z)eB$ as well as 
$-\text{Re}~\Pi_\pm$ gives strong negative contributions. 
For the other combinations of $n$ and $S_z$, there is a 
competition between the trivial term and the self energy; the term $(2n+1-2S_z)eB$ gives strong positive contribution, 
whereas the term $-\text{Re}~\Pi_\pm$ always gives small negative contribution which is subleading as compared to the 
trivial term. 
This leads to the fact that $m^*(n,S_z)$ increases monotonically with $eB$ except the case when $n=0$ and $S_z=1$. 
Moreover, the graphs for $n=0,S_z=0$ and $n=1,S_z=1$ (both having equal trivial contributions) 
lie almost on top of each other. The same behaviour is seen in the graphs for $n=0,S_z=-1$ and $n=1,S_z=0$ as well. 
It is to be noted that, the real part of the self energy in 
Eq.~(\ref{eq.disp.1}) consists of two parts, the ``B'' part and the ``BT'' part. The $T$ and $\phi$ dependent part (``BT'') is 
very small as compared to the magnetic field dependent vacuum part (``B''). For this reason, the effect of $T$ and $\phi$ on 
$m^*(n,S_z)$ will be negligibly small and thus we have not shown its variation with $T$ and $\phi$.

We conclude this section with the result on the decay width of charged rho meson in presence of the magnetic field. 
We define the decay width of $\rho^\pm$ in its lowest energy state as, 
\begin{eqnarray}
\Gamma_{\rho^\pm\rightarrow\pi^0\pi^\pm}(eB,T) = \frac{1}{ m^*_\pm(0,1)}\int_{0}^{2\pi}\frac{d\phi}{2\pi} ~\text{Im}~\Pi_\pm\left(q^0=m^*_\pm(0,1),q_x=\sqrt{eB}\cos\phi,
q_y=\sqrt{eB}\sin\phi,q_z=0\right)~,
\end{eqnarray}
where we have made an average over the azimuthal angle $\phi$. 
The decay width is shown in Fig.~\ref{fig.decay_mstar_spinav}(b) 
as a function of $eB$ at three different values of temperature (100, 130 and 160 MeV respectively). With the increase in $eB$, 
the decay width decreases and this can be understood from Fig.~\ref{fig.imaginary_LU}(g) where the magnitude of the Unitary 
cut contribution to the self energy decreases with the increase in $eB$ around the rho mass pole. 
The oscillatory behaviour in the decay width is due to 
the oscillatory behaviour of the Landau cut contribution of the self energy as can be seen from Fig.~\ref{fig.imaginary_LU}(c). 
At very high value of $eB$, the value of the decay width becomes close to zero and the $\rho^\pm$ becomes stable against 
decay into two pions. 
The decay width increases with the increase in temperature and can be understood from Fig.~\ref{fig.imaginary_LU}(d) and 
(h) where the imaginary part increases with the increase in temperature. This physically means that $\rho^\pm$ is more unstable 
at a higher temperature and at lower $eB$.

\section{Summary and Discussions}\label{sec.summary}
We have made comprehensive study of the self energies of $\rho$ meson using effective $\rho\pi\pi$ interaction at finite temperature
and arbitrary external magnetic field including all the Landau levels in the propagators of the loop particle in our calculations. 
We have also explicitly worked out the analytic structure of the self energy at finite temperature and arbitrary non-zero 
external magnetic field. The kinematic domains of 
the imaginary part of the self energy in the complex $q^0$ plane are found to be different for $\rho^0$ and $\rho^\pm$ 
as well as from the $eB=0$ case. For vanishing three momentum of the $\rho$, we have observed Landau cut contributions
to the imaginary part of the self energy at non-zero magnetic field which is absent at zero magnetic field. 
While calculating the real part of the self energy, we have taken the magnetic field dependent vacuum contributions which is 
usually ignored in most of the works in the literature and this term produces dominating contribution to 
the effective mass and dispersion relations.
The in-medium spectral functions obtained from the spin-averaged self energy is found to be quite different for $\rho^0$ and $\rho^\pm$.
The ``Threshold Singularities" in the $\rho^0$ spectral function give rise to spike like structures which is absent in case of $\rho^\pm$.
It is also shown quantitatively that, the $\rho^0$ meson ``melts" at high magnetic field whereas $\rho^\pm$ does not.
From the pole of the complete $\rho$ propagator, we have evaluated its effective mass and dispersion relations. 
First we have presented results neglecting the trivial Landau shift to the $\rho^\pm$ transverse momenta in order to see the effect of 
self energy correction which is a function of both $T$ and $eB$. This kind of situation may occure in principle when different spices of 
charged particle are present in the system all of which undergoing trivial Landau shifts. 
In that case, the real part of the self energy will 
play the deciding role in the characterization of effective masses.
We find two distinct
physical modes for the effective mass and dispersion relation. The effective mass of $\rho$ is seen to decrease with the increase in
 magnetic field. 
 We have also taken the trivial coupling between the magnetic moment of charged rho with the external magnetic 
 field along with the trivial Landau shifts. Incorporation of this term has made the effective mass of $\rho^\pm$ to be dependent 
 on the spin projection along with the Landau level $n$. For a particular combination, $n=0$ and $S_z=1$, the effective mass 
 decreases very rapidly with the increase in $eB$ and charged rho condensation may occur. The decay width of $\rho^\pm$ is 
 found to be decreasing with the increase in $eB$ and at certain higher value of $eB$, the $\rho^\pm$ becomes stable against 
 decay into pions.

\section*{Acknowledgement}
Snigdha Ghosh acknowledges Center for Nuclear Theory, Variable Energy Cyclotron Centre and Department of Atomic Energy, Government of India for support.
%

% ###############################################################################################################################################

\appendix

\section{Calculation of $\text{Re}(\Pi^{\mu\nu}_0)_{B}$}\label{appendix.a}
In this appendix we shall briefly sketch how to obtain Eq.~(\ref{eq.re.pibar0.eb}). We have
\begin{eqnarray}
(\Pi^{\mu\nu}_0)_{B} &=& i\int\frac{d^4k}{(2\pi)^4}\mathcal{N}^{\mu\nu}\Delta_B(k)\Delta_B(p=q-k) \nonumber \\
&=& i\int\frac{d^4k}{(2\pi)^4}\mathcal{N}^{\mu\nu} \sum\limits_{l=0}^{\infty}\sum\limits_{n=0}^{\infty} \left[ \frac{\phi_l(\alpha_k)\phi_n(\alpha_k)}{\left(k_\parallel^2-m_l^2+i\epsilon\right)\left(p_\parallel^2-m_n^2+i\epsilon\right)}\right]. \nn
\end{eqnarray}
Using standard Feynman parametrization, we combine the denominators of the two propagators and obtain
\begin{eqnarray}
(\Pi^{\mu\nu}_0)_{B} &=& i\sum\limits_{l=0}^{\infty}\sum\limits_{n=0}^{\infty} \int\limits_{0}^{1}dx\int
\frac{d^2k_\perp}{(2\pi)^2}\phi_l(\alpha_k)\phi_n(\alpha_k)\int\frac{d^2k_\parallel}{(2\pi)^2}
\frac{\mathcal{N}^{\mu\nu}}{\left[(k_\parallel-xq_\parallel)^2-\Delta_{n,l}\right]^2}, \nn
\end{eqnarray}
where, $\Delta_{n,l} = \Delta + eB(2l+1-2xl+2xn)$. Shifting monemta $k_\parallel\rightarrow(k_\parallel+xq_\parallel)$ 
and performing the $k_\parallel$ integration using dimensional regularization we get,
\begin{eqnarray}
(\Pi^{\mu\nu}_0)_{B} &=& \left(\frac{g_{\rho\pi\pi}^2q_\parallel^2}{4\pi}\right)i\sum\limits_{l=0}^{\infty}\sum\limits_{n=0}^{\infty} \int\limits_{0}^{1}dx\int\frac{d^2k_\perp}{(2\pi)^2}\phi_l(\alpha_k)\phi_n(\alpha_k)\left[\left( q_\parallel^2g_\parallel^{\mu\nu}-q^\mu_\parallel q^\nu_\parallel \right)\frac{1}{2}\Gamma\left(1-\frac{d}{2}\right)\left(\frac{1}{\Delta_{n,l}}\right)^{1-d/2}\right. \nonumber \\
&& \hspace{5cm} \left. -~ q_\parallel^2k_\perp^\mu k_\perp^\nu\Gamma\left(2-\frac{d}{2}\right) \left(\frac{1}{\Delta_{n,l}}\right)^{2-d/2} \right]_{d\rightarrow2}. \nn
\end{eqnarray}
Now using the $\overline{\text{MS}} $ scheme we subtract out the divergence arising from the pole of the Gamma function.
 The remaining $k_\perp$ integration is convergent and can be evaluated using the orthogonality properties of the 
 Laguerre polynomials present in the numerator which also remove one sum from the double sum. Then the $(\Pi^{\mu\nu}_0)_{B}$ 
 can be written in the following compact form
\begin{eqnarray}
(\Pi^{\mu\nu}_0)_{B} = \left(\frac{g_{\rho\pi\pi}^2q_\parallel^2}{32\pi^2}\right)\int_{0}^{1}dx \left[ \left( q_\parallel^2g_\parallel^{\mu\nu}-q^\mu_\parallel q^\nu_\parallel \right)S_\parallel + q_\parallel^2g_\perp^{\mu\nu}S_\perp \right] \label{eq.pi0vac_appendix}
\end{eqnarray}
where, $S_\parallel$ and $S_\perp$ are given by,
\begin{eqnarray}
S_\parallel &=& \sum\limits_{n=0}^{\infty} \left[2eB\ln\left(\frac{\Delta_{n,n}}{\mu_0}\right)\right] \label{eq.S_pll}\\
S_\perp &=& \sum\limits_{n=0}^{\infty}eB\left[ \frac{2n+1}{\Delta_{n,n}}+\frac{2n+2}{\Delta_{n+1,n}} \label{eq.S_perp} \right].
\end{eqnarray}
The infinite sums in Eq.~(\ref{eq.S_pll}) and (\ref{eq.S_perp}) are divergent and can be regularized using 
derivative regularization technique \cite{Schwartz:2013pla}. To do this we differentiate them with respect to $M=q_\parallel^2$ twice and 
obtain convergent sums, which are expressible in terms of polygamma functions ($\Psi$, $\Psi^\prime$ and $\Psi^{\prime\prime}$),
\begin{eqnarray}
\left(\frac{\partial^2S_\parallel}{\partial M^2}\right) &=& \frac{-x^2(1-x)^2}{2eB} \Psi^\prime\left( \frac{\Delta}{2eB}+\frac{1}{2}\right) \label{eq.S_pll2} \\
\left(\frac{\partial^2S_\perp}{\partial M^2}\right) &=& \frac{x^2(1-x)^2}{2(eB)^2} \left[ 
\Psi^\prime \left(\frac{\Delta}{2eB}+\frac{1}{2}\right) + \left(\frac{\Delta}{4eB}\right)\Psi^{\prime\prime}
\left(\frac{\Delta}{2eB}+\frac{1}{2}\right) + \right. \nonumber \\ && \left. \Psi^\prime\left(\frac{\Delta}{2eB}+\frac{1}{2}+x\right) + \left(\frac{\Delta}{4eB}+\frac{2x-1}{4}\right)\Psi^{\prime\prime}\left(\frac{\Delta}{2eB}+\frac{1}{2}+x\right) \right] \label{eq.S_perp2}.
\end{eqnarray}
Integating Eq.~(\ref{eq.S_pll2}) and (\ref{eq.S_perp2}), and substituting $S_\parallel$ and $S_\perp$ into Eq.~(\ref{eq.pi0vac_appendix}), we obtain,
\begin{eqnarray}
(\Pi^{\mu\nu}_0)_{B} &=& \left(\frac{g_{\rho\pi\pi}^2q_\parallel^2}{32\pi^2}\right)\int_{0}^{1} dx \left[ \left( q_\parallel^2g_\parallel^{\mu\nu}-q^\mu_\parallel q^\nu_\parallel \right)\left\{2eB\ln\Gamma\left(\frac{\Delta}{2eB}+\frac{1}{2}\right)+C_1q_\parallel^2+C_2 \right\} \right. \nonumber \\
&& \left. +~q_\parallel^2g_\perp^{\mu\nu}\left\{ \frac{\Delta}{2eB} \Psi\left( \frac{\Delta}{2eB}+\frac{1}{2} \right) 
+ \left( \frac{\Delta}{2eB}-\frac{1}{2}+x \right)\Psi\left( \frac{\Delta}{2eB}+\frac{1}{2}+x \right)+C_3q_\parallel^2+C_4 \right\} \right] \label{eq.pi0vac_appendix2}
\end{eqnarray}
where $C_1$, $C_2$, $C_3$ and $C_4$ are integration constants and independent of $q_\parallel^2$ but can be 
functions of $x$ and $eB$. These constants have to be chosen such that in the limit of zero magnetic field, 
we get back the vacuum self energy as given in Eq.~(\ref{eq.vacself.rho}) i.e.
\begin{eqnarray}
\lim\limits_{eB\rightarrow0} (\Pi^{\mu\nu}_0)_{B} = (\Pi^{\mu\nu})_{vac} \label{eq.eb-0limit_rho0}.
\end{eqnarray}
From Eq.~(\ref{eq.eb-0limit_rho0}), we get $C_1 = C_3 = -x(1-x)\ln\left(\frac{2eB}{\mu_0}\right)$, 
$C_2 = m^2\ln\left(\frac{2eB}{\mu_0}\right)$ and $C_4 = m^2\ln\left(\frac{2eB}{\mu_0}\right)-m^2$. Substituting 
the values of $C_1$,$C_2$,$C_3$ and $C_4$ into Eq.~(\ref{eq.pi0vac_appendix2}) and taking the real part, we get Eq.~(\ref{eq.re.pibar0.eb}).
%
%
%++++++++++++++++++++++++++++++++++++++++++++++++++++++++++++++++++++++++++++++++++++++++++++++++++++++++++++++++++++++++++++++
%
%
\section{Calculation of $\text{Re}(\Pi^{\mu\nu}_\pm)_{B}$}\label{appendix.b}
We have 
\begin{eqnarray}
(\Pi^{\mu\nu}_\pm)_{B} &=& i\int\frac{d^4k}{(2\pi)^4}\mathcal{N}^{\mu\nu}\Delta_0(k)\Delta_B(p=q-k) \nonumber \\
&=& i\int\frac{d^4k}{(2\pi)^4}\mathcal{N}^{\mu\nu} \sum\limits_{n=0}^{\infty} \left[ \frac{\phi_n(\alpha_p)}
{\left(k^2-m_\pi^2+i\epsilon\right)\left(p_\parallel^2-m_n^2+i\epsilon\right)}\right]. \nn
\end{eqnarray}
Using standard Feynman parametrization, we combine the denominators of the two propagators and obtain
\begin{eqnarray}
(\Pi^{\mu\nu}_\pm)_{B} &=& i\sum\limits_{n=0}^{\infty} \int\limits_{0}^{1}dx\int
\frac{d^2k_\perp}{(2\pi)^2}\phi_n(\alpha_p)\int\frac{d^2k_\parallel}{(2\pi)^2}
\frac{\mathcal{N}^{\mu\nu}}{\left[(k_\parallel-xq_\parallel)^2-\Delta_n\right]^2}, \nn
\end{eqnarray}
where, $\Delta_n = \Delta + x(2n+1)eB-(1-x)k_\perp^2$. Shifting momenta $k_\parallel\rightarrow(k_\parallel+xq_\parallel)$ and performing the $k_\parallel$ integration using dimensional regularization we get,
\begin{eqnarray}
(\Pi^{\mu\nu}_\pm)_{B} = \left(\frac{g_{\rho\pi\pi}^2}{4\pi}\right)\sum\limits_{n=0}^{\infty} \int\limits_{0}^{1}dx\int\frac{d^2k_\perp}{(2\pi)^2}\phi_n(\alpha_p)\left[A^{\mu\nu}~\Gamma\left(1-\frac{d}{2}\right)\left(\frac{1}{\Delta_n}\right)^{1-d/2}
 -B^{\mu\nu}~\Gamma\left(2-\frac{d}{2}\right) \left(\frac{1}{\Delta_n}\right)^{2-d/2} \right]_{d\rightarrow2}, \nn
\end{eqnarray}
where $A^{\mu\nu}$ and $B^{\mu\nu}$ are given by,
\begin{eqnarray}
A^{\mu\nu} &=& \frac{1}{2} \left[ q^4g_\parallel^{\mu\nu}+q_\parallel^2q^\mu q^\nu-q^2\left(q^\mu q_\parallel^\nu+q^\nu q_\parallel^\mu\right) \right] \\
B^{\mu\nu} &=& q^4\left(x^2q_\parallel^\mu q_\parallel^\nu+k_\perp^\mu k_\perp^\nu + xq_\parallel^\mu k_\perp^\nu+xq_\parallel^\nu k_\perp^\mu\right) + \left(xq_\parallel^2+q_\perp.k_\perp\right)^2q^\mu q^\nu \nonumber \\
&& -~q^2\left(xq_\parallel^2+q_\perp.k_\perp\right)\left(xq^\mu q_\parallel^\nu+ xq^\nu q_\parallel^\mu + q^\mu k_\perp^\nu+q^\nu k_\perp^\mu\right).\nn
\end{eqnarray}
Now as before we subtract out the divergences arising from the pole of the Gamma function and obtain,
\begin{eqnarray}
(\Pi^{\mu\nu}_\pm)_{B} = \left(\frac{-g_{\rho\pi\pi}^2}{4\pi}\right)\int\limits_{0}^{1}dx\int\frac{d^2k_\perp}{(2\pi)^2} \left[ A^{\mu\nu}\tilde{S}_\parallel + B^{\mu\nu}\tilde{S}_\perp \right] \label{eq.pipmvac_appendix}
\end{eqnarray}
where, $\tilde{S}_\parallel$ and $\tilde{S}_\perp$ are given by,
\begin{eqnarray}
\tilde{S}_\parallel &=& \sum\limits_{n=0}^{\infty} \left[\phi_n(\alpha_p)\ln\left(\frac{\Delta_n}{\mu_0}\right)\right] \label{eq.Stilde_pll}\\
\tilde{S}_\perp &=& \sum\limits_{n=0}^{\infty} \left[\frac{\phi_n(\alpha_p)}{\Delta_n} \label{eq.Stilde_perp} \right].
\end{eqnarray}
Differentiating Eq.~(\ref{eq.Stilde_pll}) with respect to $q_\parallel^2$, we get,
\begin{eqnarray}
\frac{\partial\tilde{S}_\parallel}{\partial q_\parallel^2} &=& \sum\limits_{n=0}^{\infty} \left[ -x(1-x)
\frac{\phi_n(\alpha_p)}{\Delta_n} \label{eq.Stilde'_pll} \right].
\end{eqnarray}
To evaluate the infinite sums in Eq.~(\ref{eq.Stilde_perp}) and (\ref{eq.Stilde'_pll}), we introduce a new parameter $z$ and write 
\begin{eqnarray}
\frac{1}{\Delta_n} = \int\limits_{0}^{1}\frac{dz}{m_\pi^2}z^{\Delta_n/m_\pi^2-1} \label{eq.z1}.
\end{eqnarray}
Substituting Eq.~(\ref{eq.z1}) into Eq.~(\ref{eq.Stilde_perp}) and (\ref{eq.Stilde'_pll}) and using the identity 
$\sum\limits_{n=0}^{\infty}L_n(t)z^n = (1-z)^{-1}\exp\left(\frac{tz}{z-1}\right)$ for $\left|z\right|\leq1$, we obtain,
\begin{eqnarray}
\tilde{S}_\parallel &=& \int\limits_{0}^{1}\frac{dz}{\ln z}z^{\Delta/m_\pi^2-1}\text{sech}\left(x\frac{eB}{m_\pi^2}\ln z\right)e^\zeta + \tilde{C}_1 \label{eq.Stilde_pll2} \\
\tilde{S}_\perp &=& \int\limits_{0}^{1}\frac{dz}{m_\pi^2}z^{\Delta/m_\pi^2-1}\text{sech}\left(x\frac{eB}{m_\pi^2}\ln z\right)e^\zeta \label{eq.Stilde_perp2}
\end{eqnarray}
where, $\tilde{C}_1$ is the constant of the $q_\parallel^2$-integration and is independent of $q_\parallel^2$ which is chosen to be zero and 
\begin{eqnarray}
\zeta = -(1-x)\frac{k_\perp^2}{m_\pi^2}\ln z + \alpha_p\tanh\left(x\frac{eB}{m_\pi^2}\ln z\right). \nn
\end{eqnarray}
Substituting Eq.~(\ref{eq.Stilde_pll2}) and (\ref{eq.Stilde_perp2}) into Eq.~(\ref{eq.pipmvac_appendix}) we get,
\begin{eqnarray}
(\Pi^{\mu\nu}_\pm)_{B} = \left(\frac{-g_{\rho\pi\pi}^2}{4\pi}\right)\int\limits_{0}^{1}\int\limits_{0}^{1} dxdz 
z^{\Delta/m_\pi^2-1} \text{sech}\left(x\frac{eB}{m_\pi^2}\ln z\right) \int\frac{d^2k_\perp}{(2\pi)^2}e^\zeta
\left[ \frac{A^{\mu\nu}}{\ln z} + \frac{B^{\mu\nu}}{m_\pi^2} \right] . \nn
\end{eqnarray}
In order to perform the $d^2k_\perp$ integration, we rewrite the $\zeta$ by completing the square in $k_\perp$ as bellow,
\begin{eqnarray}
\zeta = -\tilde{\zeta}(k_\perp-yq_\perp)^2-y(1-x)\frac{q_\perp^2}{m_\pi^2}\ln z \nn
\end{eqnarray}
where $\tilde{\zeta} = (1-x)\frac{\ln z}{m_\pi^2} + \frac{1}{eB}\tanh\left(x\frac{eB}{m_\pi^2}\ln z\right)$ and 
$y=\frac{1}{\tilde{\zeta} eB}\tanh\left(x\frac{eB}{m_\pi^2}\ln z\right)$. Now shifting momenta $k_\perp\rightarrow(k_\perp+yq_\perp)$ and performning the remaining Gaussian integration of the variable $\left|\vec{k}_\perp\right|$ we finally obtain, 
\begin{eqnarray}
(\Pi^{\mu\nu}_\pm)_{B} = \left(\frac{g_{\rho\pi\pi}^2}{32\pi^2}\right)\int\limits_{0}^{1}\int\limits_{0}^{1}dxdz
z^{\Delta/m_\pi^2-y(1-x)q_\perp^2/m_\pi^2-1}\left(\frac{1}{\tilde{\zeta}}\right)\text{sech}\left(x\frac{eB}{m_\pi^2}\ln z\right)
\left[ \frac{P^{\mu\nu}}{\ln z} + \frac{2Q^{\mu\nu}}{m_\pi^2} + \frac{R^{\mu\nu}}{\tilde{\zeta} m_\pi^2} \right] \label{eq.pipmvac_appendix2},
\end{eqnarray}
where, 
\begin{eqnarray}
P^{\mu\nu} &=& q^4g_\parallel^{\mu\nu} + q_\parallel^2 q^\mu q^\nu - q^2\left(q^\mu q_\parallel^\nu+q^\nu q_\parallel^\mu\right) \label{eq.app.P} \\
Q^{\mu\nu} &=& q^4\left[ x^2q_\parallel^\mu q_\parallel^\nu + y^2q_\perp^\mu q_\perp^\nu 
+ xy\left( q_\parallel^\mu q_\perp^\nu + q_\parallel^\nu q_\perp^\mu \right) \right] 
+ q^\mu q^\nu \left( xq_\parallel^2+yq_\perp^2 \right)^2 \nonumber \\
&&- q^2\left( xq_\parallel^2+yq_\perp^2 \right)
\left\{ x\left(q^\mu q_\parallel^\nu + q^\nu q_\parallel^\mu\right)+y\left(q^\mu q_\perp^\nu + q^\nu q_\perp^\mu\right)\right\} 
\label{eq.app.Q} \\
R^{\mu\nu} &=& q^4g_\perp^{\mu\nu} + q_\perp^2 q^\mu q^\nu - q^2\left(q^\mu q_\perp^\nu+q^\nu q_\perp^\mu\right)  \label{eq.app.R}
\end{eqnarray}
In the limit of zero magnetic field, we get back the vacuum self energy as given in Eq.~(\ref{eq.vacself.rho}), i.e. 
\begin{eqnarray}
\lim\limits_{eB\rightarrow0} (\Pi^{\mu\nu}_\pm)_{B} = \left(\frac{g_{\rho\pi\pi}^2q^2}{32\pi^2}\right)
\left( q^2g^{\mu\nu}-q^\mu q^\nu \right)\int\limits_{0}^{1}\int\limits_{0}^{1}dxdzz^{\Delta/m_\pi^2-1}\frac{m_\pi^2}{(\ln z)^2} = (\Pi^{\mu\nu})_{vac}. \label{eq.eb0limitpm}
\end{eqnarray}
Using Eq.~(\ref{eq.pipmvac_appendix2}) and (\ref{eq.eb0limitpm}), we arrive at Eq.~(\ref{eq.re.pibarpm.eb}). 
\section{Kinematic Domains of the Imaginary Parts}\label{appendix.c}
The imaginary part of the in-medium $\rho$ self energy function at \textit{zero magnetic field} in Eq.~(\ref{eq.im.pibar.therm}) contains four Dirac delta functions namely
$\delta\left(q^0\mp\omega_k\mp\omega_p\right) = \delta\left( q^0\mp E\right)$ and $\delta\left(q^0\mp\omega_k\pm\omega_p\right) = \delta\left( q^0\mp E^\prime\right)$, 
where $E=\omega_k+\omega_p$ and $E^\prime=\omega_k-\omega_p$. The functions $E=E\left(|\vec{k}|,\cos\theta\right)$ and $E^\prime=E^\prime\left(|\vec{k}|,\cos\theta\right)$ both are defined in the domain $0\le|\vec{k}|<\infty$ and $|\cos\theta|\le1$. In this domain the ranges of this two functions are found to be,
\begin{eqnarray}
\sqrt{\vec{q}^2+4m_\pi^2}\le E<\infty ~~~~~ \text{and}~~~~~ -|\vec{q}|\le E^\prime \le 0 \label{eq.kinematic.domain.appendix}
\end{eqnarray}
respectively. So, from Eq.~(\ref{eq.kinematic.domain.appendix}) it is evident that, for $\delta\left(q^0-E\right)$ and $\delta\left(q^0+E\right)$ to be non-vanishing, we must have $\sqrt{\vec{q}^2+4m_\pi^2} \le q^0 <\infty$ and $-\infty<q^0\le-\sqrt{\vec{q}^2+4m_\pi^2}$ respectively. Similarly for $\delta\left(q^0-E^\prime\right)$ and $\delta\left(q^0+E^\prime\right)$ to be non-vanishing, we must have $-|\vec{q}|\le q^0\le0$ and $0\le q^0\le|\vec{q}|$ respectively.

The imaginary part of the in-medium $\rho^0$ self energy function at \textit{non-zero magnetic field} in Eq.~(\ref{eq.im.pibar0.eb.t}) contains four Dirac delta functions for a particular set of 
Landau levels $\{n,l\}$ namely $\delta\left(q^0\mp\omega_k^l\mp\omega_p^n\right) = \delta\left( q^0\mp E_{n,l}\right)$ and $\delta\left(q^0\mp\omega_k^l\pm\omega_p^n\right) = \delta\left( q^0\mp E^\prime_{n,l}\right)$, where  $E_{n,l}=\omega_k^l+\omega_p^n$ and $E^\prime_{n,l}=\omega_k^l-\omega_p^n$. The functions $E_{n,l}=E_{n,l}\left(k_z\right)$ and $E^\prime_{n,l}=E^\prime_{n,l}\left(k_z\right)$ both are defined in the domain $-\infty<k_z<\infty$. In this given domain the ranges of this two functions are found to be,
\begin{eqnarray}
\sqrt{q_z^2+\left(m_l+m_n\right)^2}\le E_{n,l}<\infty ~~~~\text{and}~~~~\min\left[ q_z,E^\pm_{n,l}\right] \le E^\prime_{n,l} \le \max\left[ q_z,E^\pm_{n,l} \right] \label{eq.kinematic.domain.appendix2}
\end{eqnarray}
where, $E^\pm_{n,l} = \frac{\left(m_l-m_n\right)}{\left|m_l\pm m_n\right|}\sqrt{q_z^2+\left(m_l\pm m_n\right)^2}$. So from 
Eq.~(\ref{eq.kinematic.domain.appendix2}), we find that, for a particular set of Landau levels $\{n,l\}$, for 
$\delta\left(q^0-E_{n,l}\right)$ and $\delta\left(q^0+E_{n,l}\right)$ to be non-vanishing, we have 
$\sqrt{q_z^2+\left(m_l+m_n\right)^2} \le q^0 <\infty$ and $-\infty<q^0\le-\sqrt{q_z^2+\left(m_l+m_n\right)^2}$ respectively. 
Similarly for $\delta\left(q^0-E^\prime_{n,l}\right)$ and $\delta\left(q^0+E^\prime_{n,l}\right)$ to be non-vanishing, 
the corresponding ranges are  $\min\left[ q_z,E^\pm_{n,l}\right] \le q^0 \le \max\left[ q_z,E^\pm_{n,l}\right]$ 
and $-\max\left[ q_z,E^\pm_{n,l}\right]\le q^0\le -\min\left[ q_z,E^\pm_{n,l}\right]$ respectively.

In Eq.~(\ref{eq.im.pibar0.eb.t}), the indices $l$ and $n$ run from $0$ to $\infty$. However for a particular value 
of $l$, $n$ can have only three values $l-1$, $l$ and $l+1$ due to the presence of Kronecker delta function in Eq.~(\ref{eq.N_nl}). Hence, when 
these indices are summed over from $0$ $\rightarrow$ $\infty$, $\delta\left(q^0-\omega_k^l-\omega_p^n\right)$ and
$\delta\left(q^0+\omega_k^l+\omega_p^n\right)$ will be non-vanishing for $\sqrt{q_z^2+4(m_\pi^2+eB)}<q^0<\infty$ and $-\infty<q^0<-\sqrt{q_z^2+4(m_\pi^2+eB)}$ respectively whereas, both $\delta\left(q^0-\omega_k^l+\omega_p^n\right)$ and $\delta\left(q^0+\omega_k^l-\omega_p^n\right)$ will be non-vanishing at 
$|q^0|<\sqrt{q_z^2+( \sqrt{m_\pi^2+eB}-\sqrt{m_\pi^2+3eB})^2}$.

Again the imaginary part of the in-medium $\rho^\pm$ self energy function at non-zero magnetic field in Eq.~(\ref{eq.im.pibarpm.eb.t}) contains four Dirac delta functions for a particular 
Landau level $n$ namely $\delta\left(q^0\mp\omega_k\mp\omega_p^n\right) = \delta\left( q^0\mp E_n\right)$ and 
$\delta\left(q^0\mp\omega_k\pm\omega_p^n\right) = \delta\left( q^0\mp E^\prime_n\right)$, where  $E_n=\omega_k+\omega_p^n$ and $E^\prime_n=\omega_k-\omega_p^n$. 
The functions $E_n=E_n\left(|\vec{k}|,\cos\theta\right)$ and $E^\prime_n=E^\prime_n\left(|\vec{k}|,\cos\theta\right)$ both are defined in the domain
 $0\le|\vec{k}|<\infty$ and $|\cos\theta|\le1$. In this domain the ranges of these two functions are found to be,
\begin{eqnarray}
\sqrt{q_z^2+\left(m_\pi+m_n\right)^2}\le E_n<\infty ~~~~~ \text{and}~~~~~ -\sqrt{q_z^2+\left(m_\pi-m_n\right)^2}\le E^\prime_n \le 0 \label{eq.kinematic.domain.appendix3}
\end{eqnarray}
respectively. So from Eq.~(\ref{eq.kinematic.domain.appendix3}), we find that, for a particular Landau level $n$, for $\delta\left(q^0-E_n\right)$ and 
$\delta\left(q^0+E_n\right)$ to be non-zero, $q^0$ must lie in the range $\sqrt{q_z^2+\left(m_\pi+m_n\right)^2}\le q^0<\infty$ and 
$-\infty<q^0\le-\sqrt{q_z^2+\left(m_\pi+m_n\right)^2}$ respectively. Similarly, for $\delta\left(q^0-E^\prime_n\right)$ 
and $\delta\left(q^0+E^\prime_n\right)$ to be non-vanishing, the inequalities $-\sqrt{q_z^2+\left(m_\pi-m_n\right)^2}\le q^0 \le 0$ 
and $0\le q^0 \le \sqrt{q_z^2+\left(m_\pi-m_n\right)^2}$ must be satisfied.

In Eq.~(\ref{eq.im.pibarpm.eb.t}), the index $n$ runs from $0$ to $\infty$ and when it is summed over,  $\delta\left(q^0-\omega_k-\omega_p^n\right)$ and
$\delta\left(q^0+\omega_k+\omega_p^n\right)$ will be non-vanishing at $\sqrt{q_z^2+(\sqrt{m_\pi^2+eB}+m_\pi)^2}<q^0<\infty$ and $-\infty<q^0<-\sqrt{q_z^2+(\sqrt{m_\pi^2+eB}+m_\pi)^2}$ respectively whereas, $\delta\left(q^0-\omega_k+\omega_p^n\right)$ and $\delta\left(q^0+\omega_k-\omega_p^n\right)$ will be non-vanishing 
at $0<q^0<\infty$ and $-\infty<q^0<0$ respectively.
%
%
%
%++++++++++++++++++++++++++++++++++++++++++++++++++++++++++++++++++++++++++++++++++++++++++++++++++++++++++++++++++++++++++++++++++++++++++
%
\section{Simplification of the Imaginary Parts}\label{appendix.d}
The imaginary part of in-medium self energy function of $\rho$ at \textit{zero magnetic field} in Eq.~(\ref{eq.im.pibar.therm}) can be written as,
\begin{eqnarray}
\text{Im}~\bar{\Pi}^{\mu\nu}(q) &=& -\pi\epsilon(q^0)\int\limits_{0}^{\infty}\int\limits_{0}^{\pi}\int\limits_{0}^{2\pi}\frac{|\vec{k}|^2\sin\theta d|\vec{k}|d\theta d\phi}{(2\pi)^34\omega_k\omega_p} \left[ \frac{}{}U_1\left(q,\vec{k}\right)\delta\left(q^0-\omega_k-\omega_p\right) + U_2\left(q,\vec{k}\right)\delta\left(q^0+\omega_k+\omega_p\right) \right. \nonumber \\
&& \hspace{4cm} \left. +~L_1\left(q,\vec{k}\right)\delta\left(q^0+\omega_k-\omega_p\right) + L_2\left(q,\vec{k}\right)\delta\left(q^0-\omega_k+\omega_p\right)\frac{}{} \right],
\label{eq.im.pibar.therm.appendix}
\end{eqnarray}
where, 
\begin{eqnarray}
U_1 &=& \left(1+\eta^k+\eta^p\right)\mathcal{N}^{\mu\nu}\left(q,k^0=\omega_k,\vec{k}\right) \nonumber \\
U_2 &=& \left(-1-\eta^k-\eta^p\right)\mathcal{N}^{\mu\nu}\left(q,k^0=-\omega_k,\vec{k}\right) \nonumber \\
L_1 &=& \left(\eta^k-\eta^p\right)\mathcal{N}^{\mu\nu}\left(q,k^0=-\omega_k,\vec{k}\right) \nonumber \\
L_2 &=& \left(-\eta^k+\eta^p\right)\mathcal{N}^{\mu\nu}\left(q,k^0=\omega_k,\vec{k}\right) \nn
\end{eqnarray}
For simplicity, we have taken $\vec{q}=\vec{0}$ and it is to be noted that, the nonzero components of $\mathcal{N}^{\mu\nu}\left(q^0,\vec{q}=\vec{0},\vec{k}\right)$ are independent of $\theta$ and $\phi$. In Appendix~(\ref{appendix.c}) it is shown that, for $\vec{q}=\vec{0}$ only Unitary terms contribute. So Eq.~(\ref{eq.im.pibar.therm.appendix}) simplifies to,
\begin{eqnarray}
\text{Im}~\bar{\Pi}^{\mu\nu}(q^0) &=& \frac{-\epsilon(q^0)}{8\pi}\int\limits_{0}^{\infty}\frac{|\vec{k}|^2d|\vec{k}|}{\omega_k^2}
 \left[ U_1\left(q^0,|\vec{k}|\right)\delta\left(q^0-2\omega_k\right)\Theta\left(q^0-2m_\pi\right) \right. \nonumber \\
&& \hspace{3cm}\left. +~ U_2\left(q^0,|\vec{k}|\right)\delta\left(q^0+2\omega_k\right)\Theta\left(-q^0-2m_\pi\right) \right]. \nn
\end{eqnarray}
Transforming the Dirac delta functions $\delta\left(q^0\mp2\omega_k\right) = \left(\frac{\omega_k}{2|\vec{k}|}\right)\delta\left(|\vec{k}|-\tilde{k}\right)$ 
where $\tilde{k}=\left(\frac{1}{2q^0}\right)\lambda^{1/2}\left( q_0^2,m_\pi^2,m_\pi^2\right)$ and 
performing the remaining $|\vec{k}|$ integration we arrive at Eq.~(\ref{eq.impi.simple}). Here $\Theta(x)$ is the unit step function and 
$\lambda(x,y,z)=\left(x^2+y^2+z^2-2xy-2yz-2zx\right)$ is the Kallen function.

We now turn on the magnetic field. The imaginary part of in-medium self energy function of $\rho^0$ at \textit{non-zero magnetic field} in Eq.~(\ref{eq.im.pibar0.eb.t}) for $\vec{q}=\vec{0}$ can be written as,
\begin{eqnarray}
\text{Im}~\bar{\Pi}^{\mu\nu}_0 &=& -\pi\epsilon(q^0)\sum\limits_{l=0}^{\infty}\sum\limits_{n=0}^{\infty}
\int\limits_{-\infty}^{+\infty}\frac{dk_z}{(2\pi)}\frac{1}{4\omega_k^l\omega_k^n} 
 \left[ \frac{}{}U_1^{n,l}\left(q^0,k_z\right)\delta\left(q^0-\omega_k^l-\omega_k^n\right) + U_2^{n,l}\left(q^0,k_z\right)\delta\left(q^0+\omega_k^l+\omega_k^n\right) \right. \nonumber \\
&& \hspace{3cm}\left. + ~L_1^{n,l}\left(q^0,k_z\right)\delta\left(q^0+\omega_k^l-\omega_k^n\right) + L_2^{n,l}\left(q^0,k_z\right)\delta\left(q^0-\omega_k^l+\omega_k^n\right) \right] \nn
\end{eqnarray}
where, 
\begin{eqnarray}
U_1^{n,l} &=& \left(1+\eta^k_l+\eta^k_n\right)\mathcal{N}^{\mu\nu}_{n,l}\left(q,k^0=\omega_k^l,k_z\right) \nonumber \\
U_2^{n,l} &=& \left(-1-\eta^k_l-\eta^k_n\right)\mathcal{N}^{\mu\nu}_{n,l}\left(q,k^0=-\omega_k^l,k_z\right) \nonumber \\
L_1^{n,l} &=& \left(\eta^k_l-\eta^k_n\right)\mathcal{N}^{\mu\nu}_{n,l}\left(q,k^0=-\omega_k^l,k_z\right) \nonumber \\
L_2^{n,l} &=& \left(-\eta^k_l+\eta^k_n\right)\mathcal{N}^{\mu\nu}_{n,l}\left(q,k^0=\omega_k^l,k_z\right) \nn
\end{eqnarray}
Now we transform the Dirac delta functions $\delta\left(q^0\mp\omega_k^l\mp\omega_k^n\right) = \delta\left(q^0\mp\omega_k^l\pm\omega_k^n\right) =
\left(\frac{\omega_k^l\omega_p^n}{|k_zq^0|}\right)\left[\delta\left(k_z-\tilde{k}_z\right)+\delta\left(k_z+\tilde{k}_z\right)\right]$ 
where $\tilde{k}_z=\left(\frac{1}{2q^0}\right)\lambda^{1/2}\left( q_0^2,m_l^2,m_n^2\right)$ and impose the kinematic domains as obtained in Appendix~(\ref{appendix.c}). 
After performing the $k_z$ integration we arrive at Eq.~(\ref{eq.impi0.simple}).

We now look at the charged $\rho$. The imaginary part of in-medium self energy function of $\rho^\pm$ at \textit{non-zero magnetic field} in Eq.~(\ref{eq.im.pibarpm.eb.t}) for $\vec{q}=(q_x,q_y,0)$ can be written as,
\begin{eqnarray}
\text{Im}~\bar{\Pi}^{\mu\nu}_\pm &=& -\pi\epsilon(q^0)\sum\limits_{n=0}^{\infty}
\int\limits_{0}^{\infty}\int\limits_{0}^{\pi}\int\limits_{0}^{2\pi} \frac{|\vec{k}|^2\sin\theta d\theta d\phi}{(2\pi)^3} \frac{1}{4\omega_k\omega_k^n} 
\left[ \frac{}{}U_1^n\left(q^0,\vec{k}\right)\delta\left(q^0-\omega_k-\omega_k^n\right) + U_2^n\left(q^0,\vec{k}\right)\delta\left(q^0+\omega_k+\omega_k^n\right) \right. \nonumber \\
&& \hspace{3cm}\left. +~ L_1^n\left(q^0,\vec{k}\right)\delta\left(q^0+\omega_k-\omega_k^n\right) + L_2^n\left(q^0,\vec{k}\right)\delta\left(q^0-\omega_k+\omega_k^n\right) \right]
\label{eq.im.pibarpm.eb.t.appendix}
\end{eqnarray}
where, 
\begin{eqnarray}
U_1^n &=& \phi_n(\alpha_k)\left(1+\eta^k+\eta^k_n\right)\mathcal{N}^{\mu\nu}\left(q,k^0=\omega_k,\vec{k}\right) \nonumber \\
U_2^n &=& \phi_n(\alpha_k)\left(-1-\eta^k-\eta^k_n\right)\mathcal{N}^{\mu\nu}\left(q,k^0=-\omega_k,\vec{k}\right) \nonumber \\
L_1^n &=& \phi_n(\alpha_k)\left(\eta^k-\eta^k_n\right)\mathcal{N}^{\mu\nu}\left(q,k^0=-\omega_k,\vec{k}\right) \nonumber \\
L_2^n &=& \phi_n(\alpha_k)\left(-\eta^k+\eta^k_n\right)\mathcal{N}^{\mu\nu}\left(q,k^0=\omega_k,\vec{k}\right) \nn
\end{eqnarray}
Now we write the Dirac delta functions as,
\begin{eqnarray}
\delta\left(q^0-\omega_k\mp\omega_k^n\right) = \left(\frac{\omega_k^n}{|\vec{k}|^2\cos\theta_0}\right)
\left[\frac{}{}\delta\left(\cos\theta-\cos\theta_0\right)+\delta\left(\cos\theta-\cos\theta_0\right)\right] \nonumber \\
\delta\left(q^0+\omega_k\mp\omega_k^n\right) = \left(\frac{\omega_k^n}{|\vec{k}|^2\cos\theta^\prime_0}\right)
\left[\frac{}{}\delta\left(\cos\theta-\cos\theta^\prime_0\right)+\delta\left(\cos\theta-\cos\theta^\prime_0\right)\right] \nonumber \\
\end{eqnarray}
where $\cos\theta_0 = \frac{1}{|\vec{k}|}\sqrt{\left(q^0+\omega_k\right)^2-m_n^2}$ and $\cos\theta^\prime_0 = \frac{1}{|\vec{k}|}\sqrt{\left(q^0-\omega_k\right)^2-m_n^2}$. Changing the variable of integration from $|\vec{k}|$ to $\omega_k$ and performing the $\theta$ integration using the transformed Dirac delta function and imposing the kinematic domains as obtained in Appendix~(\ref{appendix.c}), Eq.~(\ref{eq.im.pibarpm.eb.t.appendix}) becomes,
\begin{eqnarray}
\text{Im}~\bar{\Pi}^{\mu\nu}_\pm &=& \frac{-\epsilon(q^0)}{32\pi^2}\sum\limits_{n=0}^{\infty}~\int\limits_{m_\pi}^{\infty}\frac{d\omega_k}{|\vec{k}|}\int\limits_{0}^{2\pi}d\phi
\left[ \frac{\Theta\left(1-\cos\theta_0\right)}{\cos\theta_0}\left\{U_1^n\left(q^0,|\vec{k}|,\theta_0,\phi\right)+U_1^n\left(q^0,|\vec{k}|,-\theta_0,\phi\right)\right\}\Theta\left(q^0-m_\pi-m_n\right) \right. \nonumber \\
&& \hspace{1.5cm}\left. +~ \frac{\Theta\left(1-\cos\theta^\prime_0\right)}{\cos\theta^\prime_0}\left\{U_2^n\left(q^0,|\vec{k}|,\theta^\prime_0,\phi\right)+U_2^n\left(q^0,|\vec{k}|,-\theta^\prime_0,\phi\right)\right\}\Theta\left(-q^0-m_\pi-m_n\right) \right. \nonumber \\ 
&& \hspace{1.5cm} \left. +~ \frac{\Theta\left(1-\cos\theta^\prime_0\right)}{\cos\theta^\prime_0}\left\{L_1^n\left(q^0,|\vec{k}|,\theta^\prime_0,\phi\right)+L_1^n\left(q^0,|\vec{k}|,-\theta^\prime_0,\phi\right)\right\}\Theta\left(-q^0-m_\pi+m_n\right)\Theta(q^0) \right. \nonumber \\ 
&&\hspace{1.5cm} \left. +~ \frac{\Theta\left(1-\cos\theta_0\right)}{\cos\theta_0}\left\{L_2^n\left(q^0,|\vec{k}|,\theta_0,\phi\right)+L_2^n\left(q^0,|\vec{k}|,-\theta_0,\phi\right)\right\}\Theta\left(q^0-m_\pi+m_n\right)\Theta(-q^0) \right].\label{eq.im.pibarpm.eb.t.appendix2}
\end{eqnarray}
In Eq.~(\ref{eq.im.pibarpm.eb.t.appendix2}), the presence of the $\Theta\left( 1-\cos\theta_0 \right)$ and $\Theta\left( 1-\cos\theta^\prime_0 \right)$ will 
further put restriction on the limits of the $\omega_k$ integration and we will get Eq.~(\ref{eq.impipm.simple}).


\begin{thebibliography}{99}

  
%\cite{Kharzeev:2012ph}
\bibitem{Kharzeev:2012ph} 
  D.~E.~Kharzeev, K.~Landsteiner, A.~Schmitt and H.~U.~Yee,
  %``'Strongly interacting matter in magnetic fields': an overview,''
  Lect.\ Notes Phys.\  {\bf 871}, 1 (2013)
%  doi:10.1007/978-3-642-37305-3_1
  [arXiv:1211.6245 [hep-ph]].
  %%CITATION = doi:10.1007/978-3-642-37305-3_1;%%
  %137 citations counted in INSPIRE as of 10 Apr 2017


%\cite{Kharzeev:2007tn}
\bibitem{Kharzeev:2007tn} 
  D.~Kharzeev and A.~Zhitnitsky,
  %``Charge separation induced by P-odd bubbles in QCD matter,''
  Nucl.\ Phys.\ A {\bf 797}, 67 (2007)
%  doi:10.1016/j.nuclphysa.2007.10.001
  [arXiv:0706.1026 [hep-ph]].
  
  
%\cite{Kharzeev:2007jp}
\bibitem{Kharzeev:2007jp} 
  D.~E.~Kharzeev, L.~D.~McLerran and H.~J.~Warringa,
  %``The Effects of topological charge change in heavy ion collisions: 'Event by event P and CP violation',''
  Nucl.\ Phys.\ A {\bf 803}, 227 (2008)
%  doi:10.1016/j.nuclphysa.2008.02.298
  [arXiv:0711.0950 [hep-ph]].
  %%CITATION = doi:10.1016/j.nuclphysa.2008.02.298;%%
  %870 citations counted in INSPIRE as of 10 Apr 2017
  
  
 %\cite{Fukushima:2008xe}
\bibitem{Fukushima:2008xe} 
    K.~Fukushima, D.~E.~Kharzeev and H.~J.~Warringa,
    %``The Chiral Magnetic Effect,''
    Phys.\ Rev.\ D {\bf 78}, 074033 (2008)
%    doi:10.1103/PhysRevD.78.074033
    [arXiv:0808.3382 [hep-ph]].
    %%CITATION = doi:10.1103/PhysRevD.78.074033;%%
    %816 citations counted in INSPIRE as of 10 Apr 2017
    
 %\cite{Gusynin:1995nb}
\bibitem{Gusynin:1995nb} 
  V.~P.~Gusynin, V.~A.~Miransky and I.~A.~Shovkovy,
   %``Dimensional reduction and catalysis of dynamical symmetry breaking by a magnetic field,''
   Nucl.\ Phys.\ B {\bf 462}, 249 (1996)
%   doi:10.1016/0550-3213(96)00021-1
    [hep-ph/9509320].
    %%CITATION = doi:10.1016/0550-3213(96)00021-1;%%
    %352 citations counted in INSPIRE as of 10 Apr 2017
      
%\cite{Gusynin:1999pq}
\bibitem{Gusynin:1999pq} 
    V.~P.~Gusynin, V.~A.~Miransky and I.~A.~Shovkovy,
    %``Theory of the magnetic catalysis of chiral symmetry breaking in QED,''
    Nucl.\ Phys.\ B {\bf 563}, 361 (1999)
%    doi:10.1016/S0550-3213(99)00573-8
    [hep-ph/9908320].
     %%CITATION = doi:10.1016/S0550-3213(99)00573-8;%%
     %98 citations counted in INSPIRE as of 10 Apr 2017
        
%\cite{Bali:2011qj}
\bibitem{Bali:2011qj} 
  G.~S.~Bali, F.~Bruckmann, G.~Endrodi, Z.~Fodor, S.~D.~Katz, S.~Krieg, A.~Schafer and K.~K.~Szabo,
  %``The QCD phase diagram for external magnetic fields,''
  JHEP {\bf 1202}, 044 (2012)
%  doi:10.1007/JHEP02(2012)044
  [arXiv:1111.4956 [hep-lat]].
  %%CITATION = doi:10.1007/JHEP02(2012)044;%%
  %314 citations counted in INSPIRE as of 10 Apr 2017

%\cite{Chernodub:2010qx}
\bibitem{Chernodub:2010qx} 
  M.~N.~Chernodub,
  %``Superconductivity of QCD vacuum in strong magnetic field,''
  Phys.\ Rev.\ D {\bf 82}, 085011 (2010)
%  doi:10.1103/PhysRevD.82.085011
  [arXiv:1008.1055 [hep-ph]].
  %%CITATION = doi:10.1103/PhysRevD.82.085011;%%
  %162 citations counted in INSPIRE as of 10 Apr 2017
  
%\cite{Chernodub:2012tf}
\bibitem{Chernodub:2012tf} 
  M.~N.~Chernodub,
  %``Electromagnetic superconductivity of vacuum induced by strong magnetic field,''
  Lect.\ Notes Phys.\  {\bf 871}, 143 (2013)
%  doi:10.1007/978-3-642-37305-3_6
  [arXiv:1208.5025 [hep-ph]].
  %%CITATION = doi:10.1007/978-3-642-37305-3_6;%%
  %18 citations counted in INSPIRE as of 10 Apr 2017
  
%\cite{Skokov:2009qp}
\bibitem{Skokov:2009qp} 
  V.~Skokov, A.~Y.~Illarionov and V.~Toneev,
  %``Estimate of the magnetic field strength in heavy-ion collisions,''
  Int.\ J.\ Mod.\ Phys.\ A {\bf 24}, 5925 (2009)
%  doi:10.1142/S0217751X09047570
  [arXiv:0907.1396 [nucl-th]].
  %%CITATION = doi:10.1142/S0217751X09047570;%%
  %446 citations counted in INSPIRE as of 10 Apr 2017
  
%\cite{Duncan:1992hi}
\bibitem{Duncan:1992hi} 
  R.~C.~Duncan and C.~Thompson,
  %``Formation of very strongly magnetized neutron stars - implications for gamma-ray bursts,''
  Astrophys.\ J.\  {\bf 392}, L9 (1992).
%  doi:10.1086/186413
  %%CITATION = doi:10.1086/186413;%%
  %1225 citations counted in INSPIRE as of 10 Apr 2017
  
%\cite{Ferrer:2005vd}
\bibitem{Ferrer:2005vd} 
  E.~J.~Ferrer, V.~de la Incera and C.~Manuel,
  %``Magnetic color flavor locking phase in high density QCD,''
  Phys.\ Rev.\ Lett.\  {\bf 95}, 152002 (2005)
%  doi:10.1103/PhysRevLett.95.152002
  [hep-ph/0503162].
  %%CITATION = doi:10.1103/PhysRevLett.95.152002;%%
  %132 citations counted in INSPIRE as of 10 Apr 2017
  
%\cite{Ferrer:2006vw}
\bibitem{Ferrer:2006vw} 
  E.~J.~Ferrer, V.~de la Incera and C.~Manuel,
  %``Color-superconducting gap in the presence of a magnetic field,''
  Nucl.\ Phys.\ B {\bf 747}, 88 (2006)
%  doi:10.1016/j.nuclphysb.2006.04.013
  [hep-ph/0603233].
  %%CITATION = doi:10.1016/j.nuclphysb.2006.04.013;%%
  %104 citations counted in INSPIRE as of 10 Apr 2017
  
  
%\cite{Ferrer:2007iw}
\bibitem{Ferrer:2007iw} 
  E.~J.~Ferrer and V.~de la Incera,
  %``Magnetic Phases in Three-Flavor Color Superconductivity,''
  Phys.\ Rev.\ D {\bf 76}, 045011 (2007)
%  doi:10.1103/PhysRevD.76.045011
  [nucl-th/0703034 [NUCL-TH]].
  %%CITATION = doi:10.1103/PhysRevD.76.045011;%%
  %88 citations counted in INSPIRE as of 10 Apr 2017
  

%\cite{Fukushima:2007fc}
\bibitem{Fukushima:2007fc} 
  K.~Fukushima and H.~J.~Warringa,
  %``Color superconducting matter in a magnetic field,''
  Phys.\ Rev.\ Lett.\  {\bf 100}, 032007 (2008)
%  doi:10.1103/PhysRevLett.100.032007
  [arXiv:0707.3785 [hep-ph]].
  %%CITATION = doi:10.1103/PhysRevLett.100.032007;%%
  %96 citations counted in INSPIRE as of 10 Apr 2017
  
  
%\cite{Feng:2009vt}
\bibitem{Feng:2009vt} 
  B.~Feng, D.~Hou, H.~c.~Ren and P.~p.~Wu,
  %``The Single Flavor Color Superconductivity in a Magnetic Field,''
  Phys.\ Rev.\ Lett.\  {\bf 105}, 042001 (2010)
%  doi:10.1103/PhysRevLett.105.042001
  [arXiv:0911.4997 [hep-ph]].
  %%CITATION = doi:10.1103/PhysRevLett.105.042001;%%
  %14 citations counted in INSPIRE as of 10 Apr 2017
  
%\cite{Fayazbakhsh:2010gc}
\bibitem{Fayazbakhsh:2010gc} 
  S.~Fayazbakhsh and N.~Sadooghi,
  %``Color neutral 2SC phase of cold and dense quark matter in the presence of constant magnetic fields,''
  Phys.\ Rev.\ D {\bf 82}, 045010 (2010)
%  doi:10.1103/PhysRevD.82.045010
  [arXiv:1005.5022 [hep-ph]].
  %%CITATION = doi:10.1103/PhysRevD.82.045010;%%
  %44 citations counted in INSPIRE as of 10 Apr 2017
  
%\cite{Fayazbakhsh:2010bh}
\bibitem{Fayazbakhsh:2010bh} 
  S.~Fayazbakhsh and N.~Sadooghi,
  %``Phase diagram of hot magnetized two-flavor color superconducting quark matter,''
  Phys.\ Rev.\ D {\bf 83}, 025026 (2011)
%  doi:10.1103/PhysRevD.83.025026
  [arXiv:1009.6125 [hep-ph]].
  %%CITATION = doi:10.1103/PhysRevD.83.025026;%%
  %55 citations counted in INSPIRE as of 10 Apr 2017
  
%+++++++++++++++++++++++++++++++++++++++++++++++++++++++++++++++++

%..................................................................................ARGHYA INCLUDED............................................
% %\bibitem{Andersen_review}
% %J. O. Andersen and W. R. Naylor, Rev. Mod. Phys. \textbf{ 88}, 025001.

%\bibitem{wangprd86}
% %K. L. Wang, S. X. Qin, Y. X. Liu, L. Chang, C. D. Roberts and S. M. Schmidt, Phys. Rev.  D \textbf{86}, 114001(2012)

%%%%%%%%%%%%%%%%%%%%%%%%%%%%%%%%%%%%%%%%%%%%%%%%%%%%%%%%%%%%%%%%%%%%%%%%%%%%%%%%%%%%%%%%

%\cite{Andersen:2014xxa}
\bibitem{Andersen_review} 
  J.~O.~Andersen, W.~R.~Naylor and A.~Tranberg,
  %``Phase diagram of QCD in a magnetic field: A review,''
  Rev.\ Mod.\ Phys.\  {\bf 88}, 025001 (2016)
%  doi:10.1103/RevModPhys.88.025001
  [arXiv:1411.7176 [hep-ph]].
  %%CITATION = doi:10.1103/RevModPhys.88.025001;%%
  %79 citations counted in INSPIRE as of 18 Apr 2017

%\cite{Wang:2012me}
\bibitem{wangprd86} 
  K.~l.~Wang, S.~x.~Qin, Y.~x.~Liu, L.~Chang, C.~D.~Roberts and S.~M.~Schmidt,
  %``Existence and stability of multiple solutions to the gap equation,''
  Phys.\ Rev.\ D {\bf 86}, 114001 (2012)
%  doi:10.1103/PhysRevD.86.114001
  [arXiv:1209.2757 [nucl-th]].
  %%CITATION = doi:10.1103/PhysRevD.86.114001;%%
  %19 citations counted in INSPIRE as of 18 Apr 2017

%...................................................................................................

%\cite{Vafa:1983tf}
\bibitem{Vafa:1983tf} 
  C.~Vafa and E.~Witten,
  %``Restrictions on Symmetry Breaking in Vector-Like Gauge Theories,''
  Nucl.\ Phys.\ B {\bf 234}, 173 (1984).
%  doi:10.1016/0550-3213(84)90230-X
  %%CITATION = doi:10.1016/0550-3213(84)90230-X;%%
  %542 citations counted in INSPIRE as of 10 Apr 2017

%\cite{Hidaka:2012mz}
\bibitem{Hidaka:2012mz} 
  Y.~Hidaka and A.~Yamamoto,
  %``Charged vector mesons in a strong magnetic field,''
  Phys.\ Rev.\ D {\bf 87}, no. 9, 094502 (2013)
%  doi:10.1103/PhysRevD.87.094502
  [arXiv:1209.0007 [hep-ph]].
  %%CITATION = doi:10.1103/PhysRevD.87.094502;%%
  %62 citations counted in INSPIRE as of 10 Apr 2017

%\cite{Chernodub:2012zx}
\bibitem{Chernodub:2012zx} 
  M.~N.~Chernodub,
  %``Vafa-Witten theorem, vector meson condensates and magnetic-field-induced electromagnetic superconductivity of vacuum,''
  Phys.\ Rev.\ D {\bf 86}, 107703 (2012)
%  doi:10.1103/PhysRevD.86.107703
  [arXiv:1209.3587 [hep-ph]].
  %%CITATION = doi:10.1103/PhysRevD.86.107703;%%
  %23 citations counted in INSPIRE as of 10 Apr 2017

%\cite{Li:2013aa}
\bibitem{Li:2013aa} 
  C.~Li and Q.~Wang,
  %``Amending the Vafa-Witten Theorem,''
  Phys.\ Lett.\ B {\bf 721}, 141 (2013)
%  doi:10.1016/j.physletb.2013.02.050
  [arXiv:1301.7009 [hep-th]].
  %%CITATION = doi:10.1016/j.physletb.2013.02.050;%%
  %11 citations counted in INSPIRE as of 10 Apr 2017

%\cite{Chernodub:2013uja}
\bibitem{Chernodub:2013uja} 
  M.~N.~Chernodub,
  %``Comment on “Charged vector mesons in a strong magnetic field”,''
  Phys.\ Rev.\ D {\bf 89}, no. 1, 018501 (2014)
%  doi:10.1103/PhysRevD.89.018501
  [arXiv:1309.4071 [hep-ph]].
  %%CITATION = doi:10.1103/PhysRevD.89.018501;%%
  %11 citations counted in INSPIRE as of 10 Apr 2017

%\cite{Liu:2014uwa}
\bibitem{Liu:2014uwa} 
  H.~Liu, L.~Yu and M.~Huang,
  %``Charged and neutral vector $\rho$ mesons in a magnetic field,''
  Phys.\ Rev.\ D {\bf 91}, no. 1, 014017 (2015)
%  doi:10.1103/PhysRevD.91.014017
  [arXiv:1408.1318 [hep-ph]].
  %%CITATION = doi:10.1103/PhysRevD.91.014017;%%
  %24 citations counted in INSPIRE as of 10 Apr 2017

%\cite{Kawaguchi:2015gpt}
\bibitem{Kawaguchi:2015gpt} 
  M.~Kawaguchi and S.~Matsuzaki,
  %``Vector meson masses from a hidden local symmetry in a constant magnetic field,''
  Phys.\ Rev.\ D {\bf 93}, no. 12, 125027 (2016)
%  doi:10.1103/PhysRevD.93.125027
  [arXiv:1511.06990 [hep-ph]].
  %%CITATION = doi:10.1103/PhysRevD.93.125027;%%
  %5 citations counted in INSPIRE as of 10 Apr 2017

%++++++++++++++++++++++++++++++++++++++++++++++++++++++++++++++++++++++++++++++++++++++++
%\cite{Rapp:1999ej}
\bibitem{Rapp:1999ej} 
  R.~Rapp and J.~Wambach,
  %``Chiral symmetry restoration and dileptons in relativistic heavy ion collisions,''
  Adv.\ Nucl.\ Phys.\  {\bf 25}, 1 (2000)
%  doi:10.1007/0-306-47101-9_1
  [hep-ph/9909229].
  %%CITATION = doi:10.1007/0-306-47101-9_1;%%
  %733 citations counted in INSPIRE as of 10 Apr 2017
  
%\cite{Alam:1999sc}
\bibitem{Alam:1999sc} 
  J.~Alam, S.~Sarkar, P.~Roy, T.~Hatsuda and B.~Sinha,
  %``Thermal photons and lepton pairs from quark gluon plasma and hot hadronic matter,''
  Annals Phys.\  {\bf 286}, 159 (2001)
%  doi:10.1006/aphy.2000.6091
  [hep-ph/9909267].
  %%CITATION = doi:10.1006/aphy.2000.6091;%%
  %136 citations counted in INSPIRE as of 10 Apr 2017  

%\cite{Mallik:2016anp}
\bibitem{Mallik:2016anp} 
  S.~Mallik and S.~Sarkar,
  \textit{``Hadrons at Finite Temperature,''} Cambridge University Press.
%  doi:10.1017/9781316535585
  %%CITATION = doi:10.1017/9781316535585;%%

%\cite{Ghosh:2009bt}
\bibitem{Ghosh:2009bt} 
  S.~Ghosh, S.~Sarkar and S.~Mallik,
  %``Analytic structure of rho meson propagator at finite temperature,''
  Eur.\ Phys.\ J.\ C {\bf 70}, 251 (2010)
%  doi:10.1140/epjc/s10052-010-1446-8
  [arXiv:0911.3504 [hep-ph]].
  %%CITATION = doi:10.1140/epjc/s10052-010-1446-8;%%
  %32 citations counted in INSPIRE as of 10 Apr 2017

%++++++++++++++++++++++++++++++++++++++++++++++++++++++++++++++++++++++++++++++++++

%..........................................................................arghya citation again


\bibitem{rhopaper}S. Ghosh, A. Mukherjee, M. Mandal, S. Sarkar and P. Roy, Phys. Rev.  D \textbf{94} 094043.




%\cite{Navarro:2010eu}
\bibitem{Navarro:2010eu} 
  J.~Navarro, A.~Sanchez, M.~E.~Tejeda-Yeomans, A.~Ayala and G.~Piccinelli,
  %``Symmetry restoration at finite temperature with weak magnetic fields,''
  Phys.\ Rev.\ D {\bf 82}, 123007 (2010)
%  doi:10.1103/PhysRevD.82.123007
  [arXiv:1007.4208 [hep-ph]].
  %%CITATION = doi:10.1103/PhysRevD.82.123007;%%
  %13 citations counted in INSPIRE as of 10 Apr 2017

%\cite{Ayala:2004dx}
\bibitem{Ayala:2004dx} 
  A.~Ayala, A.~Sanchez, G.~Piccinelli and S.~Sahu,
  %``Effective potential at finite temperature in a constant magnetic field. I. Ring diagrams in a scalar theory,''
  Phys.\ Rev.\ D {\bf 71}, 023004 (2005)
%  doi:10.1103/PhysRevD.71.023004
  [hep-ph/0412135].
  %%CITATION = doi:10.1103/PhysRevD.71.023004;%%
  %31 citations counted in INSPIRE as of 10 Apr 2017

%\cite{Bandyopadhyay:2016fyd}
\bibitem{Bandyopadhyay:2016fyd} 
  A.~Bandyopadhyay, C.~A.~Islam and M.~G.~Mustafa,
  %``Electromagnetic spectral properties and Debye screening of a strongly magnetized hot medium,''
  Phys.\ Rev.\ D {\bf 94},  114034 (2016)
%  doi:10.1103/PhysRevD.94.114034
  [arXiv:1602.06769 [hep-ph]].
  %%CITATION = doi:10.1103/PhysRevD.94.114034;%%
  %6 citations counted in INSPIRE as of 10 Apr 2017

%\cite{DOlivo:2002omk}
\bibitem{DOlivo:2002omk} 
  J.~C.~D'Olivo, J.~F.~Nieves and S.~Sahu,
  %``Field theory of the photon selfenergy in a medium with a magnetic field and the Faraday effect,''
  Phys.\ Rev.\ D {\bf 67}, 025018 (2003)
%  doi:10.1103/PhysRevD.67.025018
  [hep-ph/0208146].
  %%CITATION = doi:10.1103/PhysRevD.67.025018;%%
  %25 citations counted in INSPIRE as of 10 Apr 2017

%\cite{Hattori:2012je}
\bibitem{Hattori:2012je} 
  K.~Hattori and K.~Itakura,
  %``Vacuum birefringence in strong magnetic fields: (I) Photon polarization tensor with all the Landau levels,''
  Annals Phys.\  {\bf 330}, 23 (2013)
%  doi:10.1016/j.aop.2012.11.010
  [arXiv:1209.2663 [hep-ph]].
  %%CITATION = doi:10.1016/j.aop.2012.11.010;%%
  %32 citations counted in INSPIRE as of 10 Apr 2017

%++++++++++++++++++++++++++++++++++++++++++++++++++++++++++++++++++++++++++++++++++++++++


\bibitem{bellac} 
M. Le Bellac, 
\textit{``Thermal Field Theory"}, Cambridge University Press, Cambridge, England, 1996.

%\cite{Schwinger:1951nm}
\bibitem{Schwinger:1951nm} 
  J.~S.~Schwinger,
  %``On gauge invariance and vacuum polarization,''
  Phys.\ Rev.\  {\bf 82}, 664 (1951).
%  doi:10.1103/PhysRev.82.664
  %%CITATION = doi:10.1103/PhysRev.82.664;%%
  %4213 citations counted in INSPIRE as of 10 Apr 2017


%\cite{Ayala:2015qwa}
\bibitem{Ayala:2015qwa} 
  A.~Ayala, C.~A.~Dominguez, L.~A.~Hernandez, M.~Loewe, J.~C.~Rojas and C.~Villavicencio,
  %``Quark deconfinement and gluon condensate in a weak magnetic field from QCD sum rules,''
  Phys.\ Rev.\ D {\bf 92}, no. 1, 016006 (2015)
%  doi:10.1103/PhysRevD.92.016006
  [arXiv:1504.01308 [hep-ph]].
  %%CITATION = doi:10.1103/PhysRevD.92.016006;%%
  %12 citations counted in INSPIRE as of 10 Apr 2017


%++++++++++++++++++++++++++++++++++++++++++++++++++++++++++++++++++++++++++++++++++++++++++++++++

%\cite{Schwartz:2013pla}
\bibitem{Schwartz:2013pla} 
  M.~D.~Schwartz,
\textit{``Quantum Field Theory and the Standard Model"}, Cambridge University Press.
  %%CITATION = INSPIRE-1276589;%%
  %12 citations counted in INSPIRE as of 12 Apr 2017


\end{thebibliography}
\end{document}